\documentclass[11pt, a4paper]{article}
\pdfoutput=1

\usepackage{jheppub}

\usepackage[utf8]{inputenc}

\usepackage{mathtools}
\usepackage{dsfont}
\usepackage{mathdots}

\usepackage{tikz}
\usetikzlibrary{arrows,shapes}
\usetikzlibrary{trees}
\usetikzlibrary{patterns}
\usetikzlibrary{matrix,arrows} 				% For commutative diagram
\usetikzlibrary{positioning}				% For "above of=" commands
\usetikzlibrary{calc,through}				% For coordinates
\usetikzlibrary{decorations.pathreplacing}  % For curly braces
\usepackage{pgffor}							% For repeating patterns

\usetikzlibrary{decorations.pathmorphing}	% For Feynman Diagrams
\usetikzlibrary{decorations.markings}
\tikzset{
    vector/.style={decorate, decoration={snake}, draw},
	provector/.style={decorate, decoration={snake,amplitude=2.5pt}, draw},
	antivector/.style={decorate, decoration={snake,amplitude=-2.5pt}, draw},
        smallvector/.style={decorate, decoration={snake,amplitude=1.5pt,post length=0.5mm}, draw},
    fermion/.style={draw=black, postaction={decorate},
        decoration={markings,mark=at position .55 with {\arrow[draw=black]{>}}}},
    fermionbar/.style={draw=black, postaction={decorate},
        decoration={markings,mark=at position .55 with {\arrow[draw=black]{<}}}},
    fermionnoarrow/.style={draw=black},
    gluon/.style={decorate, draw=black,
        decoration={coil,amplitude=4pt, segment length=5pt}},
    scalar/.style={dashed,draw=black, postaction={decorate},
        decoration={markings,mark=at position .55 with {\arrow[draw=black]{>}}}},
    scalarbar/.style={dashed,draw=black, postaction={decorate},
        decoration={markings,mark=at position .55 with {\arrow[draw=black]{<}}}},
    scalarnoarrow/.style={dashed,draw=black},
    electron/.style={draw=black, postaction={decorate},
        decoration={markings,mark=at position .55 with {\arrow[draw=black]{>}}}},
    bigvector/.style={decorate, decoration={snake,amplitude=4pt}, draw},
    arrow/.style={draw=black, postaction={decorate},
        decoration={markings,mark=at position 1 with {\arrow[draw=black]{>}}}},
}

\tikzstyle{block} = [draw, rectangle, 
    minimum height=3em, minimum width=6earticlem]

\newcommand{\reef}[1]{(\ref{#1})}

\def\be{\begin{equation}}
\def\ee{\end{equation}}
\def\bea{\begin{eqnarray}}
\def\eea{\end{eqnarray}}
\def\ba{\begin{array}}
\def\ea{\end{array}}
\def\bd{\begin{displaymath}}
\def\ed{\end{displaymath}}

\def\>{\rangle} %right angle
\def\<{\langle} %left angle
\def\Dsl{D \hskip-.6em \raise1pt\hbox{$ / $ } }
\def\to{\rightarrow}

\newcommand{\eps}{\epsilon}

\title{All-Multiplicity One-Loop Amplitudes in Born-Infeld Electrodynamics from Generalized Unitarity}

\author{Henriette Elvang,}
\author{Marios Hadjiantonis,}
\author{Callum R.~T.~Jones}
\author{and Shruti Paranjape}

\affiliation{Leinweber Center for Theoretical Physics,\\ 
Department of Physics, University of Michigan,\\
450 Church St, Ann Arbor, MI 48109, USA}

\emailAdd{elvang@umich.edu}
\emailAdd{mhadjian@umich.edu}
\emailAdd{jonescal@umich.edu}
\emailAdd{shrpar@umich.edu}

\abstract{
We initiate a study of non-supersymmetric Born-Infeld electrodynamics in 4d at the quantum level. 
Explicit all-multiplicity expressions are calculated for the purely rational one-loop amplitudes in the self-dual ($++\ldots+$) and next-to-self-dual ($-+\ldots+$) helicity sectors. Using a supersymmetric decomposition, $d$-dimensional unitarity cuts of the integrand factorize into tree-amplitudes in a 
4d 
model of Born-Infeld photons coupled to a massive complex scalar. The two-scalar tree-amplitudes needed to construct the Born-Infeld integrand are computed using two complimentary approaches: (1) as a double-copy of Yang-Mills coupled to a massive adjoint scalar with a dimensionally reduced form of Chiral Perturbation Theory, and (2) by imposing consistency with low-energy theorems under a reduction from 4d to 3d and T-duality. The Born-Infeld integrand is integrated in $d=4-2\eps$ dimensions at order $\mathcal{O}(\epsilon^0)$ using the dimension-shifting formalism. 
We comment on the implications for electromagnetic duality in quantum Born-Infeld theory.}

\keywords{Global Symmetries, Scattering Amplitudes}

\begin{document}  
%%%%%%%%%%%%%%%%%%%%%%%%%%%%%%%%%%%%%%%

\maketitle

\newpage

%%%%%%%%%%%%%%%%%%%%%%%%%%%%%%
\section{Introduction}
\label{sec:Introduction}
%%%%%%%%%%%%%%%%%%%%%%%%%%%%%%

The Born-Infeld model of non-linear electrodynamics is a low-energy effective field theory of central importance in theoretical physics. Introduced long ago as an (ultimately misguided) proposed classical solution to the electron self-energy problem \cite{Born:1934gh}, it subsequently reappeared as the low-energy effective description of world-volume gauge fields on D-branes \cite{Fradkin:1985qd,Polchinski:1995mt,Polchinski:1996na}. Independently of this \textit{stringy} characterization, the Born-Infeld model has proven to be a truly exceptional example of a low-energy effective theory of non-linear electrodynamics, though perhaps at times a mysterious one. 

As a classical field theory in $d=4$ the Born-Infeld model can be described by the effective action
\begin{equation} \label{BIaction}
	S_{\text{BI}} = - \Lambda^4	\int \text{d}^4 x \left[\sqrt{-\text{det}\left(g_{\mu\nu}+\frac{1}{ \Lambda^2}F_{\mu\nu}\right)}-1\right], 
\end{equation}
where $\Lambda$ is the characteristic scale in the problem. In the D-brane picture,  $\Lambda$ is related to the brane tension.

Low-energy scattering of light-by-light in the Born-Infeld model can be calculated as a perturbative expansion in $1/\Lambda$. The tree-approximation to these scattering amplitudes has been a subject of interest recently in the context of modern on-shell approaches to quantum field theory. For example, in \cite{Cheung:2018oki} two novel on-shell approaches for calculating 4d tree-level Born-Infeld amplitudes were given: by imposing \textit{multi-chiral} low-energy theorems derived from supersymmetric relations with Goldstone fermions, and from T-duality constraints under dimensional reduction.  Also very striking is the discovery in \cite{Cachazo:2014xea}, in the context of the CHY formulation of the tree-level S-matrix, that the KLT formula relating Yang-Mills (YM) and gravity amplitudes also gives Born-Infeld tree amplitudes if one of the gauge theory factors is replaced with the flavor-ordered amplitudes of Chiral Perturbation Theory ($\chi\text{PT}$):
\begin{equation}
  \label{BIdc}
  \text{BI}_d = \text{YM}_d \otimes_{\text{KLT}} \chi\text{PT}_d.
\end{equation}
The subscript $d$ indicates the spacetime dimensions of these theories.
What all of these discoveries make clear is that there is an enormous amount of structure hidden behind the action (\ref{BIaction}) which may be leveraged to make possible previously unattainable calculations. It should also be noted that Born-Infeld plays a central role in the ever growing web of mysterious connections between gauge theories, gravity theories, and EFTs in diverse dimensions \cite{Cheung:2017yef, Cheung:2017ems}. Also of great relevance in this paper, pure Born-Infeld can be defined as a consistent truncation of $\mathcal{N} >1$ supersymmetrizations of Dirac-Born-Infeld theory. 

\begin{figure}[t!]
\begin{center}
	\begin{tikzpicture}[scale=1, line width=1 pt]
	\node[below] at (-1.8, -3) {$\chi$PT$_4$};
	\draw [->] (-1.8,-3) -- (-1.8,-2.3);
	\draw[black,fill=white] (-1.8,-2) circle (1.5ex);
	\draw [-] (-1.98,-1.82)--(-1.62,-2.18);
	\draw [-] (-1.98,-2.18)--(-1.62,-1.82);
	\node at (-1.3,-2.36) {\tiny KLT$_{4}$};
	\draw [->] (-1.8,-1) -- (-1.8,-1.7);
	\node[above] at (-1.8, -1) {YM$_4$};
	\draw [->] (-1.55,-2)-- (-0.5,-2);
	\node [right] at (-0.5,-2) {BI$_4$};
 	\draw[->] (0, -1.7) arc (-70:250:0.4);
 	\node [above] at (0,-1) {\small EM duality};
	\draw [->] (3,-2)-- (0.3,-2);
	\node [right] at (3,-2) {DBI$_3$};
	\node[above] at (1.7,-2) {\small T-duality};
	\node[below] at (1.7,-2) {\small constraints};
	 \draw[->,dashed] (0.1,-2.3) arc (210:330:2);
	 \node [below] at (2.1,-3.3) {\small dimensional reduction};
	\end{tikzpicture}
\end{center}
\caption{Some key-properties of BI amplitudes at tree-level, in particular the  double-copy construction and 4d electromagnetic duality. 
The idea behind the T-duality constraint \cite{Cheung:2018oki} is that when dimensionally reduced along one direction, a linear combination of the photon polarizations become a scalar modulus of the compactified direction,. i.e.~it is the Goldstone mode of the spontaneously broken translational symmetry and as such it must have enhanced $\mathcal{O}(p^2)$ soft behavior.}
\label{fig:BItree}
\end{figure}
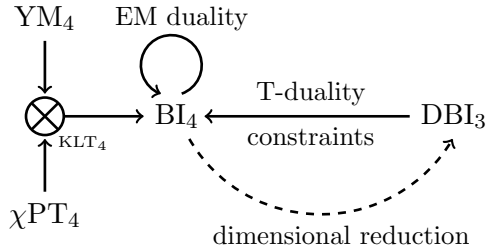

The tree amplitudes in 4d Born-Infeld theory exhibit an important and interesting feature: they vanish unless the external states have an equal number of positive and negative helicity states. This is the on-shell manifestation of electromagnetic duality of the classical theory in 4d. 
In particular, the 4-particle tree amplitude\footnote{Compared to the action \eqref{BIaction}, we have rescaled $\Lambda^4 \to \Lambda^4 / 2$, such that the 4-point amplitude has coupling $1/\Lambda^4$.}
is 
\be
 \label{BI4tree}
\mathcal{A}^{(\text{tree})\;\text{BI}_4}_4(1_\gamma^+,\, 2_\gamma^+,\, 3_\gamma^-,\, 4_\gamma^-) 
= \frac{1}{\Lambda^4}[12]^2\<34\>^2\,,
\ee
while all other helicity configurations vanish. Note that the emergence of electromagnetic duality  is highly non-trivial in the double-copy construction \reef{BIdc}. 
Some of the key properties of the BI tree amplitudes are summarized in Figure \ref{fig:BItree}. 

The recent progress in Born-Infeld scattering has so far been restricted to tree-level amplitudes. Given the development of powerful unitarity based methods for recycling trees into loops \cite{Bern:1994cg}, there is every reason to believe that interesting structures are waiting for us in the loop amplitudes. In this context almost nothing is known.\footnote{One of the few explicit calculations is the determination of the cut-constructible part of the 4-point MHV amplitude in $\mathcal{N}=4$ $\text{DBI}_4$ in \cite{Shmakova:1999ai}.} 
There are good reasons for this; the calculations in Born-Infeld electrodynamics at one-loop are challenging, in ways that are importantly different from superficially similar calculations in perturbative quantum gravity. Similar to calculations at one-loop using Feynman rules derived from expanding the linearized Einstein-Hilbert action, the first computational bottleneck in Born-Infeld is given by the problem of determining the off-shell vertex factors for the interaction terms given by expanding (\ref{BIaction})
\begin{equation} 
	S_{\text{BI}} \sim \int \text{d}^4 x \left[F^2 + \frac{c_1}{\Lambda^4}F^4 + \frac{c_2}{\Lambda^8}F^6 +\ldots\right].
\end{equation}

As the multiplicity of external states increases, more and more terms in this expansion must be kept, and so an ever growing list of increasingly long vertex factors must be calculated. At multiplicity $n$,  operators of the form $F^n$ will contribute; with vertex factors given as sums over permutations growing exponentially in $n$. Beyond the lowest multiplicity, calculating such an amplitude by hand is almost unthinkable, and even with state-of-the art computing power one soon hits a hard wall when performing such a brute force calculation. The situation here is a little different from perturbative gravity. In gravity, the vertex factors are not independent since they are not separately gauge invariant; the higher-point interactions are in principle completely determined by locality and Lorentz invariance by the three-particle ones. 
This can have dramatic consequences, for example in \cite{Bern:1998sv} all-multiplicity, rational one-loop results are obtained from the lowest multiplicity results by enforcing the correct collinear and soft limits. In Born-Infeld, however, 
these higher-valence operators are genuinely gauge invariant physical operators, the associated Wilson coefficients are not related by any inviolable field theory principle and must instead be fixed by imposing additional physical constraints. 
No analysis of soft or collinear limits could possibly determine the all-multiplicity one-loop amplitudes in Born-Infeld, unless it incorporated additional physical information beyond Lorentz invariance and locality.

 The second computational bottleneck occurs when evaluating the required loop integrals. Even if the required loop integrands can be constructed, we still have to integrate the resulting expressions. Operators of the form $F^n$ are $n$-derivative operators and the associated vertex factors have $n$ powers of momentum. The resulting loop-integrands therefore involve tensors with ranks that grow larger and larger with the multiplicity. This is unlike gravity that only has two-derivative interactions. Attempting to apply traditional Passarino-Veltman reduction algorithms to such high-rank tensor expressions again quickly leads to a confrontation with the limits of computing power. Such a direct calculation is primarily limited by the fact that the method of Feynman diagrams is completely general. It therefore makes no use of any of the aforementioned properties that make Born-Infeld electrodynamics exceptional. For example, such an approach would be equally well-suited to calculating loop corrections in the Euler-Heisenberg effective theory \cite{Heisenberg:1935qt}, another well-studied example of a model of non-linear electrodynamics.

In this paper, we initiate a study of 4d non-supersymmetric Born-Infeld theory at the loop-level. We use modern on-shell methods (supersymmetric decomposition, double-copy, T-duality\ldots) that are specialized to the particular properties of Born-Infeld and to the objects we compute. We derive results that would be impossible to obtain with traditional methods. Specifically, we derive all-multiplicity results for the one-loop amplitudes in the \textit{self-dual} (SD)  and \textit{next-to-self-dual} (NSD) sectors of 4d non-supersymmetric Born-Infeld:
\begin{equation}
\label{SDNSD}
  \mathcal{A}_n^{\text{SD}}\left(1_\gamma^+,\,2_\gamma^+,\,\ldots\,(n-1)_\gamma^+,\,n_\gamma^+\right)
  ~~~~~\text{and}~~~~~
  \mathcal{A}_n^{\text{NSD}}\left(1_\gamma^+,\,2_\gamma^+,\,\ldots\,(n-1)_\gamma^+,\,n_\gamma^-\right).
\end{equation} 
Any 4d cuts of these amplitudes vanish, hence to obtain them $d$-dimensional unitarity is used and the results are necessarily rational functions of the external momenta. 

One motivation for these calculations is to examine the fate of electromagnetic duality at loop-level in pure Born-Infeld theory. We make some observations at the end of the paper, but otherwise this will be the subject of a forthcoming paper.

\subsection*{Outline of Paper and Results}
\label{results}

In Section \ref{sec:susydec} we introduce the methods used in the paper. In particular, Section \ref{sec:unitarity} presents the unitarity methods and a very useful supersymmetric decomposition. 
At one-loop order, this allows us to trade the photon running in the loop with a complex scalar in the self-dual and next-to-self-dual helicity sectors \reef{SDNSD}.  We describe the equivalence between $d$-dimensional unitarity cuts and 4-dimensional cuts into massive scalars, and we illustrate the ideas in the context of Yang-Mills theory. 
In Section \ref{sec:massive}, we argue that an appropriate definition of the model coupling a massive scalar to a Born-Infeld photon (called $\text{mDBI}_4$), preserving eight supercharges, is given by the dimensional reduction of pure Born-Infeld in $d=6$. The outcome of this section is that the 1-loop SD and NSD integrands can be calculated from cuts on which they factorize
into  $\text{mDBI}_4$ tree amplitudes of the form 
\be
   \label{mDBIintro}
   \mathcal{A}^{\text{mDBI}_4}_n(\mathbf{1}_\phi, \, 2_\gamma, \ldots (n-1)_\gamma, \mathbf{n}_{\bar\phi})\,,
\ee
where the two scalars are massive (boldfaced) and the helicity configurations of the photons are either all-plus or all-plus-one-minus. 

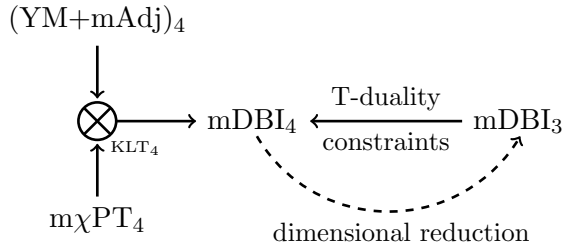
\begin{figure}[t!]
\begin{center}
	\begin{tikzpicture}[scale=1, line width=1 pt]
	\node[below] at (-1.8, -3) {m$\chi$PT$_4$};
	\draw [->] (-1.8,-3) -- (-1.8,-2.3);
	\draw[black,fill=white] (-1.8,-2) circle (1.5ex);
	\draw [-] (-1.98,-1.82)--(-1.62,-2.18);
	\draw [-] (-1.98,-2.18)--(-1.62,-1.82);
	\node at (-1.3,-2.36) {\tiny KLT$_{4}$};
	\draw [->] (-1.8,-1) -- (-1.8,-1.7);
	\node[above] at (-1.8, -1) {(YM+mAdj)$_4$};
	\draw [->] (-1.55,-2)-- (-0.5,-2);
	\node [right] at (-0.5,-2) {mDBI$_4$};
	\draw [->] (3,-2)-- (1,-2);
	\node [right] at (3,-2) {mDBI$_3$};
	\node[above] at (2,-2) {\small T-duality};
	\node[below] at (2,-2) {\small constraints};
	 \draw[->,dashed] (0.3,-2.2) arc (210:330:2);
	 \node [below] at (2.2,-3.2) {\small dimensional reduction};
	\end{tikzpicture}
\end{center}
\caption{Overview of calculational approaches to the tree amplitudes in mDBI$_4$.}
\label{fig:mDBItree}
\end{figure}

In Section \ref{sec:mDBI}, two different approaches are presented for calculating the necessary  $\text{mDBI}_4$ tree amplitudes in 
\reef{mDBIintro}. Section \ref{sec:Structure} discusses the general structure of the amplitudes.  
In Section \ref{sec:KLT}, a form of the massive KLT relations is given by dimensionally reducing the KLT product from $d=6$. The $\text{mDBI}_4$ amplitudes are then calculated numerically as the double copy of Yang-Mills coupled to a massive adjoint scalar ($\text{YM+mAdj}_4$) and a dimensional reduction of Chiral Perturbation theory ($\text{m}\chi\text{PT}_4$). The former is calculated using standard massive BCFW recursion, the latter is calculated using the \textit{soft bootstrap} method by imposing the Adler zero in $d$-dimensions. Explicit results are obtained for %he 
$n=4, 6, 8$ points, 
where it is shown that all possible contact structures are absent. In Section \ref{sec:Tdual}, an alternate method is given for calculating the needed $\text{mDBI}_4$ tree amplitudes. A dimensional reduction to 3d, followed by T-dualization, allows us to identify a DBI brane modulus satisfying an $\mathcal{O}(p^2)$ low-energy theorem. It is shown that for all multiplicities, imposing this condition is sufficient to completely fix all needed $\text{mDBI}_4$ tree amplitudes, with results that agree with the explicit numerical calculations using the KLT product. In particular, this explains the absence of contact terms. The calculational approaches to the $\text{mDBI}_4$ tree amplitudes are illustrated in Figure \ref{fig:mDBItree}.
  
In Section \ref{sec:Loop}, we introduce a set of new diagrammatic rules for constructing $d$-dimensional loop integrands. These are inferred from the results of the previous sections and ensure that the integrands have all the correct $d$-dimensional cuts. Using these rules, we derive all multiplicity integrands for the self-dual (\ref{SDall}) and next-to-self-dual 
(\ref{NSDallw})-(\ref{NSDallg}) helicity sectors. 
  
Three appendices contain technical results.  Appendix \ref{app:Contact} contains a proof  that there is only one possible contact term at each multiplicity for the  relevant $\text{mDBI}_4$ tree amplitudes \reef{mDBIintro}. Appendix \ref{app:Tdual8} contains details for the 3d moduli constraints of the 8-point amplitude. Finally, in Appendix \ref{app:Rational} the leading $\mathcal{O}(\epsilon^0)$ terms in the corresponding rational integrals in $d=4-2\epsilon$ are evaluated explicitly, giving all multiplicity expressions for the one-loop amplitudes (\ref{SDallamp}) and (\ref{NSDallamp}). 

We conclude with a discussion of the results and implications for quantum electromagnetic duality in Section \ref{sec:out}.

%%%%%%%%%%%%%%%%%%%%%%%%%%%%%%
\section{Overview of Method}
\label{sec:susydec}
%%%%%%%%%%%%%%%%%%%%%%%%%%%%%%

Our goal in this paper is to calculate SD and NSD one-loop amplitudes in non-supersymmetric Born-Infeld in $d=4$. As discussed in Section \ref{sec:Introduction} instead of traditional Feynman diagrammatics we make extensive use of modern on-shell methods to construct the amplitudes. In particular, we use $d$-dimensional generalized unitarity methods \cite{Anastasiou:2006jv} to construct the complete loop-integrand in a physically motivated dimensional scheme. We begin with a brief overview of unitarity methods and then describe in detail the approach taken in this paper. In Section \ref{sec:unitarity}, we introduce the techniques in the familiar context of Yang-Mills theory, then adapt the methods to Born-Infeld in Section \ref{sec:massive}. 

%%%%%%%%%%%%%%%%%%%%%%%%%%%%%%
\subsection{Generalized Unitarity and Supersymmetric Decomposition}
\label{sec:unitarity}
%%%%%%%%%%%%%%%%%%%%%%%%%%%%%%
The main idea of unitarity based methods \cite{Bern:2011qt} is to exploit that the loop integrand is a complex rational function of the loop momentum with singularity structure constrained by factorization into on-shell tree amplitudes. Here we focus specifically on one-loop order and all calculations are made in a given dimensional regularization scheme. This means that while the external momenta and polarizations are strictly $d=4$-dimensional, the loop momentum is formally regarded as $d=(4-2\epsilon)$-dimensional. 

\subsubsection*{4-Dimensional Unitarity Methods}
Expanding the loop-\emph{integrand} around $\epsilon=0$, the leading $\mathcal{O}(\epsilon^0)$ component has an unambiguous physical meaning related to unitarity of the S-matrix. 
Via the Cutkosky theorem \cite{Cutkosky:1960sp}, the factorization of the integrand into on-shell tree amplitudes on 4d cuts 
\begin{equation}
  l^2_1\ l^2_2\ \cdots\ l^2_k\ \mathcal{I}_n[l]\biggr\vert_{l_1^2=\cdots=l_k^2=0} = \sum_{\text{states}}\mathcal{A}^{\text{tree}_4}_{(1)}\ldots\mathcal{A}^{\text{tree}_4}_{(k)},
\end{equation}
where $l_i^\mu$, for $i=1,\cdots k$ are 4d momenta, ensures that the integrated amplitude has the correct branch cut discontinuities required by the optical theorem. A rational function with all the correct 4d cuts (and no spurious cuts) then yields the correct amplitude after integration up to a function with no branch cuts, i.e.~a rational function, up to and including terms of $\mathcal{O}(\epsilon^0)$.  This is the idea of the \textit{4-dimensional unitarity} approach: the cut-constructible part of the amplitude is completely fixed by the physical tree amplitudes. Due to a complete understanding of integrand reduction to a basis of master scalar integrals at one-loop this procedure can be completely automated \cite{Britto:2010xq}. The remaining rational function ambiguity must then be determined by imposing additional physical constraints, such as cancellation of spurious singularities in the cut-constructible part or by imposing known behavior in soft or collinear limits \cite{Bern:1997sc}. One advantage of calculating the 4d-cut-constructible part and the rational part separately in this way is that at all stages of the calculation we make use of regularization scheme-independent, physical objects (on-shell 4d tree-amplitudes). The primary disadvantage to this approach is the relative difficulty in calculating the rational terms separately.  

\subsubsection*{$d$-Dimensional Unitarity Methods}
In certain cases, the cut-constructible part vanishes and the integrated loop-amplitude is purely rational.  In that case, the method outlined above for determining the rational part is not applicable. 
This, in particular, will be the situation for the amplitudes \reef{SDNSD} of interest in this paper. 

A  more familiar example is the SD and NSD sectors of pure Yang-Mills theory (i.e.~the all-plus and all-plus-one-minus gluon amplitudes): at one-loop, any 4d cut has factors of tree amplitudes of the SD and NSD helicity configurations and those vanish \cite{Grisaru:1977px}, hence all the 4d cuts vanish. According to the discussion above, the absence of 4d cuts implies that the resulting integrand is zero at $\mathcal{O}(\epsilon^0)$  (vanishes in $d=4$), but may have non-zero contributions at $\mathcal{O}(\epsilon)$. As a result, SD and NSD one-loop amplitudes have no branch cut discontinuities and are instead  purely rational functions.  These rational contributions arise from subtle $\epsilon/\epsilon$ cancellations after integration; the same mechanism gives rise to the chiral anomaly in dimensional regularization \cite{tHooft:1972tcz}. Since the SD and NSD sectors of YM and BI theory are very similar, we introduce the method here for YM , then adapt it to BI theory in the Section \ref{sec:massive}. 

The method of \textit{d-dimensional unitarity} \cite{Anastasiou:2006jv} does not separate the 4d-cuts and rational terms. In the d-dimensional unitarity approach, we must first define a suitable dimensional regularization scheme in which $d$-dimensional integrand cuts have the form
\begin{equation}
  l^2_1l^2_2\ldots l^2_k\mathcal{I}_n[l]\biggr\vert_{l_1^2=\ldots =l_k^2=0} = \sum_{\text{states}}\mathcal{A}^{\text{tree}_d}_{(1)}\ldots \mathcal{A}^{\text{tree}_d}_{(k)}\,,
\end{equation}
where the on-shell cut momenta $l_i$ are $d$-dimensional. The additional constraint of correct cuts in $d$-dimensions is sufficient to construct the integrand to all orders in $\epsilon$, allowing us to determine both the 4d cut-constructible and rational parts at the same time. This approach is therefore well-suited to the purely rational SD and NSD one-loop amplitudes of Yang-Mills. The difficulty of this approach is that we are forced to work with regularization scheme-dependent quantities, which are therefore non-unique, and furthermore since the cuts are in $d$-dimensions, we lose the simplicity of spinor-helicity variables. 

In certain special cases, such as pure Yang-Mills and pure Born-Infeld in $d=4$, we can maneuver around these difficulties and define a regularization scheme in which both the $d$-dimensional-cut structure is quite simple \textit{and} we can still make use of spinor-helicity variables. This simplified implementation of $d$-dimensional unitarity is sometimes referred to as \textit{supersymmetric decomposition} and this is what we describe next.  

\subsubsection*{Consistent Truncation and Supersymmetric Decomposition}

It is instructive to first review the concept of supersymmetric consistent truncation at tree-level. In general we say that model A is a \textit{consistent truncation} of model B if the on-shell states of A form a subset of the on-shell states of B and (when restricted to the A-states) the S-matrices are identical at tree-level.\footnote{This is equivalent to the statement that solutions to the classical equations of motion for model A are also solutions to the equations of motion of model B with the fields in B/A turned off.} This occurs in any model in which the states of B/A (B-states that are not A-states) carry an independent charge or parity; such states can give no contribution to state-sums on factorization singularities and hence no contribution to the tree-level S-matrix elements with all external A-states. A simple example of this occurs in any model containing both Bosonic and Fermionic states; since the quantity $(-1)^F$ is conserved we can always construct a consistent truncation by restricting to the Bosonic sector. If there are additional conserved quantities in the Bosonic sector, then it may be possible to give a further truncation.

As a relevant example, consider $\mathcal{N}=2$ super Yang-Mills (without matter hypermultiplets) in $d=4$. The spectrum consists of a massless vector multiplet containing a gauge boson $g^\pm$, two Weyl fermions $\psi_{1,2}^\pm$ and a complex scalar $\phi,\; \overline{\phi}$. Restricting to the Bosonic sector gives a consistent truncation, the resulting model is non-supersymmetric and describes Yang-Mills coupled to a massless (adjoint) complex scalar. In this model there is an additional \textit{global} symmetry, descended from R-symmetry, under which the states are charged as
\begin{equation} \label{Qphi}
Q[g^\pm] = 0, \;\;\;\;\;  Q[\phi]=1, \;\;\;\;\; Q[\overline{\phi}]=-1.
\end{equation} 
Consequently, we can define a further truncation to the purely \textit{gluonic sector}, the resulting model is precisely pure non-supersymmetric Yang-Mills. The statement of consistent truncation in this example is then
\begin{equation} \label{consitent}
	\mathcal{A}_n^{(\text{tree})\;\mathcal{N}=2\;\text{SYM}}\left[1_g,\ldots n_g\right] = \mathcal{A}_n^{(\text{tree})\;\text{YM+Adj}}\left[1_g,\ldots n_g\right] = \mathcal{A}_n^{(\text{tree})\;\text{YM}}\left[1_g,\ldots n_g\right].
\end{equation}
Since gluonic amplitudes in $\mathcal{N}=2$ SYM in the SD and NSD helicity sectors vanish at all orders of perturbation theory, these same helicity sectors must likewise vanish in tree-level non-supersymmetric Yang-Mills. 

The notion of consistent truncation in the form of equalities such as (\ref{consitent}) does not continue to hold at loop-level. We can, however, make use of supersymmetric truncations at one-loop to form a \textit{supersymmetric decomposition}. Let us illustrate this in the context of Yang-Mills. At one-loop, all states in the model generically run in every loop, for $\mathcal{N}=0,1$ and $2$ SYM we can schematically represent the contributions to purely gluonic amplitudes as
\begin{align} \label{VFS}
\mathcal{A}_n^{(\text{1-loop})\;\text{YM}}\left[1_g\ldots n_g\right] 
&= \mathcal{A}_n^{[V]}\left[1_g\ldots n_g\right] \nonumber\\
\mathcal{A}_n^{(\text{1-loop})\;\mathcal{N}=1\;\text{SYM}}\left[1_g\ldots n_g\right] 
&= \mathcal{A}_n^{[V]}\left[1_g\ldots n_g\right]+\mathcal{A}_n^{[F]}\left[1_g\ldots ,n_g\right]\nonumber\\
\mathcal{A}_n^{(\text{1-loop})\;\mathcal{N}=2\;\text{SYM}}\left[1_g\ldots n_g\right] 
&= \mathcal{A}_n^{[V]}\left[1_g\ldots n_g\right]+2\mathcal{A}_n^{[F]}\left[1_g\ldots n_g\right]+\mathcal{A}_n^{[S]}\left[1_g\ldots n_g\right],
\end{align}
where $V$, $F$, and $S$ represent contributions from vector bosons, Weyl fermions, and complex scalars, respectively. The contributions on the right-hand-side have no invariant physical meaning, even in the context of a Feynman diagram expansion, as a grouping of terms they depend on the choice of regularization scheme. One can, however, give invariant physical meaning to these expressions on 4d-unitarity cuts: the decomposition reflects the contributions to the state sums. Note that it is the existence of the same conservation laws that allowed us to construct consistent truncations at tree-level that make this decomposition sensible. In particular, due to (\ref{Qphi}), there are no mixed scalar/gluon contributions to 4d cuts. If the amplitudes are calculated in the Four Dimensional Helicity (FDH) or similar schemes, in which the one-to-one correspondence between the (external) 4-dimensional helicity states and the (internal) $d$-dimensional states is preserved \cite{Bern:1991aq} then the relations (\ref{VFS}) are well-defined on $d$-dimensional cuts. 

The notion of a \textit{supersymmetric decomposition} is a rearrangement of (\ref{VFS}) such that one-loop amplitudes in non-supersymmetric Yang-Mills can be given as sums over contributions from $\mathcal{N}=1,2$ vector multiplets and adjoint scalars
\begin{align}
&\mathcal{A}_n^{(\text{1-loop})\;\text{YM}}\left[1_g\ldots n_g\right]  \nonumber\\
&\hspace{5mm}
= -\mathcal{A}_n^{(\text{1-loop})\;\mathcal{N}=2\;\text{SYM}}\left[1_g\ldots n_g\right] 
+ 2\mathcal{A}_n^{(\text{1-loop})\;\mathcal{N}=1\;\text{SYM}}\left[1_g\ldots n_g\right] 
+\mathcal{A}_n^{[S]}\left[1_g\ldots n_g\right].
\end{align}

Next, we assume that our regularization scheme is supersymmetric (for example FDH \cite{Bern:2002zk}), and therefore the one-loop amplitudes satisfy the same supersymmetry Ward identities as the tree-level amplitudes.\footnote{In a non-supersymmetric scheme such as conventional dimensional regularization (CDR) the result of the loop integrals will typically not satisfy the supersymmetry Ward identities. Supersymmetry must be restored by adding finite local counterterms which modify the rational part of the one-loop amplitudes.}  This dramatically simplifies in the SD and NSD sectors, since the contributions from the $\mathcal{N}>0$ components vanish. In these sectors the supersymmetric decomposition simplifies to
\begin{equation} \label{scalarloopYM}
  \mathcal{A}_n^{(\text{1-loop})\;\text{YM}}\left[1^+_g\ldots (n-1)_g^+,\,n^\pm_g\right] = \mathcal{A}_n^{[S]}\left[1^+_g\ldots (n-1)_g^+,\,n^\pm_g\right].
\end{equation}
We refer to this as the \textit{scalar-loop representation} of the one-loop amplitude. Again, in the context of $d$-dimensional unitarity we can interpret this statement unambiguously as a statement about the $d$-dimensional unitarity cuts of the loop-integrand. 
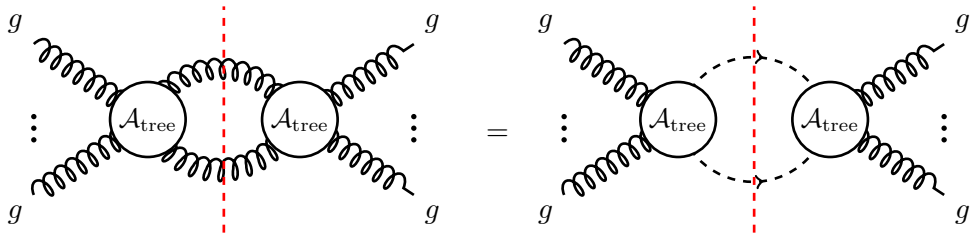
\begin{center}
	\begin{tikzpicture}[scale=1, line width=1 pt]
	\draw [gluon] (-1.5,1)--(0,0);
	\node at (-1.5,0) {$\huge\vdots$};
	\draw [gluon] (-1.5,-1)--(0,0);
	\draw [gluon] (0,0) arc (160:20:1);
	\draw [gluon] (0,0) arc (200:340:1);
	\draw [gluon] (2,0)--(3.5,1);
	\draw [gluon] (2,0)--(3.5,-1);
	\node at (3.5,0) {$\huge\vdots$};
	\draw[black,fill=white] (0,0) circle (3ex);
	\draw[black,fill=white] (2,0) circle (3ex);
	\node at (0,0) {\small $\mathcal{A}_\text{tree}$};
	\node at (2,0) {\small $\mathcal{A}_\text{tree}$};
	\node [above left] at (-1.5,1) {$g$};
	\node [ below left] at (-1.5,-1) {$g$};
	\node [ above right] at (3.5,1) {$g$};
	\node [ below right] at (3.5,-1) {$g$};
	\draw[red,dashed] (1,-1.5)--(1,1.5);
	\node at (4.6,-0.2) {$=$};
	\end{tikzpicture}
	\begin{tikzpicture}[scale=1, line width=1 pt]
	\draw [gluon] (-1.5,1)--(0,0);
	\node at (-1.5,0) {$\huge\vdots$};
	\draw [gluon] (-1.5,-1)--(0,0);
	\draw [scalar] (0,0) arc (170:10:1);
	\draw [scalar] (0,0) arc (190:350:1);
	\draw [gluon] (2,0)--(3.5,1);
	\draw [gluon] (2,0)--(3.5,-1);
	\node at (3.5,0) {$\huge\vdots$};
	\draw[black,fill=white] (0,0) circle (3ex);
	\draw[black,fill=white] (2,0) circle (3ex);
	\node at (0,0) {\small $\mathcal{A}_\text{tree}$};
	\node at (2,0) {\small $\mathcal{A}_\text{tree}$};
	\node [above left] at (-1.5,1) {$g$};
	\node [ below left] at (-1.5,-1) {$g$};
	\node [ above right] at (3.5,1) {$g$};
	\node [ below right] at (3.5,-1) {$g$};
	\draw[red,dashed] (1,-1.5)--(1,1.5);
	\end{tikzpicture}
\end{center}

As a consequence, the complete one-loop integrand can be reconstructed by requiring the correct $d$-dimensional unitarity cuts into $d$-dimensional tree-amplitudes of the form
\begin{equation} \label{treecut}
  \mathcal{A}^{(\text{tree})}_n\big[1_\phi,2_g\ldots (n-1)_g,n_{\overline{\phi}}\big].
\end{equation}
Here only the momenta of the scalars are $d$-dimensional, while the momenta and polarizations of the gluons are 4-dimensional. 

We rewrite the $d$-dimensional momenta in terms of 4-dimensional momenta as
\begin{equation}
 l^\mu = l^\mu_{[4]} + l^\mu_{[-2\epsilon]}.
\end{equation}
Due to the orthogonality of 4-dimensional and $(-2\epsilon)$-dimensional subspaces, we can rewrite the various Lorentz singlets that appear in the amplitude as
\begin{equation}
 q\cdot l = q \cdot l_{[4]}, \hspace{7mm} l^2 = l_{[4]}^2 + l_{[-2\epsilon]}^2 \equiv l_{[4]}^2 + \mu^2,
\end{equation}
where $q^\mu$ is any 4-dimensional vector and $\mu^2\equiv l_{[-2\epsilon]}^2$. Using these relations we find that we can rewrite all $d$-dimensional amplitudes (\ref{treecut}) as 4-dimensional amplitudes with a massive scalar of mass $\mu^2$. 

Up to this point we have not explicitly defined the regularization scheme, we have only made use of some general properties that it should have. We could give such a precise definition and then calculate the $d$-dimensional scalar amplitudes (\ref{treecut}). Instead, we shall define the massive scalar amplitudes directly in 4d, requiring all of the standard tree-level properties of Lorentz invariance, locality and unitarity, in addition to the requirement  
\begin{equation} \label{mu2zero}
 \mathcal{A}^{\text{tree}}_n\big[1_\phi,2_g,\ldots ,(n-1)_g,n_{\overline{\phi}}\big]\xrightarrow{\mu^2\rightarrow 0} \mathcal{A}^{\text{tree}\; (\mathcal{N}=2)}_n\big[1_\phi,2_g,\ldots ,(n-1)_g,n_{\overline{\phi}}\big].
\end{equation}  
Even though the 4d cuts vanish in the SD and NSD amplitudes of consideration, the relations (\ref{VFS}) make sense for all helicity amplitudes, and for those with non-vanishing 4d cuts the $\mathcal{A}_n^{[S]}$ cuts must be equal to products
of tree-amplitudes of $\mathcal{N}=2$ SYM. The problem of constructing the integrand in the scalar loop representation then has two parts:

\begin{enumerate}
  \item Define a model of a massive adjoint scalar coupled to Yang-Mills which reduces to the Bosonic sector of $\mathcal{N}=2$ SYM in the massless limit.
  \item Construct a complex rational function of 4d momenta with correct cuts into the massive scalar tree amplitudes and no spurious cuts.
\end{enumerate} 

The required massless limit \reef{mu2zero} is not sufficient to determine the massive scalar model described in Step 1. In addition to the \textit{minimal} coupling,\footnote{This includes the $|\phi|^4$ term in the scalar potential required to satisfy the requirement of $\mathcal{N}=2$ supersymmetry in the massless limit.} we could also add generic terms to the scalar potential or higher-derivative couplings, for example we might consider a model described by the action
\begin{equation}
  S[A_\mu,\phi,\overline{\phi}] = S_{\text{minimal}}[A_\mu,\phi,\overline{\phi}] + \int \text{d}^4x \left[\frac{\mu^2}{\Lambda_1^4}|\phi|^6+\frac{\mu^2}{\Lambda_2^4}|\phi|^2\text{Tr}[F^2]\right],
\end{equation}
where $\Lambda_1$ and $\Lambda_2$ are independent mass scales. Such a model clearly satisfies the correct massless limit. The presence of independent dimensionful parameters however makes this physically unacceptable, these would appear in the integrand we construct according to Step 2, and consequently the integrated amplitude. To ensure the absence of such spurious parameters we  impose:

\begin{enumerate}
  \addtocounter{enumi}{2}
  \item The result we calculate should agree with the parametric dependence on couplings expected from a full Feynman diagram calculation, therefore an acceptable massive scalar extension of Yang-Mills theory should depend only on the dimensionless Yang-Mills coupling $g_{\text{YM}}$.
\end{enumerate}

 By this simple argument all such higher dimension couplings must be absent, the correct model is given by the minimally coupled massive adjoint scalar with the supersymmetric scalar potential. Such tree amplitudes can be generated efficiently by using massive BCFW recursion, which we will review in Section \ref{sec:YM}. 

The strategy described above has been used successfully to calculate all-multiplicity one-loop amplitudes in the SD and NSD sectors of pure Yang-Mills \cite{Bern:1994cg}. It has also been implemented in pure Einstein gravity \cite{Bern:1998sv} and also recently Einstein Yang-Mills \cite{Nandan:2018ody}. The purpose of this paper is to implement this approach in non-supersymmetric Born-Infeld electrodynamics in $d=4$. In the following subsection we will describe the novelties that appear in this model compared to Yang-Mills. 

%%%%%%%%%%%%%%%%%%%%%%%%%%%%%%
\subsection{Massive Scalar Extension of Born-Infeld}
\label{sec:massive}
%%%%%%%%%%%%%%%%%%%%%%%%%%%%%%

Almost everything we described in Section~\ref{sec:unitarity} for pure Yang-Mills in $d=4$ 
applies to pure Born-Infeld in $d=4$. At tree-level, non-supersymmetric Born-Infeld is a consistent truncation of $\mathcal{N}=2$ super Born-Infeld. Consequently, the SD and NSD amplitudes vanish at tree-level.
Moreover, in a supersymmetric regularization scheme, the SD and NSD one-loop amplitudes have a scalar-loop representation
\begin{equation} \label{scalarloop}
  \mathcal{A}_n^{(\text{1-loop})\;\text{BI}_4}
  \left(1^+_\gamma,\ldots, (n-1)_\gamma^+,n^\pm_\gamma\right) 
  = \mathcal{A}_n^{[S]}\left(1^+_\gamma,\ldots, (n-1)_\gamma^+,n^\pm_\gamma\right).
\end{equation}
These one-loop amplitudes have no $d=4$ cuts, so are purely rational. We compute the integrand using $d$-dimensional unitarity in which the cuts factor into tree amplitudes with two massive scalars coupled to the Born-Infeld photons.

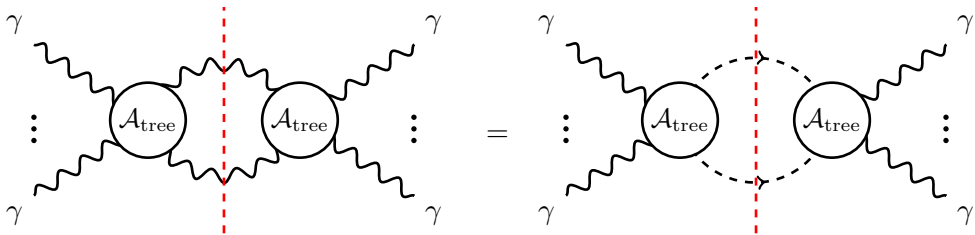
\begin{center}
	\begin{tikzpicture}[scale=1, line width=1 pt]
	\draw [vector] (-1.5,1)--(0,0);
	\node at (-1.5,0) {$\huge\vdots$};
	\draw [vector] (-1.5,-1)--(0,0);
	\draw [vector] (0,0) arc (160:20:1);
	\draw [vector] (0,0) arc (200:340:1);
	\draw [vector] (2,0)--(3.5,1);
	\draw [vector] (2,0)--(3.5,-1);
	\node at (3.5,0) {$\huge\vdots$};
	\draw[black,fill=white] (0,0) circle (3ex);
	\draw[black,fill=white] (2,0) circle (3ex);
	\node at (0,0) {\small $\mathcal{A}_\text{tree}$};
	\node at (2,0) {\small $\mathcal{A}_\text{tree}$};
	\node [above left] at (-1.5,1) {$\gamma$};
	\node [ below left] at (-1.5,-1) {$\gamma$};
	\node [ above right] at (3.5,1) {$\gamma$};
	\node [ below right] at (3.5,-1) {$\gamma$};
	\draw[red,dashed] (1,-1.5)--(1,1.5);
	\node at (4.6,-0.2) {$=$};
	\end{tikzpicture}
	\begin{tikzpicture}[scale=1, line width=1 pt]
	\draw [vector] (-1.5,1)--(0,0);
	\node at (-1.5,0) {$\huge\vdots$};
	\draw [vector] (-1.5,-1)--(0,0);
	\draw [scalar] (0,0) arc (170:10:1);
	\draw [scalar] (0,0) arc (190:350:1);
	\draw [vector] (2,0)--(3.5,1);
	\draw [vector] (2,0)--(3.5,-1);
	\node at (3.5,0) {$\huge\vdots$};
	\draw[black,fill=white] (0,0) circle (3ex);
	\draw[black,fill=white] (2,0) circle (3ex);
	\node at (0,0) {\small $\mathcal{A}_\text{tree}$};
	\node at (2,0) {\small $\mathcal{A}_\text{tree}$};
	\node [above left] at (-1.5,1) {$\gamma$};
	\node [ below left] at (-1.5,-1) {$\gamma$};
	\node [ above right] at (3.5,1) {$\gamma$};
	\node [ below right] at (3.5,-1) {$\gamma$};
	\draw[red,dashed] (1,-1.5)--(1,1.5);
	\end{tikzpicture}
\end{center}

 Here the massive scalar model should reduce to $\mathcal{N}=2$ super Born-Infeld in the massless limit, analogously to \reef{mu2zero}. Since there are independent gauge-invariant local operators coupling the Born-Infeld photon and a massive scalar which vanish in the massless limit, this is not sufficient to determine the massive model. Unlike Yang-Mills, we can construct an infinite number of such operators \textit{without} introducing spurious dimensionful parameters. To proceed, additional physical constraints must be applied to uniquely define the massive scalar extension of Born-Infeld. In the remainder of this section, we describe the model, which we call $\text{mDBI}_4$ (massive DBI in 4d), and argue from two points of view why it is an appropriate definition. In Section \ref{sec:mDBI} we then calculate the $\text{mDBI}_4$ tree amplitudes  
\be\label{1stmDBI}
   \mathcal{A}^{\text{mDBI}_4}_n\big(1_\phi, 2^+_\gamma, \ldots,(n-2)^+_\gamma, (n-1)^\pm_\gamma, n_{\bar\phi}\big)\,,
\ee
needed for the unitarity cuts, where the complex scalar has mass $\mu^2\equiv l_{[-2\epsilon]}^2$ in  $d=4$. As stated, these tree amplitudes must satisfy 
\begin{equation} \label{mu2zero2}
 \mathcal{A}^{\text{mDBI}_4}_n\big(1_\phi,2_\gamma,\ldots ,(n-1)_\gamma,n_{\overline{\phi}}\big) \xrightarrow{\mu^2\rightarrow 0} 
 \mathcal{A}^{\mathcal{N}=2\, \text{BI}_4}_n\big(1_\phi,2_\gamma,\ldots, (n-1)_\gamma,n_{\overline{\phi}}\big).
\end{equation}  

The two approaches to define $\text{mDBI}_4$ are dimensional reduction and the double-copy; we now describe each in turn.

\subsubsection*{Dimensional Reduction and Supersymmetry}

We define $\text{mDBI}_4$ as the dimensional reduction of pure Born-Infeld from $d=6$ ($\text{BI}_6$). Specifically we take 6d tree-amplitudes with momenta and polarizations in the configuration described in Table \ref{helicitytable}, i.e.~the photon momenta and polarizations lie in a 4d subspace for lines $2,3,\ldots,n\!-\!1$ while lines $1$ and $n$ have genuinely 6d momenta but polarizations orthogonal to the 4d subspace, so in the 4d setting they are scalars. 
This is an appropriate definition because the amplitudes \reef{1stmDBI} arise from $d$-dimensional cuts of a loop-integrand in a supersymmetric regularization scheme.

\begin{table}[t!]
\begin{center}
\begin{tabular}{|c|c|c|c|c|c|} 
  \hline
  & 1 & 2 & 3 & 4 & 5 \\ 
  \hline\hline
  $\vec{p}_{1,n}$ & x & x & x & x & x \\
  \hline
  $\vec{\epsilon}_{1,n}$ & & & & x & x \\
  \hline
  $\vec{p}_{2,3,\ldots ,n-1}$ & x & x & x & & \\
  \hline
  $\vec{\epsilon}_{2,3,\ldots ,n-1}$  & x & x & x & &  \\
  \hline
\end{tabular}
\caption{Kinematic configuration of momenta and polarizations of $\text{BI}_6$ defining $\text{mDBI}_4$ and for $\text{YM}_6$ defining $(\text{YM}+\text{mAdj})_4$.}
\label{helicitytable}
\end{center}
\end{table}

As in the previous subsection, it is instructive to first describe the case of pure Yang-Mills. In any scheme, on 4d cuts the integrand factors into tree-amplitudes of $\text{YM}_4$, which by virtue of being a consistent truncation of $\mathcal{N}=2$ $\text{SYM}_4$ satisfy the supersymmetry Ward identities for 8 supercharges. On $d$-dimensional cuts, however, we would generically expect the action of the supersymmetry algebra to be explicitly broken. To construct a supersymmetric regularization scheme, we want to define a dimensional continuation from $d=4$ in which the action of the 8 supercharges of $\mathcal{N}=2$ is unbroken. 

A natural way to do this is to recognize that the Yang-Mills-scalar tree amplitudes (\ref{treecut}) can be obtained from pure Yang-Mills in $d=6$ ($\text{YM}_6$) with momenta and polarizations in the configuration given in Table \ref{helicitytable}. Since $\text{YM}_6$ is a consistent truncation of $\mathcal{N}=(1,0)$ $\text{SYM}_6$, the $\text{YM}_6$ tree amplitudes must satisfy the full set of $\mathcal{N}=(1,0)$ supersymmetry Ward identities. It therefore follows that in the configuration given in Table \ref{helicitytable}, the 6d amplitudes written in a 4d language, must satisfy (some version of) the supersymmetry Ward identities for 8 supercharges. We should therefore expect a regularization scheme with a scalar-loop representation (\ref{scalarloopYM}), with massive scalar amplitudes defined by this dimensional reduction from 6d, to preserve (some version of) the full $\mathcal{N}=2$ supersymmetry on \textit{d-dimensional cuts}, and it is therefore a supersymmetric scheme. This definition of the Yang-Mills-scalar amplitudes satisfies the criteria we gave in the previous subsection of absence of spurious parametric dependence. The massive scalar extension of 4d Yang-Mills theory defined this way will be denoted $(\text{YM}+\text{mAdj})_4$; as it turns out, it will be useful in our amplitude constructions. 

The same argument applies essentially verbatim to Born-Infeld. $\text{BI}_6$ is a consistent truncation of $\mathcal{N}=(1,0)$ super Born-Infeld ($\text{SBI}_6$), so the tree-amplitudes of $\text{mDBI}_4$ defined by the configuration given in Table \ref{helicitytable} must preserve the action of 8 supercharges. Hence the SD and NSD one-loop integrands of $\text{BI}_4$ in the scalar loop representation (\ref{sec:unitarity}) preserve the action of $\mathcal{N}=2$ supersymmetry on $d$-dimensional cuts, and therefore define a scheme that we expect to be supersymmetric. We do not have a formal proof of this statement.

\subsubsection*{BCJ Double-Copy}

A complimentary argument, with the same conclusion, is given by considering the BCJ double copy. It was shown in \cite{Cachazo:2014xea}, in the context of the CHY formalism \cite{Cachazo:2013hca,Cachazo:2013iea}, that the field theory KLT formulae which give gravity tree amplitudes as the double-copy of gauge theory tree amplitudes also give Born-Infeld if one of the gauge theory factors is replaced by Chiral Perturbation Theory ($\chi$PT). $\chi$PT is a non-linear sigma model with target space $\frac{SU(N)\times SU(N)}{SU(N)}$. This double-copy statement applies at tree-level in $d$-dimensions
\begin{equation}
  \text{BI}_d = \text{YM}_d \otimes_{\text{KLT}} \chi\text{PT}_d\,.
\end{equation}

It has been conjectured by BCJ that the double-copy could be extended to loop integrands \cite{Bern:2010ue}. This remains a conjecture, though it has been successfully applied in many examples and represents the current state of the art for high loop order calculations in maximal supergravity \cite{Bern:2018jmv}. In this spirit we conjecture that the tree-level double-copy construction of Born-Infeld extends to a complete loop-level double copy following BCJ. 

In this paper we do not make use of explicit color-kinematics dual BCJ integrands. Rather, we proceed by assuming that such a representation of the $\text{BI}_4$ integrand exists in a supersymmetric regularization scheme which admits a scalar-loop representation (\ref{scalarloop}). Then on $d$-dimensional cuts, the integrand factors into tree amplitudes in a model coupling Born-Infeld photons to a massive scalar. Furthermore, these tree amplitudes should be given by the tree-level double-copy of $\text{YM}_4$ coupled to a massive scalar and $\chi\text{PT}_4$ coupled to a massive scalar. The existence of such double-copy compatible massive scalar models is quite non-trivial. 

We now want to show that the proposed definition of $\text{mDBI}_4$ is indeed generated by the tree-level double copy. The key to this is that the KLT product is valid in $d$-dimensions, it therefore commutes with dimensional reduction\footnote{The dimensional reduction of $\chi\text{PT}_6$ to $d=4$ is defined by the momentum configuration in Table \ref{helicitytable}.} in the sense described by the configuration in Table \ref{helicitytable}:

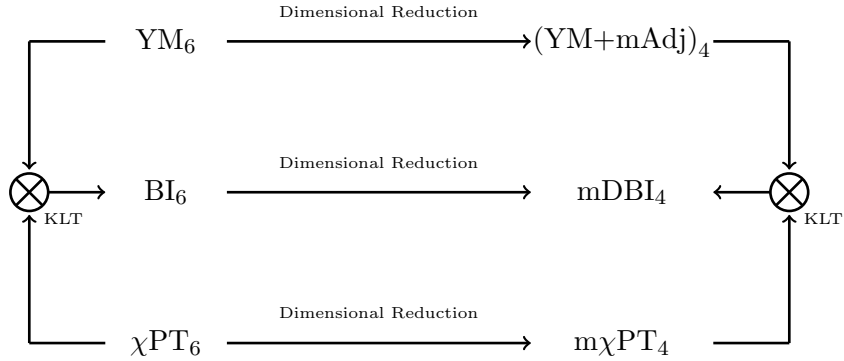
 \begin{center}
{\begin{tikzpicture}[scale=1, line width=1 pt]
\node at (0,0) {$\text{YM}_6$};
\node at (6,0) {$\text{(YM+mAdj)}_4$};
\node at (0,-4) {$\chi\text{PT}_6$};
\node at (6,-4) {$\text{m}\chi\text{PT}_4$};
\node at (0,-2) {$\text{BI}_6$};
\node at (6,-2) {$\text{mDBI}_4$};
\draw [->] (0.8,0)--(4.8,0); 
\draw [->] (0.8,-2)--(4.8,-2); 
\draw [->] (0.8,-4)--(4.8,-4); 
\draw [-] (-0.8,0)--(-1.8,0);
\draw [<-] (-0.8,-2)--(-1.8,-2);
\draw [-] (-0.8,-4)--(-1.8,-4);
\draw [->] (-1.8,0)--(-1.8,-1.7);
\draw [<-] (-1.8,-2.3)--(-1.8,-4);
\draw[black,fill=white] (-1.8,-2) circle (1.5ex);
\draw [-] (-1.98,-1.82)--(-1.62,-2.18);
\draw [-] (-1.98,-2.18)--(-1.62,-1.82);
\node at (-1.35,-2.35) {\tiny KLT};
\draw [-] (7.2,0)--(8.2,0);
\draw [<-] (7.2,-2)--(8.2,-2);
\draw [-] (7.2,-4)--(8.2,-4);
\draw [->] (8.2,0)--(8.2,-1.7);
\draw [<-] (8.2,-2.3)--(8.2,-4);
\draw[black,fill=white] (8.2,-2) circle (1.5ex);
\draw [-] (8.02,-1.82)--(8.38,-2.18);
\draw [-] (8.02,-2.18)--(8.38,-1.82);
\node at (8.65,-2.35) {\tiny KLT};
\node at (2.8,0.4) {\tiny Dimensional Reduction};
\node at (2.8,-1.6) {\tiny Dimensional Reduction};
\node at (2.8,-3.6) {\tiny Dimensional Reduction};
\end{tikzpicture}}
\end{center}
 
Since both Yang-Mills and $\chi$PT satisfy the conditions necessary for the double-copy to be well-defined in $d$-dimensions, we can begin with these models in $d=6$. As illustrated in the diagram above we have two choices, either take the 6d double-copy first and then dimensionally reduce to 4d, or dimensionally reduce to the 4d massive scalar models first and then take the 4d double-copy; it is clear these choices will agree. In the first case, the validity of the $d$-dimensional double copy gives precisely the definition of $\text{mDBI}_4$ given above, the second case gives us exactly the massive scalar double-copy we expect on $d$-dimensional cuts if the loop BCJ conjecture is correct. 
The advantage of working in the 4d formulation is that we can take advantage of the 4d spinor helicity formalism.

%%%%%%%%%%%%%%%%%%%%%%%%%%%%%%
\section{Calculating \texorpdfstring{$\text{mDBI}_4$}{mDBI4} Tree Amplitudes}
\label{sec:mDBI}
%%%%%%%%%%%%%%%%%%%%%%%%%%%%%%

%%%%%%%%%%%%%%%%%%%%%%%%%%%%%%
\subsection{General Structure}
\label{sec:Structure}
%%%%%%%%%%%%%%%%%%%%%%%%%%%%%%

As described in the previous section, the input required for constructing the (N)SD loop integrands using $d$-dimensional unitarity are tree amplitudes in some model (which we call $\text{mDBI}_4$) describing a massless Born-Infeld photon coupled to a massive complex scalar. We need two types of tree amplitudes:
\begin{itemize}
	\item mDBI$_4$ NSD amplitudes: These are of the form $\mathcal{A}_n^{\text{mDBI}_4}\left(1_\phi,2_\gamma^+,\ldots ,(n-1)_\gamma^+,n_{\overline{\phi}}\right)$ and will be used to calculate BI$_4$ SD and NSD amplitudes in Sections \ref{sec:SD} and \ref{sec:NSD} respectively. 	
	\item mDBI$_4$ MHV amplitudes: These are of the form $\mathcal{A}_n^{\text{mDBI}_4}\left(1_\phi,2_\gamma^+,\ldots ,(n-1)_\gamma^-,n_{\overline{\phi}}\right)$ and will be used to calculate BI$_4$ NSD amplitudes in Section \ref{sec:NSD}.
\end{itemize}
First we will give a general parametrization of such tree amplitudes, then in the following section we will fix all ambiguities using two complimentary approaches. 

The analytic structure of the $\text{mDBI}_4$ amplitudes have the general form of a rational function of external kinematic data and can be split into contributions 

\begin{align}
  &\mathcal{A}_n^{\text{mDBI}_4}\left(1_\phi,2_\gamma^+,\ldots ,(n-1)_\gamma^\pm,n_{\overline{\phi}}\right) \nonumber\\
  &= \mathcal{A}_n^{\text{mDBI}_4}\left(1_\phi,2_\gamma^+,\ldots ,(n-1)_\gamma^\pm,n_{\overline{\phi}}\right)\biggr\vert_{\text{factoring}}+ \mathcal{A}_n^{\text{mDBI}_4}\left(1_\phi,2_\gamma^+,\ldots ,(n-1)_\gamma^\pm,n_{\overline{\phi}}\right)\biggr\vert_{\text{contact}}.
\end{align}
The \textit{factoring} terms contain all kinematic singularities, which are required to be simple poles on invariant masses of subsets of external momenta, and have residues given by sums of products of lower point amplitudes. In this sense the factoring terms are recursively determined by amplitudes at lower multiplicity. In EFTs (such as $\text{mDBI}_4$) the resulting rational function is incompletely determined by factorization, and there is some remaining polynomial ambiguity. These ambiguities are contained in the \textit{contact} contribution, which encodes all independent local operators compatible with the assumed properties of the model. We can give a general parametrization of these contact contributions for $\text{mDBI}_4$ through a combination of dimensional analysis, little group scaling and analysis of the massless limit. 

In $d=4$ the amplitudes have mass dimension $[\mathcal{A}_n] = 4-n$, this includes both dimensionful coupling constants and kinematic dependence. The contact contribution is then a sum over terms of the schematic form
\begin{equation}
  \mathcal{A}_n^{\text{mDBI}_4}\left(1_\phi,2_\gamma^+,\ldots ,(n-1)_\gamma^\pm,n_{\overline{\phi}}\right)\biggr\vert_{\text{contact}} \sim \frac{1}{\Lambda^m}F^{\pm}_n\left(\{|i\rangle,|i]\},p^{1,n}_{[4]},\mu^2\right),
\end{equation}
where $[\Lambda]=1$ is the dimensionful scale appearing in the Born-Infeld action (\ref{BIaction}) and $[F_n]-m = 4-n$. Since this is a contact contribution $F_n$ must be a polynomial in the Lorentz invariant spinor contractions and the mass of the scalar $\mu^2$. These polynomials must have the correct little group scaling dictated by their helicity configurations. This sets a lower-bound on the mass dimension of $F_n^\pm$ since we must have
\begin{align}
  F_n^+\left(\{|i\rangle,|i]\},p^{1,n}_{[4]},\mu^2\right) &\sim |2]^2|3]^2\ldots |n-1]^2 G_n^+\left(\{|i\rangle,|i]\},p^{1,n}_{[4]},\mu^2\right) \nonumber\\
  F_n^-\left(\{|i\rangle,|i]\},p^{1,n}_{[4]},\mu^2\right) &\sim |2]^2|3]^2\ldots |n-1\rangle^2 G_n^-\left(\{|i\rangle,|i]\},p^{1,n}_{[4]},\mu^2\right).
\end{align}
Here $G^\pm$ are again polynomials in helicity spinors, but with zero little group weight. Since $[G^\pm]\geq 0$ we must have $[F_n^\pm]\geq n-2$. 

Next we impose that the complete $\text{mDBI}_4$ amplitudes should agree with $\mathcal{N}=2$ $\text{BI}_4$ in the limit $\mu^2\rightarrow 0$. This constraint is quite powerful due to the conservation of a $U(1)_R$ duality charge in $\mathcal{N}=2$ $\text{BI}_4$. Up to an arbitrary normalization, the states of the $\mathcal{N}=2$ massless vector multiplet can be assigned the following additive quantum numbers

\begin{equation} \label{duality}
Q[\gamma^\pm] = \pm 1, \hspace{5mm} Q[\psi_{1,2}^\pm] = \pm 1/2, \hspace{5mm} Q[\phi] = Q[\overline{\phi}] = 0.
\end{equation}

It is straightforward to show that these charges are conserved at tree-level since they are conserved by the leading $n=4$ interactions and the entire tree-level S-matrix is constructible by on-shell subtracted recursion \cite{Elvang:2018dco}. Note that this $U(1)_R$ is \textit{not} a subgroup of the $SU(2)_R$ symmetry group under which the fermions $\psi_A$ transform as a doublet. It is an independent symmetry which enhances the full R-symmetry group of $\mathcal{N}=2$ $\text{BI}_4$ to $U(2)_R$. The analogous enhancement of R-symmetry in maximally supersymmetric Born-Infeld was first discussed in \cite{Heydeman:2017yww}. As a consequence of the conservation of the duality charges (\ref{duality}), in the NSD and MHV sectors of $\text{mDBI}_4$ the massless limits are given by 

\begin{align}
&\mathcal{A}_4^{\text{mDBI}_4}\left(1_\phi,2_\gamma^+,3_\gamma^+,4_{\overline{\phi}}\right) \xrightarrow{\mu^2\rightarrow 0} 0, \nonumber\\
&\mathcal{A}_4^{\text{mDBI}_4}\left(1_\phi,2_\gamma^+,3_\gamma^-,4_{\overline{\phi}}\right) \xrightarrow{\mu^2\rightarrow 0} -\langle 3|p_1|2]^2, \nonumber\\
&\mathcal{A}_n^{\text{mDBI}_4}\left(1_\phi,2_\gamma^+,\ldots ,(n-1)_\gamma^\pm,n_{\overline{\phi}}\right) \xrightarrow{\mu^2\rightarrow 0} 0, \hspace{5mm} n>4.
\end{align}

Due to the different singularity structure, the factoring and contact terms cannot interfere in this limit, and so the contact terms must vanish independently. For this to happen the contact terms must be proportional to some positive power of $\mu^2$, which further increases the minimal dimension to $[F_n^\pm]\geq n$. The contact terms must then have the schematic form 
\begin{align}
  \mathcal{A}_n^{\text{mDBI}_4}\left(1_\phi,2_\gamma^+,\ldots ,(n-1)_\gamma^+,n_{\overline{\phi}}\right)\biggr\vert_{\text{contact}} &\sim \frac{\mu^2}{\Lambda^{2n-4}}|2]^2|3]^2\ldots |n-1]^2 + \mathcal{O}\left(\frac{1}{\Lambda^{2n-3}}\right) \nonumber\\
  \mathcal{A}_n^{\text{mDBI}_4}\left(1_\phi,2_\gamma^+,\ldots ,(n-1)_\gamma^-,n_{\overline{\phi}}\right)\biggr\vert_{\text{contact}} &\sim \frac{\mu^2}{\Lambda^{2n-4}}|2]^2|3]^2\ldots |n-1\rangle^2 + \mathcal{O}\left(\frac{1}{\Lambda^{2n-3}}\right).
\end{align}

It is easy to see that in the $(n-1)^-$ (MHV) case no contact term of this leading mass dimension can exist since there is no non-vanishing way to contract the angle spinors.

Next we recall our discussion from Section \ref{sec:susydec}, such contact contributions should not introduce any spurious dimensionful parameters which might appear in the final integrated amplitude. We should not consider contributions with more inverse powers of $\Lambda$ at a fixed multiplicity $n$. In Appendix \ref{app:Contact} we give a short proof that at each multiplicity $n$ there is a unique contact term, the final result can be parametrized as
\begin{equation}
  \mathcal{A}_{n}^{\text{mDBI}_4}\left(1_\phi,2_\gamma^+,\ldots ,(n-1)_\gamma^+,n_{\overline{\phi}}\right)\biggr\vert_{\text{contact}} = \frac{c_n\mu^2}{\Lambda^{2n-4}} \left([23]^2[45]^2\ldots [n-2,n-1]^2+\ldots \right),
\end{equation}
where $+\ldots $ denotes the sum over all ways of partitioning the set $\{2,\ldots ,n-1\}$ into subsets of length 2. Such local matrix elements can be generated from local operators of the form

\begin{equation}
  \mathcal{L}_{\text{mDBI}_4} \supset \frac{c_{2n} \mu^2}{\Lambda^{4n-4}}|\phi|^2 \left(F^+_{\alpha\beta} F^{+\alpha\beta}\right)^{n-1}.
\end{equation}
In subsequent sections the $\Lambda$ dependence of the scattering amplitudes will be suppressed, they can trivially be restored by dimensional analysis.

The remarkable result, which we will verify using two complimentary approaches in the following sections, is that if we define $\text{mDBI}_4$ as the dimensional reduction of $\text{BI}_6$ as described above, then $c_n=0$ for $n>4$. The complete tree amplitudes are then completely fixed by recursive factorization into the fundamental 4-point $\text{mDBI}_4$ amplitudes. 

%%%%%%%%%%%%%%%%%%%%%%%%%%%%%%
\subsection{First Method: Massive KLT Relations}
\label{sec:KLT}
%%%%%%%%%%%%%%%%%%%%%%%%%%%%%%

As discussed in Section \ref{sec:massive}, the tree-level amplitudes of Born-Infeld in $d$-dimensions are given by the KLT product 
\begin{equation}
  \text{BI}_d = \text{YM}_d \otimes_{\text{KLT}} \chi \text{PT}_d,
\end{equation}
where $\chi\text{PT}_d$ denotes the $\frac{SU(N)\times SU(N)}{SU(N)}$ non-linear sigma model in $d$-dimensions. Beginning with $d=6$ we can (formally) calculate tree amplitudes in $\text{BI}_6$ from the tree amplitudes for $\text{YM}_6$ and $\chi\text{PT}_6$ using the dimension independent form of the KLT product. Since we do not require the completely general 6d Born-Infeld amplitudes, only the configuration in Figure \ref{helicitytable}, we can dimensionally reduce the 6d KLT relations into a form of \textit{massive} KLT relations by separating the 4d and extra-dimensional components of the momenta. This amounts to taking the dimension independent form the KLT relations and making the replacements
\begin{equation} \label{dred}
  s_{1i} \rightarrow s_{1i}+\mu^2, \;\;\;\; s_{nj} \rightarrow s_{nj}+\mu^2,
\end{equation}
where $i\neq n$ and $j \neq 1$ (Note that we are defining our Mandelstam invariants as $s_{ij} \equiv (p_i+p_j)^2$). Using this prescription the needed KLT relations 
\begin{equation}
  \text{mDBI}_4 = \text{YM+mAdj}_4 \otimes_{\text{KLT}} \text{m}\chi \text{PT}_4,
\end{equation}
up to $n=8$ take the explicit form \cite{Bern:1998sv}
\begin{align} \label{4mKLT}
  &\mathcal{A}_4^{\text{mDBI}_4}\left(1_\phi,2_\gamma,3_\gamma,4_{\overline{\phi}}\right) = (s_{12}+\mu^2)\mathcal{A}_4^{\text{YM+mAdj}_4}[1_\phi,2_g,3_g,4_{\overline{\phi}}]\mathcal{A}_4^{\text{m}\chi\text{PT}_4}\left[\mathbf{1},2,\mathbf{4},3\right], \\
  &\label{6mKLT}\mathcal{A}_6^{\text{mDBI}_4}\left(1_\phi,2_\gamma,3_\gamma,4_\gamma,5_\gamma,6_{\overline{\phi}}\right) \nonumber\\
  &\hspace{5mm}= (s_{12}+\mu^2)s_{45}\mathcal{A}_6^{\text{YM+mAdj}_4}[1_\phi,2_g,3_g,4_g,5_g,6_{\overline{\phi}}]\nonumber\\
  &\hspace{10mm}\times\left(s_{35}\mathcal{A}_6^{\text{m}\chi\text{PT}_4}\left[\mathbf{1},5,3,4,\mathbf{6},2\right]+(s_{34}+s_{35})\mathcal{A}_6^{\text{m}\chi\text{PT}_4}\left[\mathbf{1},5,4,3,\mathbf{6},2\right]\right) \nonumber\\
  &\hspace{10mm}+ \mathcal{P}\left(2,3,4\right),\\
&\label{8mKLT}\mathcal{A}_8^{\text{mDBI}_4}\left(1_\phi,2_\gamma,3_\gamma,4_\gamma,5_\gamma,6_\gamma,7_\gamma,8_{\overline{\phi}}\right) \nonumber\\
&\hspace{5mm}= (s_{12}+\mu^2)s_{67}\mathcal{A}_8^{\text{YM+mAdj}_4}[1_\phi,2_g,3_g,4_g,5_g,6_g,7_g,8_{\overline{\phi}}] \nonumber\\
&\hspace{10mm}\times \left[(s_{13}+\mu^2)s_{14}\left(s_{57}\mathcal{A}^{\text{m}\chi\text{PT}_4}_8[\mathbf{1},7,5,6,\mathbf{8},2,3,4] \right . \right . \nonumber \\
&\hspace{65mm} \left. + (s_{57}+s_{56})\mathcal{A}^{\text{m}\chi\text{PT}_4}_8[\mathbf{1},7,6,5,\mathbf{8},2,3,4]\right) \nonumber\\
&\hspace{15mm}+(s_{13}+\mu^2)(s_{14}+s_{34}+\mu^2)\left(s_{57}\mathcal{A}^{\text{m}\chi\text{PT}_4}_8[\mathbf{1},7,5,6,\mathbf{8},2,4,3]\right.\nonumber\\
&\hspace{65mm}\left. + (s_{57}+s_{56})\mathcal{A}^{\text{m}\chi\text{PT}_4}_8[\mathbf{1},7,6,5,\mathbf{8},2,4,3]\right)  \nonumber\\
&\hspace{15mm}+(s_{14}+\mu^2)(s_{13}+s_{23}+\mu^2)\left(s_{57}\mathcal{A}^{\text{m}\chi\text{PT}_4}_8[\mathbf{1},7,5,6,\mathbf{8},3,2,4]\right.\nonumber\\
&\left. \hspace{65mm}+ (s_{57}+s_{56})\mathcal{A}^{\text{m}\chi\text{PT}_4}_8[\mathbf{1},7,6,5,\mathbf{8},3,2,4]\right)  \nonumber\\
&\hspace{15mm}+(s_{13}+s_{23}+\mu^2)(s_{14}+s_{24}+\mu^2)\left(s_{57}\mathcal{A}^{\text{m}\chi\text{PT}_4}_8[\mathbf{1},7,5,6,\mathbf{8},3,4,2]\right. \nonumber\\
&\hspace{65mm}\left. + (s_{57}+s_{56})\mathcal{A}^{\text{m}\chi\text{PT}_4}_8[\mathbf{1},7,6,5,\mathbf{8},3,4,2]\right)  \nonumber\\
&\hspace{15mm}+(s_{13}+\mu^2)(s_{14}+s_{24}+s_{34}+\mu^2)\left(s_{57}\mathcal{A}^{\text{m}\chi\text{PT}_4}_8[\mathbf{1},7,5,6,\mathbf{8},4,2,3] \right.\nonumber\\
&\left. \hspace{65mm}+ (s_{57}+s_{56})\mathcal{A}^{\text{m}\chi\text{PT}_4}_8[\mathbf{1},7,6,5,\mathbf{8},4,2,3]\right)  \nonumber\\
&\hspace{15mm}+(s_{13}+s_{23}+\mu^2)(s_{14}+s_{34}+s_{24}+\mu^2)\left(s_{57}\mathcal{A}^{\text{m}\chi\text{PT}_4}_8[\mathbf{1},7,5,6,\mathbf{8},4,3,2] \right. \nonumber\\
&\hspace{65mm}\left.\left.+ (s_{57}+s_{56})\mathcal{A}^{\text{m}\chi\text{PT}_4}_8[\mathbf{1},7,6,5,\mathbf{8},4,3,2]\right) \right] \nonumber\\
&\hspace{10mm}+ \mathcal{P}\left(2,3,4,5,6\right).
\end{align}
In the m$\chi$PT amplitudes bolded momenta denote massive particles.

Note that these expressions differ by an overall sign from the expressions given in \cite{Elvang:2015rqa} due to our conventions for the Mandelstam invariants. Below we will describe the calculation of both $\text{YM}+\text{mAdj}_4$ and $\text{m}\chi \text{PT}_4$ amplitudes and then give the result of the double copy.

%%%%%%%%%%%%%%%%%%%%%%%%%%%%%%
\subsubsection{\texorpdfstring{$\text{YM+mAdj}_4$}{YM+mAdj4} from Massive BCFW}
\label{sec:YM}
%%%%%%%%%%%%%%%%%%%%%%%%%%%%%%

The needed tree-level amplitudes of YM+mAdj can be calculated using BCFW recursion from 3-point input. Since this model should have only marginal couplings between the gluons and massive adjoint scalar, the tree-level amplitudes are completely fixed by gauge invariance. This approach was first used in \cite{Badger:2005zh}, below we give a brief review.

The seed amplitudes for the recursion are
\begin{equation}
  \mathcal{A}_3^{\text{YM+mAdj}_4}[1_\phi,2_g^+,3_{\overline{\phi}}] = -\frac{[2|p_1|q\rangle}{\langle 2 q \rangle}, \hspace{5mm} \mathcal{A}_3^{\text{YM+mAdj}_4}[1_\phi,2_g^-,3_{\overline{\phi}}] = \frac{[\tilde{q}|p_1|2\rangle}{[\tilde{q}\; 2]},
\end{equation}
where $|q\rangle$ and $|\tilde{q}]$ are arbitrary. We want to calculate NSD amplitudes
\begin{equation}
  \mathcal{A}_n^{\text{YM+mAdj}_4}[1_\phi,2_g^+,3_g^+\ldots ,(n-1)_g^+,n_{\overline{\phi}}],
\end{equation}
using a BCFW shift
\begin{equation}
  |\hat{2}\rangle = |2\rangle - z |3\rangle, \hspace{5mm} |\hat{3}] = |3] + z |2].
\end{equation}
With the given color-ordering (and the fact that the shifted lines must sit on opposite sides of the factorization diagram) there are two types of factorization channel which could contribute:
 
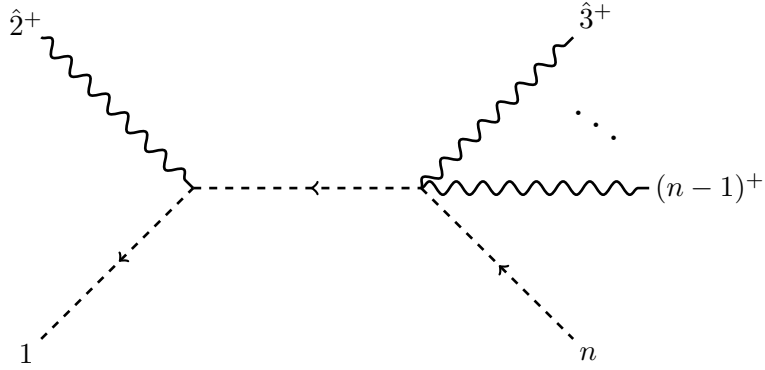
\begin{center}
{\begin{tikzpicture}[scale=1, line width=1 pt]
\draw[vector] (-2,2)--(0,0);
\draw[scalarbar] (-2,-2)--(0,0);
\draw[scalarbar] (0,0)--(3,0);
\draw[scalarbar] (3,0)--(5,-2);
\draw[vector] (3,0)--(5,2);
\draw[vector] (3,0)--(6,0);
\node at (-2.2,2.2) {$\hat{2}^+$};
\node at (-2.2,-2.2) {$1$};
\node at (5.3,2.3) {$\hat{3}^+$};
\node at (6.8,0) {$(n-1)^+$}; 
\node at (5.2,-2.2) {$n$};
\node at (5.3,1) {\LARGE $\mathbf{\ddots}$};
\end{tikzpicture}}
\end{center}
and
 
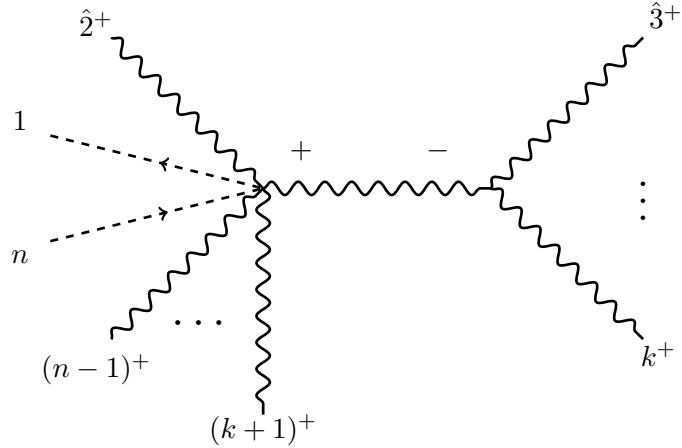
\begin{center}
{\begin{tikzpicture}[scale=1, line width=1 pt]
\draw[vector] (-2,2)--(0,0);
\draw[vector] (-2,-2)--(0,0);
\draw[scalarbar] (-2.8,0.7)--(0,0);
\draw[scalar] (-2.8,-0.7)--(0,0);
\draw[vector] (0,0)--(3,0);
\draw[vector] (3,0)--(5,-2);
\draw[vector] (3,0)--(5,2);
\draw[vector] (0,0)--(0,-3);
\node at (-3.2,0.9) {$1$};
\node at (-3.2,-0.9) {$n$};
\node at (-2.2,2.2) {$\hat{2}^+$};
\node at (-2.2,-2.4) {$(n-1)^+$};
\node at (5.3,2.3) {$\hat{3}^+$};
\node at (5.2,-2.2) {$k^+$};
\node at (-0.8,-1.8) {\LARGE $\cdots$};
\node at (5,0) {\LARGE $\mathbf{\vdots}$};
\node at (0,-3.2) {$(k+1)^+$};
\node at (0.5,0.5) {$+$};
\node at (2.3,0.5) {$-$};
\end{tikzpicture}}
\end{center}
Interestingly, the second diagram never contributes. The argument for this is has two parts, first we consider diagrams with $k >4$. In this case the right-hand amplitude is of the form $\mathcal{A}_{k-1}^{\text{YM+mAdj}_4}[-,+,\ldots ,+]$ which vanishes at tree-level. For the case $k=4$ the right-hand amplitude is simply the pure Yang-Mills 3-point amplitude\footnote{\label{foot:convention}Here and subsequently, we use the convention $| - p ] = i | p ]$ and $| - p \rangle = i | p \rangle$. This is because the prescription for dimensional reduction to $3d$ we use in Section~\ref{sec:Tdual} requires that we treat the angle and square spinors ``democratically''. A consequence of this convention choice is that the Parke-Taylor amplitudes acquire an additional factor of $-1$ for an even number of external states.}
\begin{equation}
  \mathcal{A}_{3}^{\text{YM+mAdj}_4}\left[(-\hat{p}_{34})_g^-,\hat{3}_g^+,4_g^+\right] = \frac{[\hat{3}4]^3}{[4,-\hat{p}_{34}][-\hat{p}_{34},\hat{3}]}.
\end{equation}
On the factorization channel we have $[\hat{3}4]=0$ and therefore this amplitude vanishes. So we see that only a single factorization channel contributes at each recursive step. Explicitly the BCFW recursion relation takes the form
\begin{align} \label{BCFW}
  &\mathcal{A}_{n}^{\text{YM+mAdj}_4}\left[1_\phi,2_g^+,3_g^+,\ldots ,(n-1)_g^+,n_{\overline{\phi}}\right]\nonumber\\
  &\hspace{5mm}= \frac{\mathcal{A}_3^{\text{YM+mAdj}_4}[1_\phi,\hat{2}_g^+,(-\hat{p}_{12})_{\overline{\phi}}]\mathcal{A}_{n-1}^{\text{YM+mAdj}_4}[(\hat{p}_{12})_\phi,\hat{3}_g^+,4_g^+,\ldots ,(n-2)_g^+,(n-1)_{\overline{\phi}}]}{s_{12}+\mu^2}.
\end{align}
We will now use this to calculate the amplitudes up to $n=8$. Here (and subsequently) we will use the convenient shorthand notation
\begin{equation}
  p_{1,k} \equiv p_{12\ldots k}, \hspace{5mm} D_n \equiv \langle 23\rangle \langle 34 \rangle \ldots  \langle n-2,n-1\rangle (s_{12}+\mu^2)(s_{123}+\mu^2)\ldots (s_{12\ldots n-2}+\mu^2).
\end{equation}

At 4-point we need both the NSD and MHV amplitudes
\begin{align}
  \mathcal{A}_4^{\text{YM+mAdj}_4}[1_\phi,2_g^+,3_g^+,4_{\overline{\phi}}] &=  -\frac{\mu^2[23]}{\langle 23\rangle (s_{12}+\mu^2)},
\end{align}
and 
\begin{align}
  \mathcal{A}_4^{\text{YM+mAdj}_4}[1_\phi,2_g^+,3_g^-,4_{\overline{\phi}}] &= -\frac{\langle 3|p_1|2]^2}{s_{23}(s_{12}+\mu^2)}.
\end{align}
At 6-point we will only need amplitudes in the NSD sector
\begin{equation}
   \mathcal{A}_{6}^{\text{YM+mAdj}_4}\left[1_\phi,2_g^+,3_g^+,4_g^+,5_g^+,6_{\overline{\phi}}\right] = -\frac{\mu^2[2|p_1\cdot p_{23}\cdot p_{45}\cdot p_6|5]}{D_6}. 
\end{equation}
Similarly at 8-point
\begin{align}
  &\mathcal{A}_{8}^{\text{YM+mAdj}_4}\left[1_\phi,2_g^+,3_g^+,4_g^+,5_g^+,6_g^+,7_g^+,8_{\overline{\phi}}\right]  \nonumber\\
   &\hspace{5mm} = \frac{1}{D_8}\left[-(\mu^2)^3[2|p_1\cdot p_{23} \cdot p_{67}\cdot p_8|7]+(\mu^2)^2[2|p_1\cdot p_{23}\cdot p_{4,8}\cdot p_{5,8} \cdot p_{67}\cdot p_8|7]\right. \nonumber\\
  &\hspace{20mm}+(\mu^2)^2[2|p_1\cdot p_{23}\cdot p_{5,8}\cdot p_{6,8} \cdot p_{67}\cdot p_8|7]\nonumber\\
  &\hspace{20mm}\left. -\mu^2[2|p_1\cdot p_{23}\cdot p_{4,8}\cdot p_{5,8}\cdot p_{5,8}\cdot p_{6,8} \cdot p_{67}\cdot p_8|7]\right].
\end{align}
All multiplicity results for these amplitudes have been calculated in \cite{Forde:2005ue}, but we will not need explicit expressions beyond 8-points.

%%%%%%%%%%%%%%%%%%%%%%%%%%%%%%
\subsubsection{\texorpdfstring{$\text{m}\chi \text{PT}_4$}{mChiPT4} from Soft Limits and Dimensional Reduction}
\label{sec:ChPT}
%%%%%%%%%%%%%%%%%%%%%%%%%%%%%%

The needed tree level amplitudes for $\chi\text{PT}_d$ can be calculated using the \textit{soft bootstrap} approach \cite{Cheung:2014dqa, Cheung:2016drk, Elvang:2018dco}. While it is certainly possible to setup formal recursion relations analogous to the BCFW recursion used above (this is the so-called \textit{subtracted recursion} \cite{Kampf:2013vha, Cheung:2015ota}), in practice since this is such a simple model there is a more efficient approach. We note that locality is manifest in the $\chi$PT amplitudes, and so we can treat the contact terms of lower-point amplitudes as ``vertex rules'', gluing them together in a diagrammatic expansion. This will automatically generate expressions with the correct factorization properties (which can be verified straightforwardly \textit{post hoc} by computing residues), the remaining ambiguity is contained in the contact terms. These ambiguous contributions can then be determined by imposing the Adler zero, that is, single soft limit which vanish at $\mathcal{O}\left(p\right)$ \cite{Adler:1964um}. 

We start with the flavor-ordered 4-point amplitude
\begin{equation}
  \mathcal{A}_4^{\chi\text{PT}_d}\left[1,2,3,4\right] = s_{13}. 
\end{equation}
With the dimensionful coupling suppressed, the $\chi\text{PT}_d$ tree-amplitudes take a dimension independent form. Similar to the definition of $\text{mDBI}_4$ we define the model $\text{m}\chi\text{PT}_4$ as the tree amplitudes of $\chi\text{PT}_6$ with momenta in the configuration given in Figure \ref{helicitytable}. Operationally these amplitudes are calculated using the replacement rules (\ref{dred}), on the $\chi\text{PT}_d$ amplitudes, similar to the way we derived the massive KLT relations above.

Now we turn to the explicit calculation of the 6-point $\chi\text{PT}_d$ amplitude. In this case the factoring part of the amplitude corresponds to diagrams with a unique topology
 
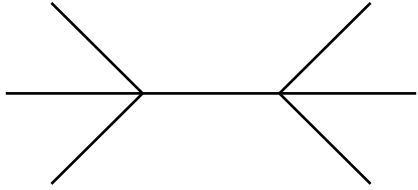
\begin{center}
{\begin{tikzpicture}[scale=0.6, line width=1 pt]
    \draw (-2,2)--(0,0);
    \draw (-3,0)--(0,0);
    \draw (-2,-2)--(0,0);
    \draw (0,0)--(3,0);
    \draw (3,0)--(5,2);
    \draw (3,0)--(6,0);
    \draw (3,0)--(5,-2);
    \node at (0,2.5) {};
\end{tikzpicture}}
\end{center}
There are three inequivalent cyclic permutations of the external labels $[1,2,3,4,5,6]$, so the factoring part of the six point amplitude has the form
\begin{equation}
  \mathcal{A}_6^{\chi\text{PT}_d}\left[1,2,3,4,5,6\right]\biggr\vert_{\text{factoring}} = \frac{s_{13}s_{46}}{s_{123}}+\frac{s_{24}s_{51}}{s_{234}}+\frac{s_{35}s_{62}}{s_{345}}. 
\end{equation}
This differs from the full answer by a possible contact term. Such a contact contribution is fixed by demanding that the amplitude vanishes in the soft limit of each particle. It is straightforward to verify that the following expression satisfies all of the aforementioned properties
\begin{equation}
  \mathcal{A}_6^{\chi\text{PT}_d}\left[1,2,3,4,5,6\right] = \frac{s_{13}s_{46}}{s_{123}}+\frac{s_{24}s_{51}}{s_{234}}+\frac{s_{35}s_{62}}{s_{345}}- s_{135}.
\end{equation}
We can then convert this into an $\text{m}\chi\text{PT}_4$ amplitude with particles 1 and 5 massive for later use in the KLT product
\begin{equation}
  \mathcal{A}_6^{\text{m}\chi\text{PT}_4}\left[\mathbf{1},2,3,4,\mathbf{5},6\right] = \frac{(s_{13}+\mu^2)s_{46}}{s_{123}+\mu^2}+\frac{s_{24}s_{51}}{s_{234}}+\frac{(s_{35}+\mu^2)s_{62}}{s_{345}+\mu^2}- s_{135}. 
\end{equation}

For $n=8$ there are three distinct factorization topologies we need to consider, two constructed from 4-point vertices
 
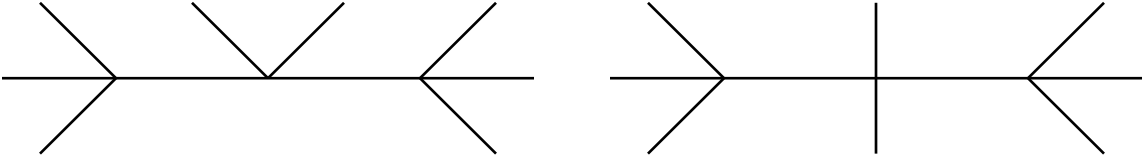
\begin{center}
{\begin{tikzpicture}[scale=1, line width=1 pt]
    \draw (-1,1)--(0,0);
    \draw (-1.5,0)--(0,0);
    \draw (-1,-1)--(0,0);
    \draw (0,0)--(4,0);
    \draw (2,0)--(3,1);
    \draw (2,0)--(1,1);
    \draw (4,0)--(5,1);
    \draw (4,0)--(5.5,0);
    \draw (4,0)--(5,-1);
    %%%%%%%%%%
    \draw (8,0)--(7,1);
    \draw (6.5,0)--(13.5,0);
    \draw (8,0)--(7,-1);
    \draw (12,0)--(13,1);
    \draw (12,0)--(13,-1);
    \draw (10,1)--(10,-1);
    \node at (0,1.5) {};
\end{tikzpicture}}
\end{center}
and one from a 4-point and a 6-point vertex
 
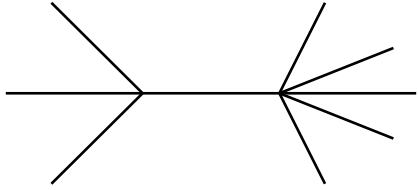
\begin{center}
{\begin{tikzpicture}[scale=0.6, line width=1 pt]
    \draw (-2,2)--(0,0);
    \draw (-3,0)--(0,0);
    \draw (-2,-2)--(0,0);
    \draw (0,0)--(3,0);
    \draw (3,0)--(5.5,1);
    \draw (3,0)--(6,0);
    \draw (3,0)--(5.5,-1);
    \draw (3,0)--(4,2);
    \draw (3,0)--(4,-2);
    \node at (0,2.5) {};
\end{tikzpicture}}
\end{center}
It is straightforward to write down the factoring part of this amplitude
\begin{align}
  &\mathcal{A}_8^{\chi\text{PT}_d}\left[1,2,3,4,5,6,7,8\right]\biggr\vert_{\text{factoring}} \nonumber\\
  &\hspace{10mm}= \frac{s_{13}s_{1235}s_{68}}{s_{123}s_{678}} + \frac{1}{2}\left(\frac{s_{13}s_{48}s_{57}}{s_{123}s_{567}}\right)-\frac{s_{13}s_{468}}{s_{123}} +\mathcal{C}\left(1,2,3,4,5,6,7,8\right).
\end{align}
where $\mathcal{C}$ denotes the sum over all \textit{cyclic} permutations. The contact terms we need to add can be found straightforwardly by taking soft limits, the result is
\begin{align}
  &\mathcal{A}_8^{\chi\text{PT}_d}\left[1,2,3,4,5,6,7,8\right]\nonumber\\
  &\hspace{10mm}= \left[ \frac{s_{13}s_{1235}s_{68}}{s_{123}s_{678}} + \frac{1}{2}\left(\frac{s_{13}s_{48}s_{57}}{s_{123}s_{567}}\right)-\frac{s_{13}s_{468}}{s_{123}} +\mathcal{C}\left(1,2,3,4,5,6,7,8\right)\right] + s_{2468}.
\end{align}
Constructing the $\text{m}\chi\text{PT}_4$ amplitude with particle 1 and 5 massive gives
\begin{align}
  &\mathcal{A}_8^{\text{m}\chi\text{PT}_4}\left[\mathbf{1},2,3,4,\mathbf{5},6,7,8\right]  \nonumber\\
  &= \frac{(s_{13}+\mu^2)s_{1235}s_{68}}{(s_{123}+\mu^2)s_{678}} + \frac{(s_{13}+\mu^2)s_{48}(s_{57}+\mu^2)}{(s_{123}+\mu^2)(s_{567}+\mu^2)}-\frac{(s_{13}+\mu^2)s_{468}}{s_{123}+\mu^2} + \frac{s_{24}s_{2346}(s_{71}+\mu^2)}{s_{234}(s_{781}+\mu^2)} \nonumber\\
  &\hspace{5mm}+ \frac{s_{24}s_{51}s_{68}}{s_{234}s_{678}}-\frac{s_{24}s_{571}}{s_{234}}+ \frac{(s_{35}+\mu^2)(s_{3457}+\mu^2)s_{82}}{(s_{345}+\mu^2)(s_{812}+\mu^2)} +\frac{(s_{35}+\mu^2)s_{62}(s_{71}+\mu^2)}{(s_{345}+\mu^2)(s_{781}+\mu^2)}\nonumber\\
 &\hspace{5mm}-\frac{(s_{35}+\mu^2)s_{682}}{s_{345}+\mu^2}+ \frac{s_{46}(s_{4568}+\mu^2)(s_{13}+\mu^2)}{(s_{456}+\mu^2)(s_{123}+\mu^2)}+ \frac{s_{46}s_{73}s_{82}}{(s_{456}+\mu^2)(s_{812}+\mu^2)}-\frac{s_{46}(s_{713}+\mu^2)}{s_{456}+\mu^2} \nonumber\\
  &\hspace{5mm} + \frac{(s_{57}+\mu^2)s_{5671}s_{24}}{(s_{567}+\mu^2)s_{234}}-\frac{(s_{57}+\mu^2)s_{824}}{s_{567}+\mu^2} + \frac{s_{68}(s_{6781}+\mu^2)(s_{35}+\mu^2)}{s_{678}(s_{345}+\mu^2)}-\frac{s_{68}s_{135}}{s_{678}} \nonumber\\
  &\hspace{5mm} + \frac{(s_{71}+\mu^2)(s_{7812}+\mu^2)s_{46}}{(s_{781}+\mu^2)(s_{456}+\mu^2)} -\frac{(s_{71}+\mu^2)s_{246}}{s_{781}+\mu^2}+ \frac{s_{82}(s_{8123}+\mu^2)(s_{57}+\mu^2)}{(s_{812}+\mu^2)(s_{567}+\mu^2)}  \nonumber \\
  &\hspace{5mm} -\frac{s_{82}(s_{357}+\mu^2)}{s_{812}+\mu^2} +s_{2468}.
\end{align}

Simple closed form expressions for all $\chi \text{PT}_d$ amplitudes are not known, but this procedure is simple enough that it can be implemented efficiently to calculate amplitudes up to the desired multiplicity. As in the previous section we will only need explicit expressions up to $n=8$. 

%%%%%%%%%%%%%%%%%%%%%%%%%%%%%%
\subsubsection{Result of Double Copy}
\label{sec:Result}
%%%%%%%%%%%%%%%%%%%%%%%%%%%%%%

We can begin with the calculation of the 4-point amplitudes of $\text{mDBI}_4$, which are simple enough to be evaluated by hand without difficulty
\begin{align} \label{KLT4NSD}
  \mathcal{A}_4^{\text{mDBI}_4}\left(1_\phi,2_\gamma^+,3_\gamma^+,4_{\overline{\phi}}\right) &= (s_{12}+\mu^2)\mathcal{A}_4^{\text{YM+mAdj}}[1_\phi,2^+_g,3^+_g,4_{\overline{\phi}}]\mathcal{A}_4^{\text{m}\chi\text{PT}}\left[\mathbf{1},2,\mathbf{4},3\right] \nonumber\\
                                                                                           &=  (s_{12}+\mu^2)\left[ -\frac{\mu^2 [23]}{\langle 23\rangle (s_{12}+\mu^2)}\right] \left[s_{23}\right] \nonumber\\
                                                                                           &= -\mu^2 [23]^2,
\end{align}
and
\begin{align}  \label{KLT4MHV}
  \mathcal{A}_4^{\text{mDBI}_4}\left(1_\phi,2_\gamma^+,3_\gamma^-,4_{\overline{\phi}}\right) &= (s_{12}+\mu^2)\mathcal{A}_4^{\text{YM+mAdj}}[1_\phi,2^+_g,3^-_g,4_{\overline{\phi}}]\mathcal{A}_4^{\text{m}\chi\text{PT}}\left[\mathbf{1},2,\mathbf{4},3\right] \nonumber\\
                                                                                           &= (s_{12}+\mu^2)\left[ -\frac{\langle 3|p_1|2]^2}{s_{23}(s_{12}+\mu^2)}\right] \left[s_{23}\right] \nonumber\\
                                                                                           &= -\langle 3|p_1|2]^2.
\end{align}
We will also need the 4-point pure Born-Infeld amplitude. This can also be calculated with the (massless) KLT product using the 4-point Parke-Taylor gluon amplitude 
\begin{align}
  \mathcal{A}_4^{\text{mDBI}_4}\left(1_\gamma^+,2_\gamma^+,3_\gamma^-,4_\gamma^-\right) &= s_{12}\left[-\frac{[12]^3}{[23][34][41]}\right][s_{23}]\nonumber\\
                                                                                      &= [12]^2\langle 34\rangle^2.
\end{align}
Notice that due to our convention choice (see comments in footnote~\ref{foot:convention}), the Parke-Taylor amplitude above has an additional factor of $-1$.

Simplifying the massive KLT relations algebraically beyond 4-point is a daunting task. Fortunately it is straightforward to construct a general Ansatz for the higher-multiplicity amplitudes. Beginning with the NSD 6-point amplitude we know the answer should have the form 
\begin{align}
  &\mathcal{A}_6^{\text{mDBI}_4}\left(1_\phi,2_\gamma^+,3_\gamma^+,4_\gamma^+,5_\gamma^+,6_{\overline{\phi}}\right) \nonumber\\
  &\hspace{15mm} =  \frac{1}{4}\left[\frac{(\mu^2)^2[23]^2[45]^2}{s_{123}+\mu^2}+\mathcal{P}\left(2,3,4,5\right)\right] + c_6\mu^2\left([23]^2[45]^2+[24]^2[35]^2+[25]^2[34]^2\right).
\end{align}
This expression has the correct factorization singularities consistent with the known 4-point amplitudes, and a polynomial ambiguity parametrized by a single coefficient $c_6$, as discussed above. To determine the coefficient $c_6$ we numerically evaluate the KLT sum (\ref{6mKLT}) on several sets of randomly generated kinematic variables and compare with a numerical evaluation of the Ansatz. For more than one choice of kinematics this overconstrains the problem and allows us to both verify the validity of the Ansatz and determine the value of the coefficient. Doing so we find that the Ansatz is valid and $c_6=0$;  the amplitude is simply
\begin{equation}
\label{KLT6NSD}
  \mathcal{A}_6^{\text{mDBI}_4}\left(1_\phi,2_\gamma^+,3_\gamma^+,4_\gamma^+,5_\gamma^+,6_{\overline{\phi}}\right) = \frac{1}{4}\left[\frac{(\mu^2)^2[23]^2[45]^2}{s_{123}+\mu^2}\right]+\mathcal{P}\left(2,3,4,5\right).
\end{equation}

Next we calculate the MHV 6-point amplitude. As discussed in Section~\ref{sec:Structure}, in this case there are \textit{no} contact terms consistent with little group scaling and Bose symmetry. There is then no ambiguity in the answer, the result of gluing together the 4-point amplitudes on factorization channels is the unique correct result. We find
\begin{align}
\label{KLT6MHV}
  &\mathcal{A}_6^{\text{mDBI}_4}\left(1_\phi,2_\gamma^+,3_\gamma^+,4_\gamma^+,5_\gamma^-,6_{\overline{\phi}}\right) \nonumber\\
  &\hspace{20mm} = \frac{\mu^2}{2}\left[\frac{[23]^2\langle 5|p_6|4]^2}{s_{123}+\mu^2}+\frac{[34]^2\langle 5|p_1|2]^2}{s_{125}+\mu^2}+\frac{[34]^2\langle 5|p_{34}|2]^2}{s_{126}}\right] +\mathcal{P}\left(2,3,4\right).
\end{align}

At 8-point the method is the same, we begin with the calculation of the NSD amplitude. Using the result $c_6=0$, we should use an Ansatz of the form
\begin{align}
  &\mathcal{A}_8^{\text{mDBI}_4}\left(1_\phi,2_\gamma^+,3_\gamma^+,4_\gamma^+,5_\gamma^+,6_\gamma^+,7_\gamma^+,8_{\overline{\phi}}\right) \nonumber\\
  & = -\frac{1}{8}\left[\frac{(\mu^2)^3[23]^2[45]^2[67]^2}{(s_{123}+\mu^2)(s_{678}+\mu^2)}\right] +c_8 \mu^2 [23]^2[45]^2[67]^2+\mathcal{P}\left(2,3,4,5,6,7\right).
\end{align}
Explicit numerical evaluation of the massive KLT relations reveals the surprising result that $c_8=0$ also! Finally, as above the MHV 8-point amplitude is completely fixed by factorization
\begin{align}
  &\mathcal{A}_8^{\text{mDBI}_4}\left(1_\phi,2_\gamma^+,3_\gamma^+,4_\gamma^+,5_\gamma^+,6_\gamma^+,7_\gamma^-,8_{\overline{\phi}}\right)\nonumber\\
  &\hspace{5mm}= -\frac{(\mu^2)^2 }{4}\left[\frac{[23]^2[45]^2\langle 7|p_{8}|6]^2}{(s_{123}+\mu^2)(s_{678}+\mu^2)}+\frac{[23]^2[45]^2\langle 7|p_{123}|6]^2}{(s_{123}+\mu^2)(s_{458}+\mu^2)}+\frac{[23]^2[45]^2\langle 7|p_{1}|6]^2}{(s_{167}+\mu^2)(s_{458}+\mu^2)}\right. \nonumber\\
  &\hspace{25mm}\left. + \frac{[34]^2[56]^2\langle 7|p_{34}|2]^2}{s_{347}(s_{568}+\mu^2)}+ \frac{[23]^2[56]^2\langle 7|p_{56}|4]^2}{s_{567}(s_{123}+\mu^2)}\right] + \mathcal{P}\left(2,3,4,5,6\right).
\end{align}

You may notice we had to work very hard just to calculate a few numbers ($c_6$ and $c_8$), both of which turned out to be zero. Continuing in this way quickly becomes computationally impractical (the number of terms in the KLT sum grows as $[(n-3)!]^2$, where $n$ is the number of external states). On the basis of these hard-won results it is tempting to conjecture that all such contact terms are zero beyond $n=4$, and so all we need is the easy part of the calculation, the construction of the factoring terms. This conjecture turns out to be correct, as we will prove in the next section from an argument based on T-duality properties of Born-Infeld, but is not at all obvious from the double copy.  

%%%%%%%%%%%%%%%%%%%%%%%%%%%%%%
\subsection{Second Method: T-Duality and Low-Energy Theorems}
\label{sec:Tdual}
%%%%%%%%%%%%%%%%%%%%%%%%%%%%%%

One of the most important and remarkable properties of D-branes (of which Born-Infeld and related models provide the low-energy effective description) is their behaviour under T-duality \cite{Bergshoeff:1996cy}. Though this is a non-perturbative stringy property, a useful remnant remains even in the tree-level scattering amplitudes of pure Born-Infeld. We will consider the configuration of momenta and polarizations described in Table \ref{Thelicitytable}.

\begin{table}
\begin{center}
 \begin{tabular}{|c|c|c|c|c|c|} 
   \hline
   & 1 & 2 & 3 & 4 & 5 \\ 
   \hline\hline
   $\vec{p}_{1,n}$ & x & x & & x & x \\
   \hline
   $\vec{\epsilon}_{1,n}$ & & & & x & x \\
   \hline
   $\vec{p}_{2,3,\ldots ,n-2}$ & x & x & & & \\
   \hline
   $\vec{\epsilon}_{2,3,\ldots ,n-2}$  & x & x & & &  \\
   \hline   
   $\vec{p}_{n-1}$ & x & x & & &  \\
   \hline
   $\vec{\epsilon}_{n-1}$ & & & x & & \\
   \hline
\end{tabular}
\end{center}
\caption{Kinematic configuration of momenta and polarizations of $\text{BI}_6$ defining the 3d dimensional reduction of $\text{mDBI}_4$. The 3-direction will be T-dualized, mapping the polarization of the photon labeled $n-1$ to a brane modulus.}
\label{Thelicitytable}
\end{table}

At tree-level all internal momenta are linear combinations of external momenta, and so in this configuration the amplitudes are independent of the 3-direction in momentum space. This means that the tree-amplitudes are invariant under compactification of the spatial 3-direction on $S^1$. T-duality in this context is the statement that a space-filling D5-brane on $\mathds{R}^{4+1}\times S^1$ with the radius of $S^1$ given by $R$, is equivalent to a codimension-1 D4-brane on $\mathds{R}^{4+1}\times S^1$, where $S^1$ is the transverse dimension with radius $\sim 1/R$. In the full string theory, T-duality relates infinite towers of KK and winding modes. In this low-energy EFT containing only the massless states as on-shell degrees of freedom, the only non-trivial mapping is between photons polarized in the compact direction on the D5-brane and the brane modulus of the D4-brane
\begin{equation}
  |\gamma^\top(\vec{p})\rangle \leftrightarrow |\Phi(\vec{p})\rangle.
\end{equation}
Since the tree-level amplitudes in Table \ref{Thelicitytable} are independent of the compactification, they must remain invariant in the limit $R\rightarrow 0$. In the T-dual configuration this corresponds to the decompactification limit in which we have a D4 brane embedded in $\mathds{R}^{5+1}$. In this limit, the spontaneous symmetry breaking pattern in the T-dual frame jumps discontinuously
\begin{equation}
\frac{\text{ISO}(4,1)\times \text{SO}(2)}{\text{ISO}(4,1)} \xrightarrow{R\rightarrow 0} \frac{\text{ISO}(5,1)}{\text{ISO}(4,1)}.
\end{equation}

The brane modulus is then identified as the Goldstone mode of both the translation symmetry in the 3-direction \textit{and} the  Lorentz transformations mixing the 3- and world-volume directions. In the physical scattering amplitudes this manifests as \textit{enhanced} soft theorems for the brane modulus 
\begin{equation}
\mathcal{A}^{\text{mDBI}_4}_n\left(1_\phi,2_\gamma^+,\ldots ,(n-2)_\gamma^+,(n-1)_\Phi,n_{\overline{\phi}}\right) \sim \mathcal{O}\left(p_{n-1}^2\right), \hspace{5mm} \text{as} \hspace{5mm} p_{n-1}\rightarrow 0,
\end{equation}
where the momenta and polarizations are as given in Table \ref{Thelicitytable}. In this section we will use this result to fix the contact term ambiguities of the $\text{mDBI}_4$ amplitudes. This momentum configuration is an effective further dimensional reduction from 4d to 3d and so we will write the explicit form of the amplitudes in 3d language. In our conventions, the dimensional reduction map takes an especially simple form
\begin{equation}
  4 d \to 3 d: \qquad 
  \langle ij \rangle \rightarrow \langle ij\rangle, \hspace{10mm} [ij] \rightarrow \langle ij\rangle,
\end{equation}
which we will then further simplify (for purely Bosonic amplitudes this means rewriting all helicity spinor contractions as Mandelstam invariants). To apply these results to the Ansatz form of the $\text{mDBI}_4$ amplitudes described above, which are in the helicity basis, we must relate the transverse polarization $\gamma^\top$ to a linear combination of helicity states. In our conventions the correct linear combination is found to be
\begin{equation}
  \label{eq:trans_photon}
  |\gamma^\top(\vec{p})\rangle = |\gamma^+(\vec{p})\rangle-|\gamma^-(\vec{p})\rangle,
\end{equation}
which for the helicity amplitudes means
\begin{multline}
  \mathcal{A}^{\text{mDBI}_4}_n\left(1_\phi,2_\gamma^+,\ldots ,(n-2)_\gamma^+,(n-1)_\gamma^\top,n_{\overline{\phi}}\right) = \\
  \mathcal{A}^{\text{mDBI}_4}_n\left(1_\phi,2_\gamma^+,\ldots ,(n-2)_\gamma^+,(n-1)_\gamma^+,n_{\overline{\phi}}\right) - \mathcal{A}^{\text{mDBI}_4}_n\left(1_\phi,2_\gamma^+,\ldots ,(n-2)_\gamma^+,(n-1)_\gamma^-,n_{\overline{\phi}}\right).
\end{multline}

The method used in this section will be to form this linear combination of Ansatze, apply the dimensional reduction map and then take the soft limit $p_{n-1}\rightarrow 0$. Compatibility with T-duality then requires that the $\mathcal{O}(p_{n-1})$ terms cancel amongst themselves, this requirement uniquely fixes the $c_n$ coefficients.

\subsubsection{Explicit Examples of T-duality Constraints}

We will begin with the 4-point amplitudes in $\text{mDBI}_4$. As described above the MHV amplitude is uniquely fixed by the $\mu^2\rightarrow 0$ limit, while the NSD amplitudes are fixed up to an overall coefficient
\begin{equation}
  \mathcal{A}_4^{\text{mDBI}_4}\left(1_\phi,2_\gamma^+,3_\gamma^+,4_{\overline{\phi}}\right) = c_4\mu^2 [23]^2.
\end{equation}
By taking the appropriate linear combination according to \eqref{eq:trans_photon} we can form an amplitude for which particle 3 is polarized in the direction \textit{transverse} to a particular 2d subspace
\begin{align}
  &\mathcal{A}_4^{\text{mDBI}_4}\left(1_\phi,2_\gamma^+,3_\gamma^\top,4_{\overline{\phi}}\right) \nonumber\\
  &\hspace{5mm}=\mathcal{A}_4^{\text{mDBI}_4}\left(1_\phi,2_\gamma^+,3_\gamma^+,4_{\overline{\phi}}\right) - \mathcal{A}_4^{\text{mDBI}_4}\left(1_\phi,2_\gamma^+,3_\gamma^-,4_{\overline{\phi}}\right) \nonumber\\
  &\hspace{5mm}= c_4\mu^2 [23]^2 +\langle 3|p_1|2]^2.
\end{align}

We then apply the dimensional reduction map, after reduction to 3d the various spinor contractions reduce to
\begin{align}
  &[23]^2 \rightarrow s_{23} \nonumber\\
  & \langle 3|p_1|2]^2 \rightarrow \text{Tr}\left[p_3 \cdot p_1 \cdot p_2 \cdot p_1\right] = 2\left(2(p_1\cdot p_3)(p_1\cdot p_2) - p_1^2 (p_2\cdot p_3)\right).
\end{align}
Applying this gives
\begin{equation}
  \mathcal{A}_4^{\text{mDBI}_4}\left(1_\phi,2_\gamma^+,3_\gamma^\top,4_{\overline{\phi}}\right) \xrightarrow[]{3d} 2(c_4+1)\mu^2(p_2\cdot p_3)+4(p_1\cdot p_3)(p_4\cdot p_3).
\end{equation}
In the limit where $p_3\rightarrow 0$ we can see that the first term vanishes at $\mathcal{O}(p_3)$ while the second term vanishes at $\mathcal{O}(p_3^2)$. The T-duality constraint then forces us to choose $c_4=-1$, which gives exactly the same relative coefficient we found from the KLT calculation (\ref{KLT4MHV}). 

At 6-point and higher it is necessary to define the soft degree more precisely. Let's quickly review the rigorous definition of a soft limit (see \cite{Elvang:2016qvq} for more details). We evaluate our amplitude on a one-parameter family of momenta of the form
\begin{equation}
  \hat{p}_{5}(\epsilon) = \epsilon p_5, \;\;\; \hat{p}_i(\epsilon) = p_i + \epsilon q_i, \;\;\;\; i\neq 5.
\end{equation}
The deformed momenta should satisfy momentum conservation and the on-shell conditions for all values of $\epsilon \in \mathds{C}$, which requires
\begin{align}
  p_5^2=0, \;\;\;\; p_i\cdot q_i = 0, \;\;\;\; q_i^2 = 0, \;\;\;\; \sum_{i\neq 5} p_i = 0, \;\;\;\; p_5 + \sum_{i\neq 5}q_i = 0.  
\end{align}
At leading order in the $\epsilon$-expansion the $q_i$ momenta do not appear. After dimensional reduction our amplitudes are trivially at least $\mathcal{O}\left(\epsilon\right)$, our goal is then to show that these leading terms are actually zero and that therefore the leading term in the expansion is $\mathcal{O}(\epsilon^2)$. For this purpose, taking the soft limit is equivalent to taking $p_i, \; i\neq 5$ to satisfy 5-particle momentum conservation, and $p_5$ as an unrelated null vector. We should bare this in mind when making algebraic manipulations involving conservation of momentum.

Let's now proceed with the calculation of the 6-point soft limit. Applying dimensional reduction to the Ansatze given above
\begin{align}
  &\mathcal{A}^{\text{mDBI}_4}_6\left(1_\phi,2_\gamma^+,3_\gamma^+,4_\gamma^+,5_\gamma^+,6_{\overline{\phi}}\right)\nonumber\\
  &\xrightarrow[]{3d+\text{soft}} \frac{(\mu^2)^2s_{23}s_{45}}{s_{123}+\mu^2}+ \frac{(\mu^2)^2s_{24}s_{35}}{s_{124}+\mu^2}+ \frac{(\mu^2)^2s_{25}s_{34}}{s_{12}+\mu^2} + c_6\mu^2\left(s_{23}s_{45}+s_{24}s_{35}+s_{25}s_{34}\right),
\end{align}
also,
\begin{align}
  &\mathcal{A}^{\text{mDBI}_4}_6\left(1_\phi,2_\gamma^+,3_\gamma^+,4_\gamma^+,5_\gamma^-,6_{\overline{\phi}}\right)\nonumber\\
  &\xrightarrow[]{3d+\text{soft}} \frac{\mu^2}{2}\left[\frac{s_{23}\left(2(p_5\cdot p_6)(s_{46}+\mu^2) +\mu^2s_{45}\right)}{s_{123}+\mu^2}+\frac{s_{34}\left(2(p_5\cdot p_1)(s_{12}+\mu^2)+\mu^2s_{25}\right)}{s_{12}+\mu^2}\right.\nonumber\\
  &\hspace{20mm}\left.+\frac{s_{34}\left(4(p_5\cdot p_{34})( p_2 \cdot p_{34})+2\mu^2 (p_2\cdot p_5)\right)}{s_{126}}\right]+\mathcal{P}\left(2,3,4\right).
\end{align}
Taking the difference we find that the $(\mu^2)^2$ terms cancel and the remaining terms are purely local
\begin{align}
  &\mathcal{A}^{\text{mDBI}_4}_6\left(1_\phi,2_\gamma^+,3_\gamma^+,4_\gamma^+,5_\gamma^\top,6_{\overline{\phi}}\right) \nonumber\\
  &=\mathcal{A}^{\text{mDBI}_4}_6\left(1_\phi,2_\gamma^+,3_\gamma^+,4_\gamma^+,5_\gamma^+,6_{\overline{\phi}}\right)-\mathcal{A}^{\text{mDBI}_4}_6\left(1_\phi,2_\gamma^+,3_\gamma^+,4_\gamma^+,5_\gamma^-,6_{\overline{\phi}}\right) \nonumber\\
  &  \xrightarrow[]{3d+\text{soft}} \frac{1}{2}c_6\mu^2 s_{23}s_{45}-\mu^2s_{23}(p_5\cdot p_{6})-\mu^2 s_{34}(p_1\cdot p_5) -2\mu^2 (p_5\cdot p_{16})(p_2\cdot p_{16}) \nonumber\\
  &\hspace{25mm}+ \mu^2s_{16}(p_2\cdot p_5) + \mathcal{P}\left(2,3,4\right) \nonumber\\
  &\hspace{20mm} = c_6\mu^2 \left(s_{23}s_{45}+s_{24}s_{35}+s_{25}s_{34}\right) -2\mu^2 s_{12}(p_5\cdot p_{16}) +4\mu^2 s_{12}(p_5\cdot p_{16})\nonumber\\
  &\hspace{25mm}-2\mu^2 s_{12}(p_5\cdot p_{16}) \nonumber\\
  &\hspace{20mm} = c_6\mu^2 \left(s_{23}s_{45}+s_{24}s_{35}+s_{25}s_{34}\right).
\end{align}
Somewhat miraculously all of the terms cancel except for the unknown contact term. Since this is manifestly $\mathcal{O}(p_5)$, we must choose $c_6=0$ to satisfy the constraint of T-duality. This is exactly the same conclusion we reached after a long numerical calculation involving the massive KLT relations. In Appendix \ref{app:Tdual8} we give the explicit calculation of $c_8$, again we confirm the result of the numerical KLT calculation. In the next subsection we will give an explicit all-multiplicity proof that the T-duality constraints require $c_n=0$ for $n>4$. 

\subsubsection{Small Mass Expansion and the Absence of Contact Terms}

That the 6-point dimensional reduction and soft limit calculation gave $c_6=0$ is somewhat remarkable, and could not easily have been anticipated without a detailed calculation. For $n\geq 8$ the conclusion that $c_n=0$ is less mysterious and can be argued on general grounds by considering the structure of the $\text{mDBI}_4$ amplitudes \textit{as an expansion around} the $\mu^2\rightarrow 0$ limit. In Appendix \ref{app:Contact} we show that there is a unique contact term at each multiplicity of the form
\begin{equation}
  \mathcal{A}_{n}^{\text{mDBI}_4}\left(1_\phi,2_\gamma^+,\ldots ,(n-1)_\gamma^+,n_{\overline{\phi}}\right)\biggr\vert_{\text{contact}} = c_n\mu^2 \left([23]^2[45]^2\ldots [n-2,n-1]^2+\ldots \right).
\end{equation}
Dimensionally reducing to 3d this becomes 
\begin{equation}
  \xrightarrow[]{3d} c_n\mu^2 \left(s_{23}s_{45}\ldots s_{n-2,n-1}+\ldots \right),
\end{equation}
which is manifestly $\mathcal{O}(p_{n-1})$ in the soft limit of particle $n-1$. If $c_n\neq 0$ then this term must cancel against some term in the factoring part of the Ansatz to give the correct $\mathcal{O}(p_{n-1}^2)$ soft limit. To show that this can never happen we expand in the limit $\mu^2\rightarrow 0$. The contact terms clearly always contribute at $\mathcal{O}(\mu^2)$. Since $\mu^2$ is a free parameter (corresponding to our choice of momenta in the 4 and 5 directions from the 6d perspective), the T-duality constraints should apply order-by-order in the expansion. For a non-trivial cancellation between the contact and factoring terms to occur, the factoring terms must give a contribution at $\mathcal{O}(\mu^2)$. If such a contribution exists then we must be able to identify a factorization channel for which the product of the leading small mass behavior on both sides  is $\mathcal{O}(\mu^2)$. Since negative and odd powers of $\mu$ do not appear, one half of the factorization diagram must be $\mathcal{O}(\mu^0)$. At each multiplicity there are only two possible factorization channels which can give such a contribution:

 
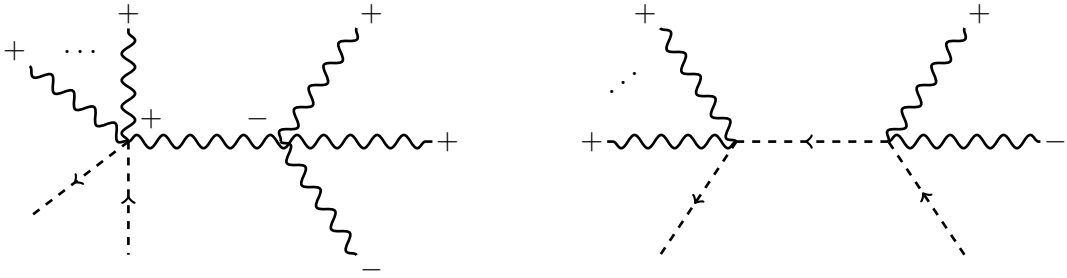
\begin{center}
{\begin{tikzpicture}[scale=1, line width=1 pt]
	\draw [vector] (0,0)--(-1.3,1);
	\draw [vector] (0,0)--(0,1.5);
	\draw [vector] (0,0)--(4,0);
	\draw [scalar] (0,0)--(-1.3,-1);
	\draw [scalarbar] (0,0)--(0,-1.5);
	\draw [vector] (2,0)--(3,1.5);
	\draw [vector] (2,0)--(3,-1.5);
	\node at (-1.5,1.2) {$+$};
	\node at (0,1.7) {$+$};
	\node at (0.3,0.3) {$+$};
	\node at (1.7,0.3) {$-$};
	\node at (4.2,0) {$+$};
	\node at (3.2,1.7) {$+$};
	\node at (3.2,-1.7) {$-$};
	\node at (-0.6,1.2) {$\dots$};
	%%%%%%%%%
	\draw [scalar] (8,0)--(7,-1.5);
	\draw [scalar] (10,0)--(8,0);
	\draw [scalar] (11,-1.5)--(10,0);
	\draw [vector] (8,0)--(6.3,0);
	\draw [vector] (8,0)--(7,1.5);
	\draw [vector] (10,0)--(11,1.5);
	\draw [vector] (10,0)--(12,0);
	\node at (6.1,0) {$+$};
	\node at (6.8,1.7) {$+$};
	\node at (11.2,1.7) {$+$};
	\node at (12.2,0) {$-$};
        \node at (6.5,0.9) {$\iddots$};
  \end{tikzpicture}}
\end{center}
both of which have the form of a lower-point NSD amplitude glued to an $\mathcal{O}(\mu^0)$ 4-point amplitude. For $n=8$, the $\mathcal{O}(\mu^2)$ contribution to the NSD amplitude arises solely from the contact term which we explicitly verified (by two different methods) was absent. So we conclude there cannot be an $\mathcal{O}(\mu^2)$ contribution to the $n=8$ MHV amplitude and hence no contact term. We can continue in this way and make an inductive argument that the absence of contact terms at $n-2$-point implies the absence of contact terms at $n$-point. Together with the explicit $n=6$ case, we find that our conjecture we made at the end of Section \ref{sec:Result} is correct. All higher point contact terms are indeed zero in $\text{mDBI}_4$, the amplitudes are (almost) as simple as possible. We will leverage this simplicity in the following section to construct all-multiplicity one-loop integrands for the SD and NSD sectors of $\text{BI}_4$.

%%%%%%%%%%%%%%%%%%%%%%%%%%%%%%
\section{All Multiplicity Rational One-Loop Amplitudes}
\label{sec:Loop}
%%%%%%%%%%%%%%%%%%%%%%%%%%%%%%

%%%%%%%%%%%%%%%%%%%%%%%%%%%%%%
\subsection{Diagrammatic Rules for Constructing Loop Integrands}
\label{sec:DiagrammaticRules}
%%%%%%%%%%%%%%%%%%%%%%%%%%%%%%

With the results in the previous section, and the discussion in Section \ref{sec:susydec}, we have in principle obtained a complete understanding of the structure of the $d$-dimensional unitarity cut structure of SD and NSD $\text{BI}_4$ one-loop integrands. Our goal is now to use this to engineer the explicit form of the integrands and then integrate them to obtain the full amplitudes. Ordinarily, gluing together on-shell tree-amplitudes into full loop integrands is a delicate business. Constructing expressions with the correct cuts in one channel may give \textit{polluting} contributions to another channel. Separating these contributions and building up loop integrands in a systematic way has been a subject of intense study over the past several decades \cite{Bern:2011qt}. 

Fortunately for us, the $\text{mDBI}_4$ tree amplitudes are of sufficiently simple form that it is straightforward to construct integrands with all of the correct cuts using a set of \textit{diagrammatic rules}. There are two properties that allow us to do this; first, locality is manifest in the $\text{mDBI}_4$ amplitudes, and second, due to the absence of contact terms above $n=4$ the number of \textit{elementary} vertex rules is strictly finite. Notice how much simpler this is than calculating loop diagrams directly from ordinary Feynman rules! If we were calculating loop amplitudes in Born-Infeld the old-fashioned way we would need to calculate new (and increasingly complicated) Feynman vertex rules at each multiplicity. 

Since we are constructing loop integrands in the scalar loop representation (\ref{sec:unitarity}) we will construct a diagrammatic representation in which each diagram consists of a scalar loop \textit{decorated} with any of the following vertex factors:
\begin{center}
{\begin{tikzpicture}[scale=1.5, line width=1 pt]
\draw [vector] (-1,1)--(0,0);
\draw [vector] (-1,-1)--(0,0);
\draw [scalar] (0,0)--(1,1);
\draw [scalar] (1,-1)--(0,0);
\draw[black,fill=black] (0,0) circle (1.5ex);
\node at (-1.2,1.2) {$i_\gamma^+$};
\node at (-1.2,-1.2) {$j_\gamma^+$};
\node at (1.3,1.2) {$(l_1)_\phi$};
\node at (1.3,-1.2) {$(l_2)_{\overline{\phi}}$};
\node at (2,0) {$=$};
\node at (3,0) {$-\mu^2 [ij]^2$};
\end{tikzpicture}}
\end{center}
\begin{center}
{\begin{tikzpicture}[scale=1.5, line width=1 pt]
\draw [vector] (-1,1)--(0,0);
\draw [vector] (-1,-1)--(0,0);
\draw [scalar] (0,0)--(1,1);
\draw [scalar] (1,-1)--(0,0);
\draw[black,fill=white] (0,0) circle (1.5ex);
\node at (-1.2,1.2) {$i_\gamma^+$};
\node at (-1.2,-1.2) {$j_\gamma^-$};
\node at (1.3,1.2) {$(l_1)_\phi$};
\node at (1.3,-1.2) {$(l_2)_{\overline{\phi}}$};
\node at (2,0) {$=$};
\node at (3,0) {$- \langle j|l_1|i]^2$};
\end{tikzpicture}}
\end{center}
\begin{center}
{\begin{tikzpicture}[scale=1.5, line width=1 pt]
\draw [vector] (-1,1)--(0,0);
\draw [vector] (-1,-1)--(0,0);
\draw [vector] (-1.4,0.6)--(0,0);
\draw [vector] (-1.4,-0.6)--(0,0);
\draw [scalar] (0,0)--(1,1);
\draw [scalar] (1,-1)--(0,0);
\draw[black,fill=lightgray] (0,0) circle (1.5ex);
\node at (-1.2,1.2) {$i_\gamma^+$};
\node at (-1.2,-1.2) {$l_\gamma^-$};
\node at (-1.6,0.6) {$j_\gamma^+$};
\node at (-1.6,-0.6) {$k_\gamma^+$};
\node at (1.3,1.2) {$(l_1)_\phi$};
\node at (1.3,-1.2) {$(l_2)_{\overline{\phi}}$};
\node at (2,0) {$=$};
\node at (4,0) {\Large $\frac{\mu^2[k|p_{ij}|l\rangle^2[ij]^2}{s_{ijl}} $\normalsize $\;+\;\mathcal{C}(i,j,k)$};
\node at (-3.4,0) {};
\end{tikzpicture}}
\end{center}
Here $+\;\mathcal{C}(i,j,k)$ denotes the sum over cyclic permutations, all of the momenta are defined to be out-going with photon lines on-shell, while the scalar lines are off-shell. These vertex rules can be glued together on scalar lines in the usual way with the standard massive scalar propagator
\begin{center}
{\begin{tikzpicture}[scale=1.5, line width=1 pt]
\draw [scalar] (-2,0)--(0,0);
\draw [->] (-1.25,0.2)--(-0.75,0.2);
\node at (-1,0.4) {$l$}; 
\node at (1,0) {$=$};
\node at (2,0) {\Large $\frac{1}{l^2+\mu^2} $};
\end{tikzpicture}}
\end{center}
These diagrammatic rules can be justified \textit{post hoc}, by verifying that the resulting loop integrands have the correct massive scalar cuts. These are not Feynman rules in the usual sense, and have not been derived from a Lagrangian. This is especially clear in the 6-point vertex rule (denoted with a gray blob), which is a non-local expression; the poles encode factorization singularities into Born-Infeld photons. Due to the helicity selection rules of $\text{BI}_4$ at tree-level arising from supersymmetric truncation, no further photonic singularities can appear in amplitudes with at most a single negative helicity external state.

In the following sections we will give explicit examples of the applications of these diagrammatic rules to 4- and 6-point SD and NSD loop integrands, and then present explicit expressions for the all-multiplicity results together with the integrated expressions at $\mathcal{O}(\epsilon^0)$. 

%%%%%%%%%%%%%%%%%%%%%%%%%%%%%%
\subsection{Self-Dual Sector}
\label{sec:SD}
%%%%%%%%%%%%%%%%%%%%%%%%%%%%%%

In the self-dual sector, since there are only positive helicity external states, at each multiplicity there is only a single topologically distinct diagram and it is constructed solely from black vertices. Beginning with $n=4$, the diagram has the form:
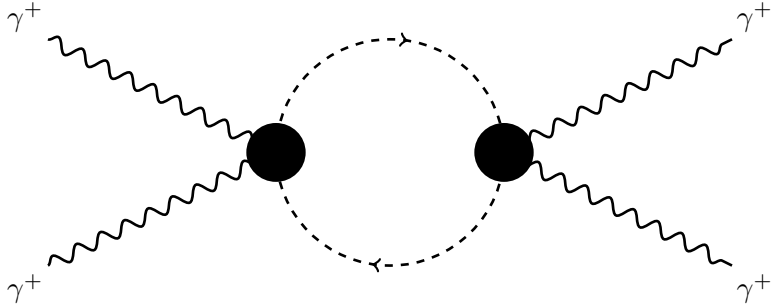
\begin{center}
	{\begin{tikzpicture}[scale=1.5, line width=1 pt]
		\draw [vector] (-2,1)--(0,0);
		\draw [vector] (-2,-1)--(0,0);
		\draw [scalar] (0,0) arc (180:0:1);
		\draw [scalar] (2,0) arc (0:-180:1);
		\draw [vector] (2,0)--(4,1);
		\draw [vector] (2,0)--(4,-1);
		\draw[black,fill=black] (0,0) circle (1.5ex);
		\draw[black,fill=black] (2,0) circle (1.5ex);
		\node at (-2.2,1.2) {$\gamma^+$};
		\node at (-2.2,-1.2) {$\gamma^+$};
		\node at (4.2,1.2) {$\gamma^+$};
		\node at (4.2,-1.2) {$\gamma^+$};
		\end{tikzpicture}}
\end{center}
There are three non-trivial permutations of the external labels. The integrand is then
\begin{align}
&\mathcal{I}^{\text{SD}}_4[l;\mu^2] = \frac{1}{2}\left[\frac{(\mu^2)^2[12]^2[34]^2}{\left[l^2+\mu^2\right]\left[(l-p_{12})^2+\mu^2\right]} +\mathcal{P}\left(2,3,4\right)\right],
\end{align}
where the factor of $\frac12$ compensates for the equivalent permutations in $\mathcal{P}\left(2,3,4\right)$ that are summed over. 

We now explicitly verify that the diagrammatic rules of Section \ref{sec:DiagrammaticRules} yield an integrand that satisfies the cut conditions. Since the integrand has only one distinct two-particle cut (all others are related by label permutations), we choose to consider the $p_{12}$-cut. When the on-shell conditions $l^2=-\mu^2$ and $(l-p_{12})^2=-\mu^2$ are imposed, the integrand yields
\begin{align}
 &\left[l^2+\mu^2\right]\left[(l-p_{12})^2+\mu^2\right]\left.\mathcal{I}^{\text{SD}}_4[l;\mu^2]\right|_{p_{12}\text{-cut}}\nonumber\\
 &\hspace{4cm}= \mathcal{A}_4\left(1_\gamma^+,2_\gamma^+,-l_\phi,(l-p_{12})_{\bar{\phi}}\right)\mathcal{A}_4\left(l_{\bar{\phi}},(p_{12}-l)_\phi,3^+_\gamma,4^+_\gamma\right)\nonumber\\
 &\hspace{4cm}=(\mu^2)^2[12]^2[34]^2
\end{align}
as expected. The NSD amplitudes above are given in \eqref{KLT4NSD}.

Using the general result for rational loop integrals (\ref{eq:1-loop-n-gon}) gives 
\begin{align}
&\mathcal{A}^{\text{BI}_4\;\text{1-loop}}_4\left(1_\gamma^+,2_\gamma^+,3_\gamma^+,4_\gamma^+\right)\nonumber\\
&\hspace{10mm}=  \frac{1}{2}\int \frac{\text{d}^4l}{(2\pi)^4}\int \frac{\text{d}^{-2\epsilon}\mu}{(2\pi)^{-2\epsilon}}
\left[\frac{(\mu^2)^2[12]^2[34]^2}{\left[l^2+\mu^2\right]\left[(l-p_{12})^2+\mu^2\right]} +\mathcal{P}\left(2,3,4\right)\right] \nonumber\\
&\hspace{10mm}=[12]^2[34]^2I^{d=4-2\epsilon}_2[(\mu^2)^2;p_{12}]+[13]^2[24]^2I^{d=4-2\epsilon}_2[(\mu^2)^2;p_{13}]\nonumber\\
&\hspace{20mm}+[14]^2[23]^2I^{d=4-2\epsilon}_2[(\mu^2)^2;p_{14}]  \nonumber\\
&\hspace{10mm}= -\frac{i}{960\pi^2}\left([12]^2[34]^2s_{12}^2+[13]^2[24]^2s_{13}^2+[14]^2[23]^2s_{14}^2\right) +\mathcal{O}(\epsilon).
\end{align}

Similarly for $n=6$ there is a unique topologically distinct class of diagram:
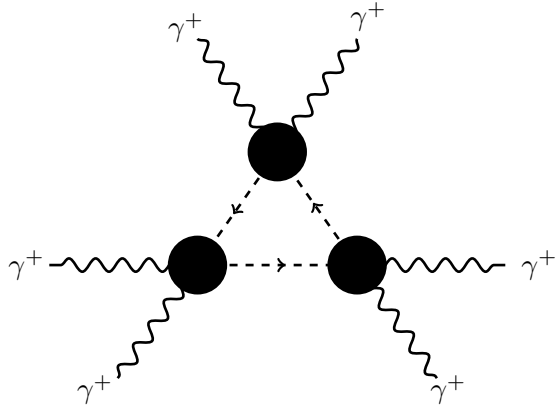
\begin{center}
	{\begin{tikzpicture}[scale=1.5, line width=1 pt]
		\draw [vector] (-0.7,1)--(0,0);
		\draw [vector] (0.7,1)--(0,0);
		\draw [scalar] (0,0)--(-0.7,-1);
		\draw [scalar] (-0.7,-1)--(0.7,-1);
		\draw [scalar] (0.7,-1)--(0,0);
		\draw [vector] (-0.7,-1)--(-2,-1);
		\draw [vector] (-0.7,-1)--(-1.4,-2);
		\draw [vector] (0.7,-1)--(2,-1);
		\draw [vector] (0.7,-1)--(1.4,-2);
		\node at (-0.8,1.1) {$\gamma^+$};
		\node at (0.8,1.2) {$\gamma^+$};
		\node at (-1.6,-2.1) {$\gamma^+$};
		\node at (1.5,-2.1) {$\gamma^+$};
		\node at (-2.2,-1) {$\gamma^+$};
		\node at (2.3,-1) {$\gamma^+$};
		\draw[black,fill=black] (0,0) circle (1.5ex);
		\draw[black,fill=black] (-0.7,-1) circle (1.5ex);
		\draw[black,fill=black] (0.7,-1) circle (1.5ex);
		\end{tikzpicture}}
\end{center}
The integrand is then given by 
\begin{equation}
\mathcal{I}^{\text{SD}}_6[l;\mu^2]  =-\frac{1}{4}\left[ \frac{(\mu^2)^3[12]^2[34]^2[56]^2}{\left[l^2+\mu^2\right]\left[(l-p_{34})^2+\mu^2\right]\left[(l+p_{12})^2+\mu^2\right]}+\mathcal{P}\left(2,3,4,5,6\right)\right].
\end{equation}

The integrand has only one distinct cut into tree-level amplitudes. Consider for example the integrand on the $p_{12}$-cut,
\begin{align}
\label{12cutNSD6}
&\left[l^2+\mu^2\right]\left[(l+p_{12})^2+\mu^2\right]\left.\mathcal{I}^{\text{SD}}_6[l;\mu^2]\right|_{p_{12}\text{-cut}}\nonumber\\
&\hspace{5mm}= \mathcal{A}_4\left(1_\gamma^+,2_\gamma^+,l_\phi,-(l+p_{12})_{\bar{\phi}}\right)\mathcal{A}_6\left(-l_{\bar{\phi}},(l+p_{12})_\phi,3^+_\gamma,4^+_\gamma,5^+_\gamma,6^+_\gamma\right)\nonumber\\
&\hspace{1.5cm}+\mathcal{A}_4\left(1_\gamma^+,2_\gamma^+,l_{\bar{\phi}},-(l+p_{12})_{\phi}\right)\mathcal{A}_6\left(-l_\phi,(l+p_{12})_{\bar{\phi}},3^+_\gamma,4^+_\gamma,5^+_\gamma,6^+_\gamma\right)\nonumber\\
&\hspace{5mm}=2\mathcal{A}_4\left(1_\gamma^+,2_\gamma^+,l_\phi,-(l+p_{12})_{\bar{\phi}}\right)\mathcal{A}_6\left(-l_{\bar{\phi}},(l+p_{12})_\phi,3^+_\gamma,4^+_\gamma,5^+_\gamma,6^+_\gamma\right).
\end{align}
where the amplitudes are given in \eqref{KLT4NSD} and \eqref{KLT6NSD} and the form of the 6-point amplitude \eqref{KLT6NSD} makes it apparent that there are no local contributions to two-scalar cuts.

The factor of 2 in \eqref{12cutNSD6} is multiplied by $\frac18$ (which compensates for the equivalent permutations in $\mathcal{P}\left(2,3,4,5,6\right)$ that are summed over). This matches the factor of $\frac14$ in the integrand and hence verifies the rules of Section \ref{sec:DiagrammaticRules}.

Integrating this using the formula \eqref{eq:1-loop-n-gon} gives
\begin{align}
&\mathcal{A}^{\text{BI}_4\;\text{1-loop}}_6\left(1_\gamma^+,2_\gamma^+,3_\gamma^+,4_\gamma^+,5_\gamma^+,6_\gamma^+\right)\nonumber\\
&\hspace{10mm}= \frac{1}{4}\bigg[\frac{i}{2880\pi^2}[12]^2[34]^2[56]^2\left(s_{12}^2+s_{34}^2+s_{56}^2+s_{12}s_{34}+s_{12}s_{56}+s_{34}s_{56}\right) \nonumber\\
&\hspace{20mm} + \mathcal{P}\left(2,3,4,5,6\right) \bigg] +\mathcal{O}(\epsilon).
\end{align}

The generalization to all multiplicity in the SD sector is now clear. There is always a single topologically distinct diagram with a corresponding scalar rational integral:
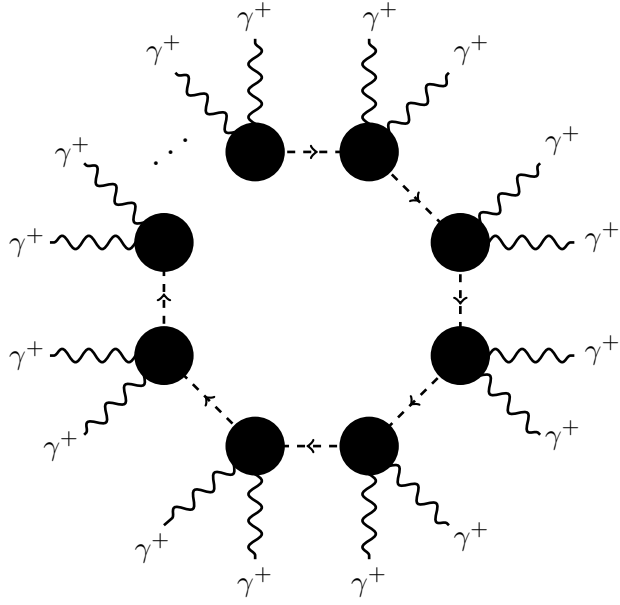
\begin{center}
	{\begin{tikzpicture}[scale=1.5, line width=1 pt]
		\draw [scalar] (0,0)--(1,0);
		\draw [scalar] (1,0)--(1.8,-0.8);
		\draw [scalar] (1.8,-0.8)--(1.8,-1.8);
		\draw [scalar] (1.8,-1.8)--(1,-2.6);
		\draw [scalar] (1,-2.6)--(0,-2.6);
		\draw [scalar] (0,-2.6)--(-0.8,-1.8);
		\draw [scalar] (-0.8,-1.8)--(-0.8,-0.8);
		\draw [vector] (0,0)--(-0.7,0.7);
		\draw [vector] (0,0)--(0,1);
		\draw [vector] (1,0)--(1,1);
		\draw [vector] (1,0)--(1.7,0.7);
		\draw [vector] (1.8,-0.8)--(2.5,-0.1);
		\draw [vector] (1.8,-0.8)--(2.8,-0.8);
		\draw [vector] (1.8,-1.8)--(2.8,-1.8);
		\draw [vector] (1.8,-1.8)--(2.5,-2.5);
		\draw [vector] (1,-2.6)--(1.7,-3.3);
		\draw [vector] (1,-2.6)--(1,-3.6);
		\draw [vector] (0,-2.6)--(0,-3.6);
		\draw [vector] (0,-2.6)--(-0.8,-3.3);
		\draw [vector] (-0.8,-1.8)--(-1.5,-2.5);
		\draw [vector] (-0.8,-1.8)--(-1.8,-1.8);
		\draw [vector] (-0.8,-0.8)--(-1.8,-0.8);
		\draw [vector] (-0.8,-0.8)--(-1.5,-0.1);
		\draw[black,fill=black] (0,0) circle (1.5ex);
		\draw[black,fill=black] (1,0) circle (1.5ex);
		\draw[black,fill=black] (1.8,-0.8) circle (1.5ex);
		\draw[black,fill=black] (1.8,-1.8) circle (1.5ex);
		\draw[black,fill=black] (1,-2.6) circle (1.5ex);
		\draw[black,fill=black] (0,-2.6) circle (1.5ex);
		\draw[black,fill=black] (-0.8,-1.8) circle (1.5ex);
		\draw[black,fill=black] (-0.8,-0.8) circle (1.5ex);
		\node at (-0.75,0.1) {\Large$\iddots$};
		\node at (-0.8,0.9) {$\gamma^+$};
		\node at (0,1.2) {$\gamma^+$};
		\node at (1.1,1.2) {$\gamma^+$};
		\node at (1.9,0.9) {$\gamma^+$};
		\node at (2.7,0.1) {$\gamma^+$};
		\node at (3.05,-0.75) {$\gamma^+$};
		\node at (3.05,-1.75) {$\gamma^+$};
		\node at (2.7,-2.5) {$\gamma^+$};
		\node at (1.9,-3.4) {$\gamma^+$};
		\node at (1,-3.8) {$\gamma^+$};
		\node at (0,-3.8) {$\gamma^+$};
		\node at (-0.9,-3.5) {$\gamma^+$};
		\node at (-1.7,-2.6) {$\gamma^+$};
		\node at (-2,-1.8) {$\gamma^+$};
		\node at (-2,-0.8) {$\gamma^+$};
		\node at (-1.6,0) {$\gamma^+$};
		\end{tikzpicture}}
\end{center}
The complete integrand is then
\begin{align} \label{SDall}
&\mathcal I^{\text{SD}}_{2n}[l;\mu^2]\nonumber\\
&= \left ( \frac 1 2 \right )^{n - 1} \left ( [ 12 ]^2 [ 34 ]^2 \ldots [2n - 1, 2n ]^2 \frac{\left ( - \mu^2 \right )^{n}}{\prod_{i = 1}^{n} \left [ \left ( l - \sum_{j = 1}^{2i} p_j \right )^2 + \mu^2 \right ]} + \mathcal P ( 2,3,\ldots,2n ) \right )\,.
\end{align}
Using the result of equation \eqref{eq:1-loop-n-gon}, we find that the integrated amplitude is
\begin{equation}
    \label{SDallamp}
    \boxed{
    \begin{aligned}
    &\mathcal A_{2n}^{\text{BI}_4\; \text{1-loop}}\left(1_\gamma^+, 2_\gamma^+,\ldots ,2n_\gamma^+\right) \\
    &\hspace{5mm}= \frac{i}{32 \pi^2} \left( - \frac 1 2 \right )^{n-1} \frac{1}{n ( n + 1 ) ( n + 2 ) ( n + 3 )} \\
    &\hspace{5mm}\times \Bigg [ [ 12 ]^2 [ 34 ]^2 \ldots [ 2n-1, 2n ]^2 \left ( \sum_{i < j}^n \sum_{k < l}^n a_{i j k l} \left ( \sum_{m = 2 i + 1}^{2 j} p_m \right )^2 \left ( \sum_{m = 2 k + 1}^{2 l} p_m \right )^2 \right ) \\
    &\hspace{10mm} + \mathcal P ( 2, 3, \ldots, 2n ) \Bigg] + \mathcal{O}(\epsilon)\,,
    \end{aligned}
    }
\end{equation}
with
\begin{equation}
a_{i j k l} = \left \{ \begin{array}{ll}
1 & \qquad \text{if all $i,j,k,l$ are different} \\
2 & \qquad \text{if exactly 2 of $i, j, k, l$ are identical} \\
4 & \qquad \text{if $i = k$ and $j = l$}
\end{array} \right .\,.
\end{equation}
It is straightforward to check that this result matches the results of the explicit calculations for the cases of $n = 2$ and $n=3$, presented above.

%%%%%%%%%%%%%%%%%%%%%%%%%%%%%%
\subsection{Next-to-Self-Dual Sector}
\label{sec:NSD}
%%%%%%%%%%%%%%%%%%%%%%%%%%%%%%

In the NSD sector the diagrams have a similar structure, consisting a single scalar loop decorated with the vertex factors. The novelty here is the appearance of a single negative helicity photon, and so each diagram contains either a single white or gray vertex. At 4-point there is only a single topologically distinct class of diagram, and contains both a black and white vertex\footnote{Note that there is no \textit{tadpole} diagram with a single gray vertex since this contributes a scaleless integral which vanishes in dimensional regularization.}:
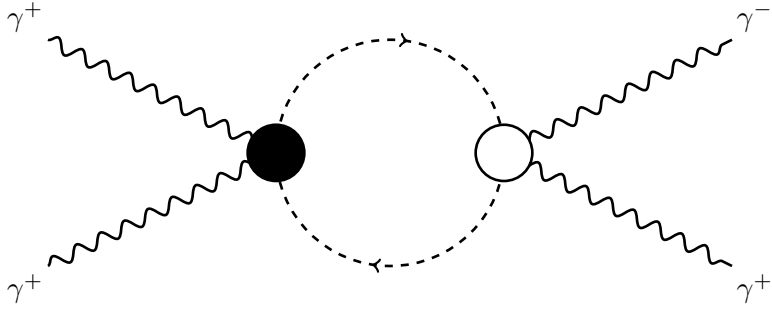
\begin{center}
	{\begin{tikzpicture}[scale=1.5, line width=1 pt]
		\draw [vector] (-2,1)--(0,0);
		\draw [vector] (-2,-1)--(0,0);
		\draw [scalar] (0,0) arc (180:0:1);
		\draw [scalar] (2,0) arc (0:-180:1);
		\draw [vector] (2,0)--(4,1);
		\draw [vector] (2,0)--(4,-1);
		\draw[black,fill=black] (0,0) circle (1.5ex);
		\draw[black,fill=white] (2,0) circle (1.5ex);
		\node at (-2.2,1.2) {$\gamma^+$};
		\node at (-2.2,-1.2) {$\gamma^+$};
		\node at (4.2,1.2) {$\gamma^-$};
		\node at (4.2,-1.2) {$\gamma^+$};
		\end{tikzpicture}}
\end{center}
There are three non-trivial permutations of the external labels. Consider a single such permuation corresponding to momenta $p_1$ and $p_2$ flowing out of the black vertex, the corresponding integrand has the form
\begin{equation}
\label{I4NSD}
\mathcal{I}^{\text{NSD}}_4\left[l;\mu^2\right]\biggr\vert_{12} = \frac{\mu^2[12]^2\langle 4|l|3]^2 }{\left[l^2+\mu^2\right]\left[(l-p_{12})^2+\mu^2\right]}.
\end{equation}
We now verify that the diagrammatic rules of Section \ref{sec:DiagrammaticRules} give an integrand with the right cuts in the NSD sector. There is only one distinct two-particle cut. As expected, the contribution to the integrand \eqref{I4NSD} on the $p_{12}$-cut is
\begin{align}
&\left[l^2+\mu^2\right]\left[(l-p_{12})^2+\mu^2\right]\left.\mathcal{I}^{\text{SD}}_4[l;\mu^2]\right|_{p_{12}\text{-cut}}\nonumber\\
&\hspace{4cm}= \mathcal{A}_4\left(1_\gamma^+,2_\gamma^+,-l_\phi,(l-p_{12})_{\bar{\phi}}\right)\mathcal{A}_4\left(l_{\bar{\phi}},(p_{12}-l)_\phi,3^+_\gamma,4^-_\gamma\right)\nonumber\\
&\hspace{4cm}=\mu^2[12]^2\langle 4|l|3]^2,
\end{align}
where the amplitudes are given in \eqref{KLT4NSD} and \eqref{KLT4MHV}.

Unlike all of the integrals in the SD sector, this is a rational tensor integral. The explicit value of an integral of this form is in (\ref{eq:1-loop-n-gon-tensor}), this gives
\begin{align}
&\int \frac{\text{d}^4l}{(2\pi)^4}\int \frac{\text{d}^{-2\epsilon}\mu}{(2\pi)^{-2\epsilon}}\left[\frac{\mu^2[12]^2\langle 4|l|3]^2 }{\left[l^2+\mu^2\right]\left[(l-p_{12})^2+\mu^2\right]}\right]\nonumber\\
&\hspace{20mm}= [12]^2I^{d=4-2\epsilon}_2[\mu^2\langle 4|l|3]^2;p_{12}]\nonumber\\
&\hspace{20mm}= \frac{-i}{1920\pi^2}[12]^2\langle 4|\sigma_\mu|3]\langle 4|\sigma_\nu|3]\left[g^{\mu\nu}s_{12}^2-6p_{12}^\mu p_{12}^\nu s_{12}\right] +\mathcal{O}(\epsilon)\nonumber\\
&\hspace{20mm} = 0+\mathcal{O}(\epsilon).
\end{align}
Since the remaining channels are simple permutations of this one we conclude
\begin{equation}
\mathcal{A}^{\text{BI}_4\;\text{ 1-loop}}_4\left(1_\gamma^+,2_\gamma^+,3_\gamma^+,4_\gamma^-\right) = 0+\mathcal{O}(\epsilon).
\end{equation}

Beginning at 6-point there are two distinct classes of diagrams, corresponding to diagrams containing a single white or gray vertex. Note that the 6-point integrand also has two distinct cuts. For instance, take the integrand on the $p_{56}$-cut,
\begin{align}
\label{56cutI6NSD}
&\left[l^2+\mu^2\right]\left[(l+p_{56})^2+\mu^2\right]\left.\mathcal{I}^{\text{SD}}_6[l;\mu^2]\right|_{p_{56}\text{-cut}}\nonumber\\
&\hspace{5mm}= \mathcal{A}_4\left(5_\gamma^+,6_\gamma^-,l_\phi,-(l+p_{56})_{\bar{\phi}}\right)\mathcal{A}_6\left(-l_{\bar{\phi}},(l+p_{56})_\phi,1^+_\gamma,2^+_\gamma,3^+_\gamma,4^+_\gamma\right)\nonumber\\
&\hspace{1.5cm}+\mathcal{A}_4\left(5_\gamma^+,6_\gamma^-,l_{\bar{\phi}},-(l+p_{56})_\phi\right)\mathcal{A}_6\left(-l_\phi,(l+p_{56})_{\bar{\phi}},1^+_\gamma,2^+_\gamma,3^+_\gamma,4^+_\gamma\right)\nonumber\\
&\hspace{5mm}=2\mathcal{A}_4\left(5_\gamma^+,6_\gamma^-,l_\phi,-(l+p_{56})_{\bar{\phi}}\right)\mathcal{A}_6\left(-l_{\bar{\phi}},(l+p_{56})_\phi,1^+_\gamma,2^+_\gamma,3^+_\gamma,4^+_\gamma\right).
\end{align}
where the explicit forms of the amplitudes are given in \eqref{KLT6NSD} and \eqref{KLT4MHV}. This generalises to any $p_{i6}$-cut, where $i\ne6$.

As a representative of the other class of cuts, consider the $p_{12}$-cut (which generalises to all $p_{ij}$-cuts where $i,j\ne 6$.),
\begin{align}
\label{12cutI6NSD}
&\left[l^2+\mu^2\right]\left[(l+p_{12})^2+\mu^2\right]\left.\mathcal{I}^{\text{SD}}_6[l;\mu^2]\right|_{p_{12}\text{-cut}}\nonumber\\
&\hspace{5mm}= \mathcal{A}_4\left(1_\gamma^+,2_\gamma^+,l_\phi,-(l+p_{12})_{\bar{\phi}}\right)\mathcal{A}_6\left(-l_{\bar{\phi}},(l+p_{12})_\phi,3^+_\gamma,4^+_\gamma,5^+_\gamma,6^-_\gamma\right)\nonumber\\
&\hspace{1.5cm}+\mathcal{A}_4\left(1_\gamma^+,2_\gamma^+,l_{\bar{\phi}},-(l+p_{12})_\phi\right)\mathcal{A}_6\left(-l_\phi,(l+p_{12})_{\bar{\phi}},3^+_\gamma,4^+_\gamma,5^+_\gamma,6^-_\gamma\right)\nonumber\\
&\hspace{5mm}=2\mathcal{A}_4\left(1_\gamma^+,2_\gamma^+,l_\phi,-(l+p_{12})_{\bar{\phi}}\right)\mathcal{A}_6\left(-l_{\bar{\phi}},(l+p_{12})_\phi,3^+_\gamma,4^+_\gamma,5^+_\gamma,6^-_\gamma\right)
\end{align}
where the amplitudes are given in \eqref{KLT4NSD} and \eqref{KLT6MHV}. Note that there are two kinds of contributions to $\mathcal{A}_6\left(-l_{\bar{\phi}},(l+p_{12})_\phi,3^+_\gamma,4^+_\gamma,5^+_\gamma,6^-_\gamma\right)$: one factorizes on an internal scalar and the other factorizes on an internal photon,
\begin{align}
	\mathcal{A}_6\left(-l_{\bar{\phi}},(l+p_{12})_\phi,3^+_\gamma,4^+_\gamma,5^+_\gamma,6^-_\gamma\right)=&\mathcal{A}^\text{scalar}_6\left(-l_{\bar{\phi}},(l+p_{12})_\phi,3^+_\gamma,4^+_\gamma,5^+_\gamma,6^-_\gamma\right)\nonumber\\
	&+\mathcal{A}^\text{photon}_6\left(-l_{\bar{\phi}},(l+p_{12})_\phi,3^+_\gamma,4^+_\gamma,5^+_\gamma,6^-_\gamma\right).
\end{align}

The first class of contributing diagrams is similar to the 4-point calculation and takes the form:
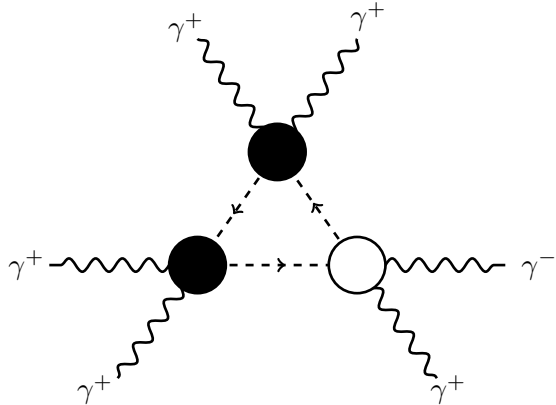
\begin{center}
	{\begin{tikzpicture}[scale=1.5, line width=1 pt]
		\draw [vector] (-0.7,1)--(0,0);
		\draw [vector] (0.7,1)--(0,0);
		\draw [scalar] (0,0)--(-0.7,-1);
		\draw [scalar] (-0.7,-1)--(0.7,-1);
		\draw [scalar] (0.7,-1)--(0,0);
		\draw [vector] (-0.7,-1)--(-2,-1);
		\draw [vector] (-0.7,-1)--(-1.4,-2);
		\draw [vector] (0.7,-1)--(2,-1);
		\draw [vector] (0.7,-1)--(1.4,-2);
		\node at (-0.8,1.1) {$\gamma^+$};
		\node at (0.8,1.2) {$\gamma^+$};
		\node at (-1.6,-2.1) {$\gamma^+$};
		\node at (1.5,-2.1) {$\gamma^+$};
		\node at (-2.2,-1) {$\gamma^+$};
		\node at (2.3,-1) {$\gamma^-$};
		\draw[black,fill=black] (0,0) circle (1.5ex);
		\draw[black,fill=black] (-0.7,-1) circle (1.5ex);
		\draw[black,fill=white] (0.7,-1) circle (1.5ex);
		\end{tikzpicture}}
\end{center}
Summing over all permutations of the external labels gives the following contribution to the integrand
\begin{align} \label{white6}
\mathcal{I}^{\text{NSD}}_6[l;\mu^2]\biggr\vert_{\text{white}} = \frac{1}{4}\left[\frac{-(\mu^2)^2[12]^2[34]^2\langle 6|l|5]^2}{\left[l^2+\mu^2\right]\left[(l-p_{12})^2+\mu^2\right]\left[(l+p_{56})^2+\mu^2\right]}+\mathcal{P}\left(1,2,3,4,5\right)\right].
\end{align}

This contribution has the correct $i6$-cuts \eqref{56cutI6NSD}. On a $p_{12}$-cut, \eqref{white6} produces 
\begin{align}
&\left[l^2+\mu^2\right]\left[(l+p_{12})^2+\mu^2\right]\left.\mathcal{I}^{\text{SD}}_6[l;\mu^2]\right|_{p_{ij}\text{-cut}}\nonumber\\
&\hspace{5mm}=2\mathcal{A}_4\left(1_\gamma^+,2_\gamma^+,l_\phi,-(l+p_{12})_{\bar{\phi}}\right)\mathcal{A}^\text{scalar}_6\left(-l_{\bar{\phi}},(l+p_{12})_\phi,3^+_\gamma,4^+_\gamma,5^+_\gamma,6^-_\gamma\right).
\end{align}
The rest of the 6-point MHV amplitude is accounted for by the second class of diagrams.

The contributions from diagrams containing a single gray vertex:
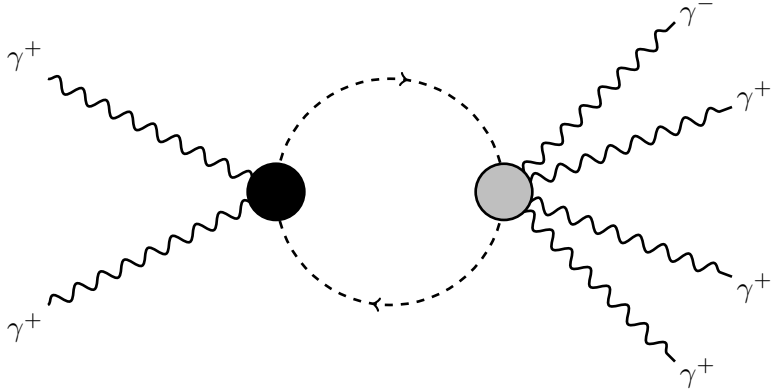
\begin{center}
	{\begin{tikzpicture}[scale=1.5, line width=1 pt]
		\draw [vector] (-2,1)--(0,0);
		\draw [vector] (-2,-1)--(0,0);
		\draw [scalar] (0,0) arc (180:0:1);
		\draw [scalar] (2,0) arc (0:-180:1);
		\draw [vector] (2,0)--(3.5,1.5);
		\draw [vector] (2,0)--(3.5,-1.5);
		\draw [vector] (2,0)--(4,0.75);
		\draw [vector] (2,0)--(4,-0.75);
		\draw[black,fill=black] (0,0) circle (1.5ex);
		\draw[black,fill=lightgray] (2,0) circle (1.5ex);
		\node at (-2.2,1.2) {$\gamma^+$};
		\node at (-2.2,-1.2) {$\gamma^+$};
		\node at (3.7,1.6) {$\gamma^-$};
		\node at (3.7,-1.6) {$\gamma^+$};
		\node at (4.2,0.85) {$\gamma^+$};
		\node at (4.2,-0.85) {$\gamma^+$};
		\end{tikzpicture}}
\end{center}
which contributes the following to the integrand
\begin{align} \label{gray6}
\mathcal{I}^{\text{NSD}}_6[l;\mu^2]\biggr\vert_{\text{gray}} = \frac{1}{2}\left[\frac{-(\mu^2)^2[12]^2[34]^2\langle 6|p_{12}|5]^2}{s_{125}\left[l^2+\mu^2\right]\left[(l-p_{12})^2+\mu^2\right]}+\mathcal{P}\left(1,2,3,4,5\right)\right].
\end{align}
Here the $p_{12}$-cut yields
\begin{align}
&\left[l^2+\mu^2\right]\left[(l+p_{12})^2+\mu^2\right]\left.\mathcal{I}^{\text{SD}}_6[l;\mu^2]\right|_{p_{12}\text{-cut}}\nonumber\\
&\hspace{5mm}=2\mathcal{A}_4\left(1_\gamma^+,2_\gamma^+,l_\phi,-(l+p_{12})_{\bar{\phi}}\right)\mathcal{A}^\text{photon}_6\left(-l_{\bar{\phi}},(l+p_{12})_\phi,3^+_\gamma,4^+_\gamma,5^+_\gamma,6^-_\gamma\right).
\end{align}
Thus the combined contributions to the integrand from both diagrams \eqref{white6} and \eqref{gray6} is verified to have the correct cuts.

The integration of (\ref{white6}) and (\ref{gray6}) can be carried out straightforwardly using the general results (\ref{eq:1-loop-n-gon}) and (\ref{eq:1-loop-n-gon-tensor})
\begin{align}
&\mathcal{A}^{\text{BI}_4\; \text{1-loop}}_6\left(1_\gamma^+,2_\gamma^+,3_\gamma^+,4_\gamma^+,5_\gamma^+,6_\gamma^-\right)\nonumber\\
&\hspace{5mm}= \frac{-i}{23040\pi^2} [12]^2[34]^2\langle 6|p_{125}|5]^2 \left(s_{56}+3s_{12}+3s_{34}-6\frac{s_{12}^2}{s_{125}}\right) + \mathcal{P}\left(1,2,3,4,5\right) +\mathcal{O}(\epsilon).
\end{align}
Unlike the cases we have seen so far, this expression is non-local. The factorization poles in the amplitude can be traced back to the non-local gray vertex factor and the associated set of gray loop diagrams. Calculating residues on these poles yields a 4-point SD amplitude times a Born-Infeld tree.

Finally we consider the all-multiplicity result in the NSD sector. Similar to the NSD 6-point example, there will be local contributions from diagrams containing a single white vertex:
\begin{center}
	{\begin{tikzpicture}[scale=1.5, line width=1 pt]
		\draw [scalar] (0,0)--(1,0);
		\draw [scalar] (1,0)--(1.8,-0.8);
		\draw [scalar] (1.8,-0.8)--(1.8,-1.8);
		\draw [scalar] (1.8,-1.8)--(1,-2.6);
		\draw [scalar] (1,-2.6)--(0,-2.6);
		\draw [scalar] (0,-2.6)--(-0.8,-1.8);
		\draw [scalar] (-0.8,-1.8)--(-0.8,-0.8);
		\draw [vector] (0,0)--(-0.7,0.7);
		\draw [vector] (0,0)--(0,1);
		\draw [vector] (1,0)--(1,1);
		\draw [vector] (1,0)--(1.7,0.7);
		\draw [vector] (1.8,-0.8)--(2.5,-0.1);
		\draw [vector] (1.8,-0.8)--(2.8,-0.8);
		\draw [vector] (1.8,-1.8)--(2.8,-1.8);
		\draw [vector] (1.8,-1.8)--(2.5,-2.5);
		\draw [vector] (1,-2.6)--(1.7,-3.3);
		\draw [vector] (1,-2.6)--(1,-3.6);
		\draw [vector] (0,-2.6)--(0,-3.6);
		\draw [vector] (0,-2.6)--(-0.8,-3.3);
		\draw [vector] (-0.8,-1.8)--(-1.5,-2.5);
		\draw [vector] (-0.8,-1.8)--(-1.8,-1.8);
		\draw [vector] (-0.8,-0.8)--(-1.8,-0.8);
		\draw [vector] (-0.8,-0.8)--(-1.5,-0.1);
		\draw[black,fill=white] (0,0) circle (1.5ex);
		\draw[black,fill=black] (1,0) circle (1.5ex);
		\draw[black,fill=black] (1.8,-0.8) circle (1.5ex);
		\draw[black,fill=black] (1.8,-1.8) circle (1.5ex);
		\draw[black,fill=black] (1,-2.6) circle (1.5ex);
		\draw[black,fill=black] (0,-2.6) circle (1.5ex);
		\draw[black,fill=black] (-0.8,-1.8) circle (1.5ex);
		\draw[black,fill=black] (-0.8,-0.8) circle (1.5ex);
		\node at (-0.75,0.1) {\Large$\iddots$};
		\node at (-0.8,0.9) {$\gamma^+$};
		\node at (0,1.2) {$\gamma^-$};
		\node at (1.1,1.2) {$\gamma^+$};
		\node at (1.9,0.9) {$\gamma^+$};
		\node at (2.7,0.1) {$\gamma^+$};
		\node at (3.05,-0.75) {$\gamma^+$};
		\node at (3.05,-1.75) {$\gamma^+$};
		\node at (2.7,-2.5) {$\gamma^+$};
		\node at (1.9,-3.4) {$\gamma^+$};
		\node at (1,-3.8) {$\gamma^+$};
		\node at (0,-3.8) {$\gamma^+$};
		\node at (-0.9,-3.5) {$\gamma^+$};
		\node at (-1.7,-2.6) {$\gamma^+$};
		\node at (-2,-1.8) {$\gamma^+$};
		\node at (-2,-0.8) {$\gamma^+$};
		\node at (-1.6,0) {$\gamma^+$};
		\end{tikzpicture}}
\end{center}
as well as non-local contributions from diagrams containing a single gray vertex:
\begin{center}
	{\begin{tikzpicture}[scale=1.5, line width=1 pt]
		\draw [scalar] (0,0)--(1,0);
		\draw [scalar] (1,0)--(1.8,-0.8);
		\draw [scalar] (1.8,-0.8)--(1.8,-1.8);
		\draw [scalar] (1.8,-1.8)--(1,-2.6);
		\draw [scalar] (1,-2.6)--(0,-2.6);
		\draw [scalar] (0,-2.6)--(-0.8,-1.8);
		\draw [scalar] (-0.8,-1.8)--(-0.8,-0.8);
		\draw [vector] (0,0)--(-0.8,0.6);
		\draw [vector] (0,0)--(0.3,1);
		\draw [vector] (0,0)--(-0.5,0.8);
		\draw [vector] (0,0)--(-0.1,1);
		\draw [vector] (1,0)--(1,1);
		\draw [vector] (1,0)--(1.7,0.7);
		\draw [vector] (1.8,-0.8)--(2.5,-0.1);
		\draw [vector] (1.8,-0.8)--(2.8,-0.8);
		\draw [vector] (1.8,-1.8)--(2.8,-1.8);
		\draw [vector] (1.8,-1.8)--(2.5,-2.5);
		\draw [vector] (1,-2.6)--(1.7,-3.3);
		\draw [vector] (1,-2.6)--(1,-3.6);
		\draw [vector] (0,-2.6)--(0,-3.6);
		\draw [vector] (0,-2.6)--(-0.8,-3.3);
		\draw [vector] (-0.8,-1.8)--(-1.5,-2.5);
		\draw [vector] (-0.8,-1.8)--(-1.8,-1.8);
		\draw [vector] (-0.8,-0.8)--(-1.8,-0.8);
		\draw [vector] (-0.8,-0.8)--(-1.5,-0.1);
		\draw[black,fill=lightgray] (0,0) circle (1.5ex);
		\draw[black,fill=black] (1,0) circle (1.5ex);
		\draw[black,fill=black] (1.8,-0.8) circle (1.5ex);
		\draw[black,fill=black] (1.8,-1.8) circle (1.5ex);
		\draw[black,fill=black] (1,-2.6) circle (1.5ex);
		\draw[black,fill=black] (0,-2.6) circle (1.5ex);
		\draw[black,fill=black] (-0.8,-1.8) circle (1.5ex);
		\draw[black,fill=black] (-0.8,-0.8) circle (1.5ex);
		\node at (-0.75,0.1) {\Large$\iddots$};
		\node at (-0.9,0.7) {$\gamma^+$};
		\node at (-0.55,1) {$\gamma^+$};
		\node at (-0.1,1.2) {$\gamma^+$};
		\node at (0.4,1.2) {$\gamma^-$};
		\node at (1.1,1.2) {$\gamma^+$};
		\node at (1.9,0.9) {$\gamma^+$};
		\node at (2.7,0.1) {$\gamma^+$};
		\node at (3.05,-0.75) {$\gamma^+$};
		\node at (3.05,-1.75) {$\gamma^+$};
		\node at (2.7,-2.5) {$\gamma^+$};
		\node at (1.9,-3.4) {$\gamma^+$};
		\node at (1,-3.8) {$\gamma^+$};
		\node at (0,-3.8) {$\gamma^+$};
		\node at (-0.9,-3.5) {$\gamma^+$};
		\node at (-1.7,-2.6) {$\gamma^+$};
		\node at (-2,-1.8) {$\gamma^+$};
		\node at (-2,-0.8) {$\gamma^+$};
		\node at (-1.6,0) {$\gamma^+$};
		\end{tikzpicture}}
\end{center}
The explicit contributions to the integrand are, respectively 
\begin{align} \label{NSDallw}
&\mathcal I^{\text{NSD}}_{2n}[l;\mu^2] \biggr|_{\text{white}}= - \left (- \frac 1 2 \right )^{n-1} [ 12 ]^2 \ldots[ 2n-3 \ 2n-2 ]^2  [ 2 n - 1|l| 2 n \rangle^2 \nonumber\\
&\hspace{30mm}\times \frac{\left ( \mu^2 \right )^{n-1}}{\prod_{i = 1}^n \left [ \left ( l - \sum_{j = 1}^{2i} p_j \right )^2 + \mu^2 \right ]} + \mathcal P ( 1,2,\ldots,2n-1 ),
\end{align}
and
\begin{align} \label{NSDallg}
& \mathcal I^{\text{NSD}}_{2n}[l;\mu^2] \biggr|_{\text{gray}} \nonumber \\
& \hspace{10mm} = - \left (- \frac 1 2 \right )^{n-1} \frac{[ 12 ]^2 \ldots[ 2n-3 \ 2n-2 ]^2  [ 2 n - 1|p_{2n}+p_{2n-2}+p_{2n-3}| 2 n \rangle^2}{s_{2n,2n-2,2n-3}} \nonumber\\ 
& \hspace{15mm}\times  \frac{\left ( \mu^2 \right )^{n-1}}{\prod_{i = 1}^{n-2} \left [ \left ( l - \sum_{j = 1}^{2i} p_j \right )^2 + \mu^2 \right ]\left(l-\sum_{j =1}^{2n}p_j\right)^2} + \mathcal P ( 1,2,\ldots,2n-1 )\,.
\end{align}
Integrating these contributions separately using (\ref{eq:1-loop-n-gon}) and (\ref{eq:1-loop-n-gon-tensor}) gives the result
\begin{equation}
    \label{NSDallamp}
    \boxed{
    \begin{aligned}
        &\mathcal{A}_{2n}^{\text{BI}_4\;\text{1-loop}}\left(1_\gamma^+,2_\gamma^+,\ldots ,(2n-1)_\gamma^+,2n_\gamma^-\right) = \\
        &\ \mathcal{A}_{2n}^{\text{BI}_4\;\text{1-loop}}\left(1_\gamma^+,2_\gamma^+,\ldots ,(2n-1)_\gamma^+,2n_\gamma^-\right) \biggr\vert_{\text{white}}+\mathcal{A}_{2n}^{\text{BI}_4\;\text{1-loop}}\left(1_\gamma^+,2_\gamma^+,\ldots ,(2n-1)_\gamma^+,2n_\gamma^-\right)\biggr\vert_{\text{gray}},
    \end{aligned}
    }
\end{equation}
where
\begin{equation}
    \boxed{
    \begin{aligned}
        &\mathcal A_{2 n}^{\text{BI}_4\;\text{1-loop}}\left(1_\gamma^+,2_\gamma^+,\ldots ,(2n-1)_\gamma^+,2n_\gamma^-\right)\biggr\vert_{\text{white}} \\
        &\hspace{10mm}= \frac{-i}{16 \pi^2} \left( - \frac 1 2 \right )^{n-1} \frac{1}{(n-1)n ( n + 1 ) ( n + 2 ) ( n + 3 )}[ 12 ]^2 \ldots[ 2n-3 \ 2n-2 ]^2 \\
        &\hspace{15mm}\times \sum_{i < j}^n \left(\sum_{m = 2i + 1}^{2j} p_m \right )^2 \Bigg[ \sum_{k < l}^n 2\ a_{i j k l} \left(\sum_{m = 1}^{2k} [ 2 n - 1|p_m| 2 n \rangle\right)\left(\sum_{m = 1}^{2l} [ 2 n - 1|p_m| 2 n \rangle\right) \\
        &\hspace{20mm} + \sum_{k=1}^n b_{ijk} \left(\sum_{m = 1}^{2k} [ 2 n - 1|p_m| 2 n \rangle\right)^2 \Bigg] + \mathcal P ( 1,2,\ldots,2n-1 )+\mathcal{O}(\epsilon),
    \end{aligned}
    }
\end{equation}
with
\begin{equation}
b_{i j k} = \left \{ \begin{array}{ll}
2 & \qquad \text{if $i\ne k$ and $j\ne k$} \\
6 & \qquad \text{if $i=k$ or $j=k$}
\end{array} \right. .
\end{equation}
And also 
\begin{equation}
    \label{NSDallamp_grey}
    \boxed{
    \begin{aligned}
        &\mathcal A_{2 n}^{\text{BI}_4\; \text{1-loop}}\left(1_\gamma^+,2_\gamma^+,\ldots ,(2n-1)_\gamma^+,2n_\gamma^-\right)\biggr\vert_{\text{gray}}\\
        & = \frac{i}{32 \pi^2}\frac{( n - 2 )!}{( n + 2 )!}\left (- \frac 1 2 \right )^{n-1} \frac{[ 12 ]^2 \ldots[ 2n-3 \ 2n-2 ]^2  [ 2 n - 1|p_{2n-2}+p_{2n-3}| 2 n \rangle^2}{s_{2n,2n-2,2n-3}} \\
        &\times \left[\sum_{i < j}^{n-2} \sum_{k < l}^{n-2} a_{i j k l} \left(\sum_{m = 2i + 1}^{2j} p_m \right )^2 \left(\sum_{m = 2k + 1}^{2l} p_m \right )^2+4\sum_{i \leq j}^{n-2} \left(\sum_{m = 1}^{2i} p_m \right )^2 \left(\sum_{m = 1}^{2j} p_m \right )^2\right. \\
        &\left.+2\sum_{i =1}^{n-2}\sum_{k < l}^{n-2}a_{i (n-1) k l} \left(\sum_{m = 1}^{2i} p_m \right )^2 \left(\sum_{m = 2k + 1}^{2l} p_m \right )^2\right] + \mathcal P ( 1,2,\ldots,2n-1 ) +\mathcal{O}(\epsilon) .
    \end{aligned}
    }
\end{equation}
It is easy to check that these generic result match the cases of $n = 2$ and $n = 3$ that were presented above.

As we have already discussed for the 6-particle case, the NSD $( 2 n )$-particle amplitudes we calculate have poles that can be traced back to the associated poles of the gray vertex factors for $n \geq 3$.
These poles are located at $s_{i,j,2 n} = 0$, for $i < j \leq 2 n - 1$, and the associated residues are products of the tree 4-particle amplitude and a SD $( 2 n - 2 )$-particle amplitude of the form \eqref{SDallamp}.
Let us now demonstrate this factorization explicitly.
Consider for example the residue of \eqref{NSDallamp} at $s_{2n-2,2n-1,2n} = 0$,
\begin{align}
	& \operatorname*{Res}_{p_f^2 = 0} \mathcal A_{2 n}^{\text{BI}_4\; \text{1-loop}}\left(1_\gamma^+,2_\gamma^+,\ldots ,(2n-1)_\gamma^+,2n_\gamma^-\right) \nonumber\\
    & = 2 \frac{1}{32 \pi^2}\frac{( n - 2 )!}{( n + 2 )!}\left (- \frac 1 2 \right )^{n-1} [ 12 ]^2 \ldots[ 2n-5 \ 2n-4 ]^2 [ 2n-2 \ 2n-1 ]^2  [ 2 n - 3|p_f| 2 n \rangle^2 \nonumber\\
	&\times \left[\sum_{i < j}^{n-2} \sum_{k < l}^{n-2} a_{i j k l} \left(\sum_{m = 2i + 1}^{2j} p_m \right )^2 \left(\sum_{m = 2k + 1}^{2l} p_m \right )^2+4\sum_{i \leq j}^{n-2} \left(\sum_{m = 1}^{2i} p_m \right )^2 \left(\sum_{m = 1}^{2j} p_m \right )^2\right. \nonumber\\
	&\left.+2\sum_{i =1}^{n-2}\sum_{k < l}^{n-2}a_{i (n-1) k l} \left(\sum_{m = 1}^{2i} p_m \right )^2 \left(\sum_{m = 2k + 1}^{2l} p_m \right )^2\right] + \mathcal P ( 1,2,\ldots,2n-3 ) +\mathcal{O}(\epsilon)\,,
\end{align}
where $p_f = p_{2n-2} + p_{2n-1} + p_{2n}$ is the momentum on the factorization channel.
Notice that not all permutations listed in \eqref{NSDallamp_grey} contribute to the residue while the additional factor of 2 in the right-hand side comes from the trivial permutation $2n-2 \leftrightarrow 2n-1$.
Now, on the factorization channel
\begin{equation}
    [ 2n - 3 | p_f | 2n \rangle = - [ 2n - 3, p_f] \langle p_f, 2n \rangle = - i [ 2n - 3, p_f] \langle -p_f, 2n \rangle\,.
\end{equation}
Also, we can use momentum conservation to write
\begin{equation}
    \sum_{m = 1}^{2 i} p_m = - p_f - \sum_{m = 2 i + 1}^{2n - 3} p_m = - \sum_{m = 2 i + 1}^{2n - 2} \tilde p_m\,,
\end{equation}
where we have defined
\begin{equation}
    \tilde p_m = \left \{ \begin{array}{cc}
        p_m & \text{if } m \leq 2n-3 \\
        p_f & \text{if } m = 2n-2 
    \end{array} \right .
\end{equation}
With this definition we can write the above residue as
\begin{align}
	& \operatorname*{Res}_{p_f^2 = 0} \mathcal A_{2 n}^{\text{BI}_4\; \text{1-loop}}\left(1_\gamma^+,2_\gamma^+,\ldots ,(2n-1)_\gamma^+,2n_\gamma^-\right) = \left ( [ 2n-2, 2n-1 ]^2 \langle -p_f, 2 n \rangle^2 \right ) \nonumber \\
	& \times \Bigg [ \frac{1}{32 \pi^2}\frac{( n - 2 )!}{( n + 2 )!}\left (- \frac 1 2 \right )^{n-2} [ 12 ]^2 \ldots[ 2n-5, 2n-4 ]^2 [ 2 n - 3, p_f ]^2 \nonumber \\
	& \hspace{10mm} \times \sum_{i < j}^{n-1} \sum_{k < l}^{n-1} a_{i j k l} \left(\sum_{m = 2i + 1}^{2j} \tilde p_m \right )^2 \left(\sum_{m = 2k + 1}^{2l} \tilde p_m \right )^2 + \mathcal P ( 1,2,\ldots,2n-3 ) +\mathcal{O}(\epsilon) \Bigg ]\,,
\end{align}
which clearly shows its factorized form.
More precisely, we can write
\begin{multline}
    \operatorname*{Res}_{p_f^2 = 0} \mathcal A_{2 n}^{\text{BI}_4\; \text{1-loop}}\left(1_\gamma^+,\ldots ,(2n-1)_\gamma^+,2n_\gamma^-\right) \\
    = \mathcal A_{2n - 2}^{\text{BI}_4\; \text{1-loop}} \left ( 1_\gamma^+,\ldots , ( 2 n - 3 )_\gamma^+, \left ( p_f \right )_\gamma^+ \right ) \times \mathcal A_4^{\text{BI}_4} \left ( \left ( - p_f \right )_\gamma^-, ( 2 n - 2 )_\gamma^+, ( 2 n - 1 )_\gamma^+, ( 2 n )_\gamma^- \right ).
\end{multline}
The fact that the pole terms of the NSD 1-loop amplitude factorize to a SD 1-loop and a tree-level MHV amplitude at all multiplicities means that if we choose to remove the SD amplitudes by introducing finite local counter-terms, then the NSD amplitudes become local and can also be set to zero with the introduction of further finite local counter-terms.
%he
The consequences will be discussed in the next section. 

%%%%%%%%%%%%%%%%%%%%%%%%%%%%%%
\section{Discussion}
\label{sec:out}
%%%%%%%%%%%%%%%%%%%%%%%%%%%%%%

The main results of this paper are (\ref{SDallamp}) and (\ref{NSDallamp}), explicit expressions for the SD and NSD amplitudes at one-loop that would have been impossible to obtain by using traditional Feynman diagrammatics. As expected, they are finite and at $\mathcal{O}(\epsilon^0)$ given by rational functions. For the SD and NSD sectors, these properties follow from the property of $\text{BI}_4$ being a consistent truncation of a supersymmetric model at tree-level. More generally however, we expect both of these properties to obtain in all helicity sectors \textit{except} the duality-conserving sector 
\begin{equation}
	\mathcal{A}_{2n}^{\text{BI}_4}\left(1_\gamma^+,\ldots,n_\gamma^+, (n+1)_\gamma^-,
    \ldots,
    (2n)_\gamma^-\right).
\end{equation}
As a consequence of an electromagnetic duality symmetry, these amplitudes which conserve a chiral charge for the photon are the only non-vanishing amplitudes at tree-level \cite{Rosly:2002jt, Boels:2008fc}. At one-loop, only amplitudes in the duality-conserving sector can have non-vanishing 4d cuts and consequently non-rational functional dependence. 

The methods of this paper do not directly extend to calculations at one-loop beyond the SD and NSD sectors. In a sense then we have explored only a small fraction of the structure of Born-Infeld at one-loop. At higher multiplicity the majority of non-duality-conserving sectors, which are expected to be rational, cannot be calculated by constructing integrands from massive scalar cuts. In the duality-conserving sector, the cut-constructible parts can be obtained using the non-vanishing 4d cuts, this will be explored in detail in a separate paper.

Having explicit forms for two infinite classes of duality-violating one-loop amplitudes, we are in a position to make an interesting observation about the fate of electromagnetic duality at the one-loop quantum-level 
(see \cite{Novotny:2018iph} for recent discussion). Recall that this is not a symmetry in the usual sense. If we insist on defining the quantum theory as a path integral weighted by the exponential factor $e^{iS}$, where $S$ is the manifestly Lorentz-invariant effective action (\ref{BIaction}), then the $U(1)$ electromagnetic duality acts on the field strength as a symmetry of the equations of motion, but \textit{not} as a symmetry of the action \cite{Gibbons:1995cv}. Alternatively, it is possible to begin with a \textit{classically} equivalent action which is invariant under duality rotations following the approach of Schwarz and Sen \cite{Schwarz:1993vs,Berman:1997iz}, at the price of sacrificing manifest Lorentz invariance. A closely related problem is that, despite the fact that duality is an \textit{ungauged} global symmetry, the Weinberg-Witten theorem forbids the existence of a conserved current as a well-defined local operator \cite{Weinberg:1980kq}. Given this state of affairs,
it is unsettled if it is possible to define a quantization of Born-Infeld electrodynamics that preserves duality, that is, it is not clear if such a symmetry is anomalous. In particular, it is unclear if it is possible to define the S-matrix at loop-level which respects the helicity selection rules associated with the conservation of duality charge. In a related context, recent explicit calculations in $\mathcal{N}=4$ supergravity in $d=4$ have revealed that the conventional understanding of chiral anomalies may be modified in the context of duality symmetries \cite{Bern:2017rjw}.

Determining if our explicit results are consistent with the existence of such a duality-respecting quantization is a little subtle. It is too naive to simply observe that the duality-violating one-loop amplitudes (\ref{SDallamp}) and (\ref{NSDallamp}) are non-zero. Similar to $U(1)$ symmetries acting on chiral fermions, duality rotations act as chiral rotations on states of spin-1, and are therefore only defined in exactly 4-dimensions. Our explicit results however were obtained in a dimensional regularization scheme which explicitly breaks the symmetry. To determine if a genuine anomaly is present, we must first recall that the classical action used to define the full quantum theory as a path integral is ambiguous up to the addition of finite local counterterms. If a consistent set of local, Lorentz-invariant counterterms can be added to the action such that their contribution \textit{cancels} the explicitly calculated rational one-loop amplitudes, then there is no anomaly and the symmetry is preserved. In the SD sector the expressions (\ref{SDallamp}) are manifestly local and Lorentz-invariant, and so can be consistently cancelled by local counterterms. In the NSD sector the expressions (\ref{SDallamp}) are non-local, here we must sum over both contact contributions from independent local operators and factoring contributions containing both counterterms and tree-level Born-Infeld vertices. The condition that these non-local contributions can be removed with finite local counterterms requires that our explicit results (\ref{SDallamp}) have the singularity and factorization properties of tree-amplitudes, and 
we verified this explicitly at the end of Section \ref{sec:NSD}. The structure of the local counterterms will be discussed further in a separate paper.

These results give an infinite number of non-trivial checks on the preservation of duality under quantization, but do not constitute a proof. Extending the results of this paper to the remaining duality-violating sectors and beyond is therefore essential to understanding the ultimate fate of electromagnetic duality in quantum Born-Infeld.

%%%%%%%%%%%%%%%%%%%%%%%%%%%%%%

\section*{Acknowledgements} 

%%%%%%%%%%%%%%%%%%%%%%%%%%%%%%

We would like to thank Aidan Herderschee, Julio Parra-Martinez, and Jaroslav Trnka for 
useful discussions. HE would like to thank the Niels Bohr International Academy for hospitality during the final stages of the project. 
This work was supported in part by the US 
Department of Energy under Grant No.~DE-SC0007859. SP had support from 
a Leinweber Summer Award. CRTJ and MH were supported in part by 
Leinweber Student Fellowships and in part by Rackham Predoctoral 
Fellowships from the University of Michigan. 

%%%%%%%%%%%%%%%%%%%%%%%%%%%%%%
%%%%%%%%%%%%%%%%%%%%%%%%%%%%%%
\appendix

%%%%%%%%%%%%%%%%%%%%%%%%%%%%%%
\section{Structure of Contact Terms}
\label{app:Contact}
%%%%%%%%%%%%%%%%%%%%%%%%%%%%%%

In Section \ref{sec:Structure} we argued, by a combination of dimensional analysis, little group scaling and requiring vanishing as $\mu^2\rightarrow 0$, that contact terms could appear in the $\text{mDBI}_4$ amplitudes in the NSD sector in the form of some contraction of the form
\begin{equation}
  \mathcal{A}_{n}^{\text{mDBI}_4}\left(1_\phi,2_\gamma^+,\ldots ,(n-1)_\gamma^+,n_{\overline{\phi}}\right)\biggr\vert_{\text{contact}} \sim \mu^2 |2]^2|3]^2\ldots |n-1]^2,
\end{equation}
where $n$ is even. In this appendix we will give a short proof that there is a unique such contact term for each $n$. We begin by noting that any candidate term has the form of a sum over terms where each term is a sum over cyclic contractions of the spinors. For example for $n=12$ typical terms might have the form
\begin{equation} \label{even}
  \left([23][34][45][56][67][72]\right)\left([89][9,10][10,11][11,8]\right),
\end{equation}
or 
\begin{equation}\label{odd}
  \left([23][34][42]\right) \left([56][67][75]\right)\left([89][9,10][10,11][11,8]\right).
\end{equation}

Neither term by itself is a candidate contact term since it does not have the appropriate Bose symmetry. We should take expression (\ref{even}) and symmetrize over each pair of spinors, beginning with 3 and 4 gives 
\begin{equation} 
  \left([23][34][45]+[24][43][35]\right)[56][67][72]\left([89][9,10][10,11][11,8]\right),
\end{equation}
applying the Schouten identity then gives
\begin{equation}
  = -[34]^2\left([25][56][67][72]\right)\left([89][9,10][10,11][11,8]\right).
\end{equation}
This has reduced a cyclic contraction of length 6 to a product of cyclic contractions of \textit{strictly shorter} length. By Bose symmetrizing over all pairs of spinors we can reduce any possible contact term to a sum over product of cyclic contractions of length 2. Terms such as (\ref{odd}) with odd cyclic contractions vanish after Bose symmetrization. The final expression then has the unique form 
\begin{equation}
  \mathcal{A}_{n}^{\text{mDBI}_4}\left(1_\phi,2_\gamma^+,\ldots ,(n-1)_\gamma^+,n_{\overline{\phi}}\right)\biggr\vert_{\text{contact}} = c_n\mu^2 \left([23]^2[45]^2\ldots [n-2,n-1]^2+\ldots \right),
\end{equation}
where $+\ldots $ denotes the sum over all ways of partitioning the set $\{2,\ldots ,n\}$ into subsets of length 2. This completes the proof that there is a unique possible contact term at each multiplicity. 

%%%%%%%%%%%%%%%%%%%%%%%%%%%%%%
\section{T-Duality Constraints on 8-point Amplitudes}
\label{app:Tdual8}
%%%%%%%%%%%%%%%%%%%%%%%%%%%%%%

Following our discussion in Section \ref{sec:Tdual}, we now investigate how T-duality constrains the 8-point amplitudes in $\text{mDBI}_4$. Begin with the dimensional reduction followed by the soft limit of particle 7 for the NSD 8-point $\text{mDBI}_4$ Ansatz
\begin{align}
  &\mathcal{A}_8^{\text{mDBI}_4}\left(1_\phi,2_\gamma^+,3_\gamma^+,4_\gamma^+,5_\gamma^+,6_\gamma^+,7_\gamma^+, 8_{\overline{\phi}}\right) \nonumber\\
  &\xrightarrow[]{3d+\text{soft}} -\frac{1}{8}\left[\frac{2(\mu^2)^3s_{23}s_{45}(p_6\cdot p_7)}{(s_{123}+\mu^2)(s_{68}+\mu^2)}\right] + c_8\mu^2s_{23}s_{45}s_{67} +\mathcal{P}\left(2,3,4,5,6,7\right).
\end{align}
The MHV amplitude has a more complicated structure, there are more factorization graphs which are not related by permutations of external lines. Explicitly
 
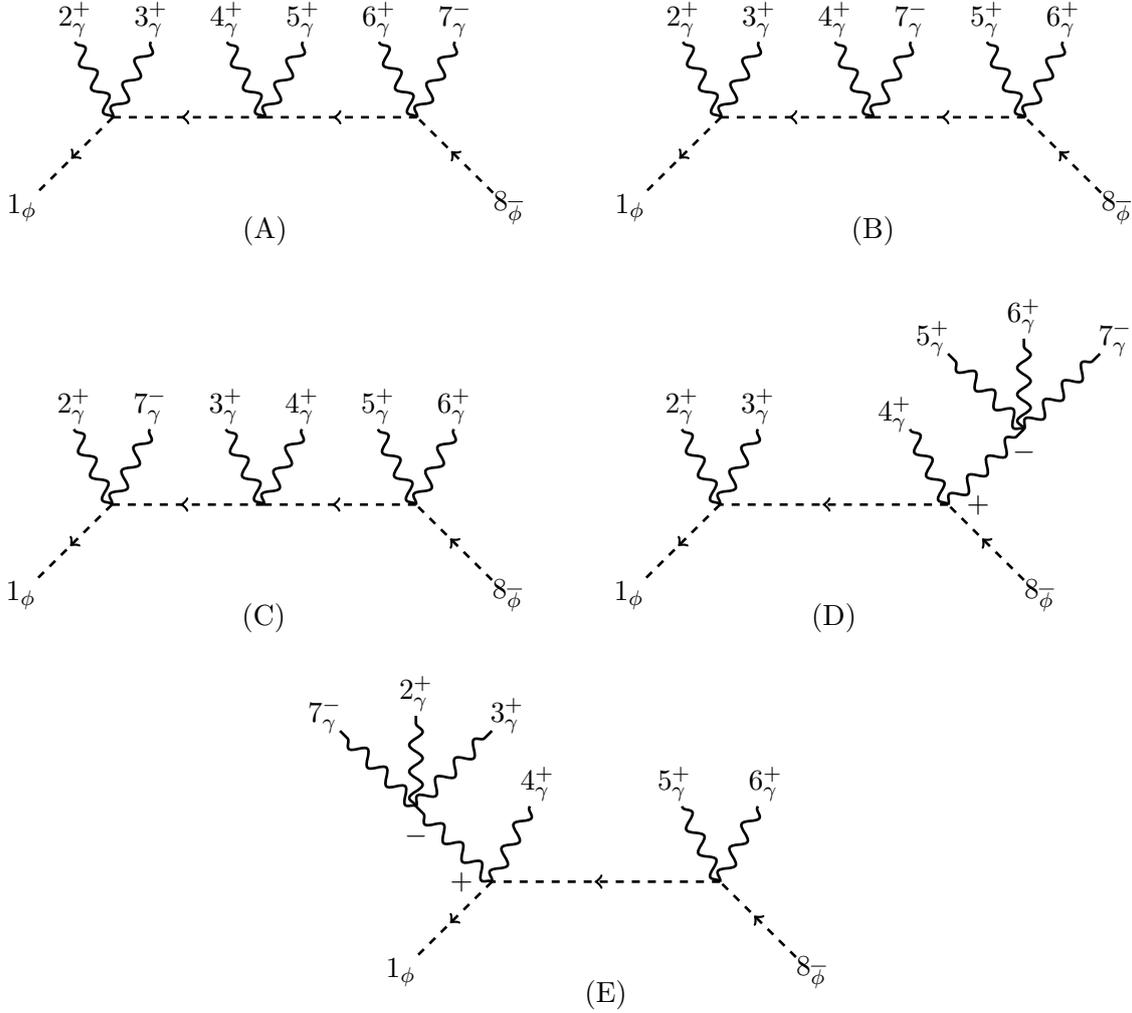
\begin{center}
{\begin{tikzpicture}[scale=1, line width=1 pt]
    \draw[scalar] (0,0)--(-1,-1);
    \draw[scalar] (2,0)--(0,0);
    \draw[scalar] (4,0)--(2,0);
    \draw[scalar] (5,-1)--(4,0);
    \draw[vector] (0,0)--(-0.5,1);
    \draw[vector] (0,0)--(0.5,1);
    \draw[vector] (2,0)--(1.5,1);
    \draw[vector] (2,0)--(2.5,1);
    \draw[vector] (4,0)--(3.5,1);
    \draw[vector] (4,0)--(4.5,1);
    \node at (-1.2,-1.2) {$1_\phi$};
    \node at (-0.5,1.3) {$2_\gamma^+$};
    \node at (0.5,1.3) {$3_\gamma^+$};
    \node at (1.5,1.3) {$4_\gamma^+$};
    \node at (2.5,1.3) {$5_\gamma^+$};
    \node at (3.5,1.3) {$6_\gamma^+$};
    \node at (4.5,1.3) {$7_\gamma^-$};
    \node at (5.2,-1.2) {$8_{\overline{\phi}}$};
    \node at (2,-1.5) {(A)};
    \begin{scope}[shift={(8,0)}]
    \draw[scalar] (0,0)--(-1,-1);
    \draw[scalar] (2,0)--(0,0);
    \draw[scalar] (4,0)--(2,0);
    \draw[scalar] (5,-1)--(4,0);
    \draw[vector] (0,0)--(-0.5,1);
    \draw[vector] (0,0)--(0.5,1);
    \draw[vector] (2,0)--(1.5,1);
    \draw[vector] (2,0)--(2.5,1);
    \draw[vector] (4,0)--(3.5,1);
    \draw[vector] (4,0)--(4.5,1);
    \node at (-1.2,-1.2) {$1_\phi$};
    \node at (-0.5,1.3) {$2_\gamma^+$};
    \node at (0.5,1.3) {$3_\gamma^+$};
    \node at (1.5,1.3) {$4_\gamma^+$};
    \node at (2.5,1.3) {$7_\gamma^-$};
    \node at (3.5,1.3) {$5_\gamma^+$};
    \node at (4.5,1.3) {$6_\gamma^+$};
    \node at (5.2,-1.2) {$8_{\overline{\phi}}$};
    \node at (2,-1.5) {(B)};
  \end{scope}
  \end{tikzpicture}}
\end{center}
\begin{center}
  \begin{tikzpicture} [scale=1, line width=1 pt]
    \draw[scalar] (0,0)--(-1,-1);
    \draw[scalar] (2,0)--(0,0);
    \draw[scalar] (4,0)--(2,0);
    \draw[scalar] (5,-1)--(4,0);
    \draw[vector] (0,0)--(-0.5,1);
    \draw[vector] (0,0)--(0.5,1);
    \draw[vector] (2,0)--(1.5,1);
    \draw[vector] (2,0)--(2.5,1);
    \draw[vector] (4,0)--(3.5,1);
    \draw[vector] (4,0)--(4.5,1);
    \node at (-1.2,-1.2) {$1_\phi$};
    \node at (-0.5,1.3) {$2_\gamma^+$};
    \node at (0.5,1.3) {$7_\gamma^-$};
    \node at (1.5,1.3) {$3_\gamma^+$};
    \node at (2.5,1.3) {$4_\gamma^+$};
    \node at (3.5,1.3) {$5_\gamma^+$};
    \node at (4.5,1.3) {$6_\gamma^+$};
    \node at (5.2,-1.2) {$8_{\overline{\phi}}$};
    \node at (2,-1.5) {(C)};
  \begin{scope}[shift={(8,0)}]
    \draw[scalar] (0,0)--(-1,-1);
    \draw[scalar] (3,0)--(0,0);
    \draw[scalar] (4,-1)--(3,0);
    \draw[vector] (0,0)--(-0.5,1);
    \draw[vector] (0,0)--(0.5,1);
    \draw[vector] (3,0)--(2.5,1);
    \draw[vector] (3,0)--(4,1);
    \draw[vector] (4,1)--(3,2);
    \draw[vector] (4,1)--(4,2.2);
    \draw[vector] (4,1)--(5,2);
    \node at (-1.2,-1.2) {$1_\phi$};
    \node at (-0.5,1.3) {$2_\gamma^+$};
    \node at (0.5,1.3) {$3_\gamma^+$};
    \node at (4.2,-1.2) {$8_{\overline{\phi}}$};
    \node at (2.3,1.2) {$4_\gamma^+$};
    \node at (2.8,2.2) {$5_\gamma^+$};
    \node at (4,2.5) {$6_\gamma^+$};
    \node at (5.2,2.2) {$7_\gamma^-$};
    \node at (3.4,0) {$+$};
    \node at (4,0.7) {$-$};
    \node at (1.5,-1.5) {(D)};
    \end{scope}
    \begin{scope}[shift={(5,-5)}]
    \draw[scalar] (0,0)--(-1,-1);
    \draw[scalar] (3,0)--(0,0);
    \draw[scalar] (4,-1)--(3,0);
    \draw[vector] (3,0)--(2.5,1);
    \draw[vector] (3,0)--(3.5,1);
    \draw[vector] (0,0)--(-1,1);
    \draw[vector] (0,0)--(0.5,1);
    \draw[vector] (-1,1)--(-2,2);
    \draw[vector] (-1,1)--(-1,2.2);
    \draw[vector] (-1,1)--(0,2);
    \node at (-1.2,-1.2) {$1_\phi$};
    \node at (-2.2,2.2) {$7_\gamma^-$};
    \node at (-1,2.5) {$2_\gamma^+$};
    \node at (0.2,2.2) {$3_\gamma^+$};
    \node at (0.6,1.3) {$4_\gamma^+$};
    \node at (2.4,1.3) {$5_\gamma^+$};
    \node at (3.6,1.3) {$6_\gamma^+$};
    \node at (-0.4,0) {$+$};
    \node at (-1,0.6) {$-$};
    \node at (4.2,-1.2) {$8_{\overline{\phi}}$};
    \node at (1.5,-1.5) {(E)};
    \end{scope}
  \end{tikzpicture}
\end{center}
In this topological decomposition the amplitude has the form
\begin{align}
  &\mathcal{A}_8^{\text{mDBI}_4}\left(1_\phi,2_\gamma^+,3_\gamma^+,4_\gamma^+,5_\gamma^+,6_\gamma^+,7_\gamma^-,8_{\overline{\phi}}\right) \nonumber \\
& \hspace{10mm}= \mathcal{A}_{8(\text{A})}^{\text{mDBI}_4} + \mathcal{A}_{8(\text{B})}^{\text{mDBI}_4} + \mathcal{A}_{8(\text{C})}^{\text{mDBI}_4} + \mathcal{A}_{8(\text{D})}^{\text{mDBI}_4} + \mathcal{A}_{8(\text{E})}^{\text{mDBI}_4} ,
\end{align}
where 
\begin{align}
  &\mathcal{A}_{8(\text{A})}^{\text{mDBI}_4} \xrightarrow[]{3d+\text{soft}} \frac{-(\mu^2)^2s_{23}s_{45}\left(2(p_7\cdot p_8)(s_{68}+\mu^2) + 2\mu^2 (p_7\cdot p_6)\right)}{(s_{123}+\mu^2)(s_{68}+\mu^2)} +\ldots  \\
  &\mathcal{A}_{8(\text{B})}^{\text{mDBI}_4} \xrightarrow[]{3d+\text{soft}} \frac{-(\mu^2)^2s_{23}s_{56}\left(4(p_7\cdot p_{123})(p_4\cdot p_{123})-2s_{123}(p_4\cdot p_{7})\right)}{(s_{123}+\mu^2)(s_{568}+\mu^2)} + \ldots  \\
  &\mathcal{A}_{8(\text{C})}^{\text{mDBI}_4} \xrightarrow[]{3d+\text{soft}} \frac{-(\mu^2)^2s_{34}s_{56}\left(2(p_7\cdot p_1)(s_{12}+\mu^2) + 2\mu^2(p_2\cdot p_7)\right)}{(s_{12}+\mu^2)(s_{568}+\mu^2)} + \ldots  \\
  &\mathcal{A}_{8(\text{D})}^{\text{mDBI}_4} \xrightarrow[]{3d+\text{soft}} -\frac{(\mu^2)^2s_{23}\left(4(p_7\cdot p_{56})(p_4\cdot p_{56})-2s_{56}(p_4\cdot p_7)\right)}{s_{123}+\mu^2} + \ldots  \\
  &\mathcal{A}_{8(\text{E})}^{\text{mDBI}_4} \xrightarrow[]{3d+\text{soft}} -\frac{(\mu^2)^2s_{56}\left(4(p_7\cdot p_{23})(p_4\cdot p_{23})-2s_{23}(p_4\cdot p_7)\right)}{s_{568}+\mu^2} + \ldots 
\end{align}
Here $+\ldots $ corresponds to summing over all topologically inequivalent relabelings of the positive helicity photons. Note that we do not include a contact contribution, as discussed in Appendix \ref{app:Contact}. 

From the singularity structure it is clear that diagrams A, B and C must cancel against the contribution of the NSD amplitude. For diagrams A and C it is easy to pick out the relevant pieces proportional to $(\mu^2)^3$. For diagram B this is a little less obvious and requires a little algebra first. The key idea is to recognize that there is something special about $p_4$ since it is the positive helicity particle in the middle of the diagram. We will see that something nice happens if we use momentum conservation and on-shellness to remove $p_4$ from the expression. That is we use
\begin{equation}
  p_4 = -p_{123}-p_{568},
\end{equation}
and the on-shell constraint
\begin{equation}
  p_4^2=0 \Rightarrow p_{123}\cdot p_{568} = -\frac{1}{2}\left(s_{123}+s_{568}\right).
\end{equation}
Using this on the numerator of B gives
\begin{align}
  &4(p_7\cdot p_{123})(p_4\cdot p_{123})-2s_{123}(p_4\cdot p_{7}) \nonumber\\
  &\hspace{10mm}= -2(p_7\cdot p_{123})\left(s_{123}-s_{568}\right)+2s_{123}(p_{123}\cdot p_{7}+p_{568}\cdot p_7) \nonumber\\
                                                                 &\hspace{10mm}= 2(p_7\cdot p_{123})(s_{568}+\mu^2) + 2(p_7\cdot p_{568})(s_{123}+\mu^2) +2\mu^2 (p_4\cdot p_7).
\end{align}
We can therefore more usefully rewrite B in the form
\begin{align}
  &\mathcal{A}_{8(\text{B})}^{\text{mDBI}}\left(1_\phi,2_\gamma^+,3_\gamma^+,4_\gamma^+,5_\gamma^+,6_\gamma^+,7_\gamma^-,8_{\overline{\phi}}\right) \nonumber\\
  &\hspace{5mm}\xrightarrow[]{3d+\text{soft}} \frac{-2(\mu^2)^3s_{23}s_{56}(p_4\cdot p_7)}{(s_{123}+\mu^2)(s_{568}+\mu^2)}-\frac{2(\mu^2)^2s_{23}s_{56}(p_7\cdot p_{123})}{s_{123}+\mu^2}-\frac{2(\mu^2)^2s_{23}s_{56}(p_7\cdot p_{568})}{s_{568}+\mu^2} + \ldots  
\end{align}
We now see explicitly that the non-local contributions from the MHV amplitude cancel completely. What remains is a sum of terms with only a single propagator. This is important since we want the remaining terms to cancel against each other, this couldn't happen unless some of the singularities disappeared upon dimensional reduction and soft limits since the topologically distinct graphs, by definition, have distinct singularity structure.

To finish the calculation we pick a singularity and verify that the sum of all contributions vanishes. Due to charge conjugation symmetry all such calculations are identical so we only need to verify a single case explicitly. We will choose the singularity associated with $s_{123}=-\mu^2$, this receives contributions from diagrams A, B and D. Summing the relevant terms
\begin{align}
  &-\frac{2(\mu^2)^2s_{23}s_{45}(p_8\cdot p_7)}{s_{123}+\mu^2}-\frac{2(\mu^2)^2s_{23}s_{56}(p_7\cdot p_{123})}{s_{123}+\mu^2}\nonumber\\
  &\hspace{10mm} -\frac{(\mu^2)^2s_{23}\left(4(p_7\cdot p_{56})(p_4\cdot p_{56})-2s_{56}(p_4\cdot p_7)\right)}{s_{123}+\mu^2} +\mathcal{C}\left(4,5,6\right) \nonumber\\
  &= -\frac{2(\mu^2)^2s_{23}s_{456}(p_8\cdot p_7)}{s_{123}+\mu^2}-\frac{2(\mu^2)^2s_{23}s_{456}(p_7\cdot p_{123})}{s_{123}+\mu^2}-\frac{2(\mu^2)^2s_{23}s_{456}(p_7\cdot p_{456})}{s_{123}+\mu^2} \nonumber\\
  &= 0.
\end{align}
As in the 6-point case we find that all of the factoring terms in the NSD and MHV $\text{mDBI}_4$ amplitudes cancel against each other and vanish in the T-dual soft configuration. Since the possible contact term is $\mathcal{O}\left(p_7\right)$, we must choose $c_8=0$ for compatibility with T-duality.

%%%%%%%%%%%%%%%%%%%%%%%%%%%%%%
\section{Evaluating Rational Integrals}
\label{app:Rational}
%%%%%%%%%%%%%%%%%%%%%%%%%%%%%%

A \textit{rational integral} in this context is defined as an integral in $d=4-2\epsilon$ dimensions, for which the integrand vanishes in $d=4$. A powerful and general method for evaluating these integrals was given in \cite{Bern:1995db} where the following \textit{dimension shifting formula} was derived
\begin{align} \label{shift}
  \int \frac{\text{d}^{4-2\epsilon}l}{(2\pi)^{4-2\epsilon}}(l_{-2\epsilon}^2)^p f(l) = (4\pi)^p\frac{\Gamma\left(-\epsilon+p\right)}{\Gamma\left(-\epsilon\right)} \int \frac{\text{d}^{4+2p-2\epsilon}l}{(2\pi)^{4+2p-2\epsilon}}f(l),
\end{align}
where $f(l)$ is some rational function of the $d$-dimensional loop momentum. This formula allows us to exchange integrals with explicit factors of $l_{-2\epsilon}^2$ for integrals without such factors evaluated in higher dimensions. The integral on the left-hand-side of (\ref{shift}) is formally defined as a tensor integral 
\begin{align}
  \int \frac{\text{d}^{4-2\epsilon}l}{(2\pi)^{4-2\epsilon}}(l_{-2\epsilon}^2)^p f(l) \equiv \left(\prod_{i=1}^{p}g^{[-2\epsilon]}_{\mu_i\nu_i}\right) \int \frac{\text{d}^{4-2\epsilon}l}{(2\pi)^{4-2\epsilon}}\left(\prod_{j=1}^p l^{\mu_j}l^{\nu_j}\right) f(l),
\end{align}
where $g_{\mu\nu}^{[-2\epsilon]}$ is the metric tensor projected onto the non-physical $-2\epsilon$-dimensional momentum subspace. The utility of the formula (\ref{shift}) is that it gives an efficient way to bypass calculating tensor reduction for integrands of arbitrarily high-rank; in this paper all integrals can be exchanged using this method to either scalar or rank-2 tensor integrals. Even with this simplification, obtaining explicit results to all orders in $\epsilon$ is a very difficult problem, for which only a small fraction of the necessary integrals are known. At $\mathcal{O}(\epsilon^0)$ however, the formula (\ref{shift}) simplifies significantly and the right-hand-side depends only on the \textit{divergent} part of the $d=4+2p-2\epsilon$-dimensional integral 
\begin{equation} \label{key}
  \int \frac{\text{d}^{4-2\epsilon}l}{(2\pi)^{4-2\epsilon}}(l_{-2\epsilon}^2)^p f(l) = -(p-1)!(4\pi)^p\left[\int \frac{\text{d}^{4+2p-2\epsilon}l}{(2\pi)^{4+2p-2\epsilon}}f\left(l\right)\right]_{1/\epsilon} + \mathcal{O}\left(\epsilon\right).
\end{equation}

This is the key formula for obtaining explicit expressions for one-loop rational integrals. As we will see below the simplification arises from the fact that after Feynman parametrization the divergent part of the integral can be extracted as the trivial integration of a polynomial in Feynman parameters. 

%%%%%%%%%%%%%%%%%%%%%%%%%%%%%%
\subsection{Rational Scalar \texorpdfstring{$n$-gon}{n-gon} Integral}
\label{app:Scalarn}
%%%%%%%%%%%%%%%%%%%%%%%%%%%%%%

In this section we present the explicit calculation of the rational scalar $n$-gon integral
\begin{equation} \label{scalarn}
  I^{d=4-2\epsilon}_n\left[(\mu^2)^n;\{p_i\}\right] \equiv \int \frac{\text{d}^{4 - 2 \epsilon} l}{( 2 \pi )^{4 - 2 \epsilon}} \frac{\left ( l_{- 2 \epsilon}^2 \right )^n}{\prod_{i = 1}^n \left ( l - \sum_{j = 1}^i p_j \right )^2},
\end{equation}
where the external momenta $p_i$ are massive. Using the dimension shifting formula (\ref{shift}) this is related to the massless scalar $n$-gon integral in $d=4+2n-2\epsilon$ dimensions
  \begin{equation}
       =\left ( 4 \pi \right )^n \frac{\Gamma ( n - \epsilon )}{\Gamma ( - \epsilon )} \int \frac{\text{d}^{4 + 2 n - 2 \epsilon} l}{( 2 \pi )^{4 + 2 n - 2 \epsilon}} \frac{1}{\prod_{i = 1}^n \left ( l - \sum_{j = 1}^i p_j \right )^2}.
  \end{equation}
  The next step is to use Feynman parametrization and write the integral as
  \begin{multline} \label{feyn}
      =\left ( 4 \pi \right )^n \frac{\Gamma ( n - \epsilon )}{\Gamma ( - \epsilon )} \left ( n - 1 \right )! \\
      \times \int_0^1 \text{d} x_1 \text{d} x_2 \ldots \text{d} x_n \left[\delta \left ( \sum_{i=1}^n x_i- 1 \right ) \int \frac{\text{d}^{4 + 2 n - 2 \epsilon} l}{( 2 \pi )^{4 + 2 n - 2 \epsilon}} \frac{1}{ \left [ \sum_{i = 1}^n x_i \left ( l - \sum_{j = 1}^i p_j \right )^2 \right ]^n}\right]\,.
  \end{multline}
  After shifting the loop momentum by $l \to l + \sum_{i = 1}^{n - 1} \sum_{j = 1}^i x_i p_j$ the denominator of the above integrand can be written as $\left [ l^2 + \Delta \right ]^n$ with
  \begin{align}
      \Delta & = \sum_{i = 1}^n x_i \left ( 1 - x_i \right ) \left ( \sum_{j = 1}^i p_j \right )^2 - 2 \sum_{i < j }^n x_i x_j \left ( \sum_{k = 1}^i p_k \right ) \cdot \left ( \sum_{k = 1}^j p_k \right ) \nonumber \\
      & = - \sum_{i = 1}^n x_i \left ( 1 - x_i \right ) \left ( \sum_{j = 1}^i p_j \right ) \cdot \left ( \sum_{j = i + 1}^n p_j \right ) + 2 \sum_{i < j }^n x_i x_j \left ( \sum_{k = 1}^i p_k \right ) \cdot \left ( \sum_{k = j + 1}^n p_k \right ) \nonumber \\
      & = - \sum_{i < j}^n p_i \cdot p_j \left ( \sum_{k = i}^{j - 1} x_k \right ) \left ( 1 - \sum_{k = i}^{j - 1} x_k \right )\,.
  \end{align}
  In the second line above, we used momentum conservation to write everything in terms of scalar products of two different momenta and in the third line, we rearranged the sums, writing explicitly the coefficient of each $p_i \cdot p_j$.
  To further simplify this, we substitute $1 = \sum_{i = 1}^n x_i$ and we collect the coefficients of each product $x_i x_j$,
  \begin{equation}
      \Delta = - \sum_{i < j}^n x_i x_j \left ( \sum_{k = i + 1}^j p_k \right ) \cdot \left ( \sum_{k = 1}^i p_k + \sum_{k = j + 1}^n p_k \right ) = \sum_{i < j}^n x_i x_j \left ( \sum_{k = i + 1}^j p_k \right )^2\,,
  \end{equation}
  where in the second step we used momentum conservation to write everything in terms of Mandelstam variables of adjacent momenta.
  Going back to (\ref{feyn}) and using the standard integral
\begin{equation} \label{standard}
   \int \frac{\text{d}^{4+2n-2\epsilon}l}{(2\pi)^{4+2n-2\epsilon}}\frac{1}{[l^2+\Delta]^n} = \frac{i}{(4\pi)^{n+2-\epsilon}}\frac{\Gamma\left(-2+\epsilon\right)}{(n-1)!}\Delta^{2-\epsilon},
\end{equation}
in full generality the rational integral (\ref{scalarn}) is given by the Feynman parameter integral
\begin{align}
  &I^{d=4-2\epsilon}_n\left[(\mu^2)^n;\{p_i\}\right] = \frac{i}{(4\pi)^{2-\epsilon}}\frac{\Gamma(n-\epsilon)\Gamma(-2+\epsilon)}{\Gamma(-\epsilon)}\nonumber\\
&\hspace{10mm}\times\int_0^1 \text{d}x_1\ldots \text{d}x_n\; \delta \left ( \sum_{i=1}^n x_i- 1 \right ) \left[\sum_{i < j}^n x_i x_j \left ( \sum_{k = i + 1}^j p_k \right )^2\right]^{2-\epsilon}.
\end{align}
Only in certain special cases ($n=2$ and $n=3$) is this integral known to all orders in $\epsilon$ \cite{Davydychev:1999mq}. The leading $\mathcal{O}\left(\epsilon^0\right)$ contribution however, can be calculated explicitly for all $n$. It is given by
\begin{equation}
  =-\frac{i}{32 \pi^2} ( n - 1 )! \int_0^1 \text{d} x_1 \text{d} x_2 \ldots \text{d} x_n \delta \left ( \sum_{i=1}^nx_i - 1 \right ) \left[\sum_{i < j}^n x_i x_j \left ( \sum_{k = i + 1}^j p_k \right )^2\right]^{2} + \mathcal{O}\left(\epsilon \right).
\end{equation}

  We now have to perform the integration over the $n$ Feynman parameters.
  For this we use the general formula
  \begin{equation}
      \int_0^1 \text{d} x_1 \text{d} x_2 \ldots \text{d} x_n \delta \left (\sum_{i=1}^n x_i  - 1 \right ) x_1^{r_1} x_2^{r_2} \ldots x_n^{r_n} = \frac{\Gamma \left ( 1 + r_1 \right ) \Gamma \left ( 1 + r_2 \right ) \ldots \Gamma \left ( 1 + r_n \right )}{\Gamma \left ( n + r_1 + r_2 + \ldots + r_n \right )}\,.
  \end{equation}
  Special instances of this formula that are relevant for the calculations of this and the next subsection are the following
  \begin{align}
      \label{Feynman1} \int_0^1 \text{d} x_1 \text{d} x_2 \ldots \text{d} x_n \delta \left (  \sum_{i=1}^nx_i - 1 \right ) x_1 x_2 x_3 x_4 & = \frac{1}{( n + 3 )!}\,, \\
      \label{Feynman2} \int_0^1 \text{d} x_1 \text{d} x_2 \ldots \text{d} x_n \delta \left (  \sum_{i=1}^nx_i - 1 \right ) x_1 x_2 x_3^2 & = \frac{2}{( n + 3 )!}\,, \\
      \label{Feynman3} \int_0^1 \text{d} x_1 \text{d} x_2 \ldots \text{d} x_n \delta \left (  \sum_{i=1}^nx_i - 1 \right ) x_1^2 x_2^2 & = \frac{4}{( n + 3 )!}\,, \\
      \label{Feynman4} \int_0^1 \text{d} x_1 \text{d} x_2 \ldots \text{d} x_n \delta \left (  \sum_{i=1}^nx_i - 1 \right ) x_1^3 x_2 & = \frac{6}{( n + 3 )!}\,,
  \end{align}
  With these, we find that the integrated result takes the form
  \begin{align}
  \label{eq:1-loop-n-gon}
      &I^{d=4-2\epsilon}_n\left[(\mu^2)^n;\{p_i\}\right] \nonumber\\
      &= -\frac{i}{32 \pi^2} \frac{1}{n ( n + 1 ) ( n + 2 ) ( n + 3 )} \sum_{i < j}^n \sum_{k < l}^n a_{i j k l} \left ( \sum_{m = i + 1}^j p_m \right )^2 \left ( \sum_{m = k + 1}^l p_m \right )^2  + \mathcal{O}\left(\epsilon \right),
  \end{align}
  where
  \begin{equation}
      a_{i j k l} = \left \{ \begin{array}{ll}
          1 & \qquad \text{if all $i,j,k,l$ are different} \\
          2 & \qquad \text{if exactly 2 of $i, j, k, l$ are identical} \\
          4 & \qquad \text{if $i = k$ and $j = l$}
      \end{array} \right .\,.
  \end{equation}

%%%%%%%%%%%%%%%%%%%%%%%%%%%%%%
\subsection{Rational Rank-2 Tensor \texorpdfstring{$n$-gon}{n-gon} Integral}
\label{app:Tensorn}
%%%%%%%%%%%%%%%%%%%%%%%%%%%%%%

Similar to the case of the rational scalar $n$-gon integral, we present the explicit calculation of the rational rank-2 tensor $n$-gon integral
\begin{equation} \label{tensorn}
  I_n^{d=4-2\epsilon}\left[\left(\mu^2\right)^{n-1}(u\cdot l)^2,\{p_i\}\right] \equiv \int \frac{\text{d}^{4 - 2 \epsilon} l}{( 2 \pi )^{4 - 2 \epsilon}} \frac{\left ( l_{- 2 \epsilon}^2 \right )^{n-1}(u\cdot l)^2}{\prod_{i = 1}^n \left ( l - \sum_{j = 1}^i p_j \right )^2},
\end{equation}
where $u^\mu$ is a 4-dimensional null vector. The dimension shifting formula (\ref{shift}) gives 
\begin{equation}
= \left ( 4 \pi \right )^{n-1} \frac{\Gamma ( n -1- \epsilon )}{\Gamma ( - \epsilon )} \int \frac{\text{d}^{2 + 2 n - 2 \epsilon} l}{( 2 \pi )^{2 + 2 n - 2 \epsilon}} \frac{(u\cdot l)^2}{\prod_{i = 1}^n \left ( l - \sum_{j = 1}^i p_j \right )^2}. 
\end{equation}
We can use the same Feynman parametrization trick as before to write the integral as
\begin{align}
 =\left ( 4 \pi \right )^{n-1} \frac{\Gamma ( n -1- \epsilon )}{\Gamma ( - \epsilon )} \left ( n - 1 \right )! &\int_0^1 \text{d} x_1 \text{d} x_2 \ldots \text{d} x_n \delta \left ( \sum_{i=1}^n x_i - 1 \right )\nonumber\\
& \times\int \frac{\text{d}^{2 + 2 n - 2 \epsilon} l}{( 2 \pi )^{2 + 2 n - 2 \epsilon}} \frac{(u\cdot l)^2}{ \left [ \sum_{i = 1}^n x_i \left ( l - \sum_{j = 1}^i p_j \right )^2 \right ]^n}\,.
\end{align}
After shifting the loop momentum by $l \to l + \sum_{i = 1}^{n - 1} \sum_{j = 1}^i x_i p_j$, we get
\begin{align}
=\left ( 4 \pi \right )^{n-1} \frac{\Gamma ( n -1- \epsilon )}{\Gamma ( - \epsilon )} &\left ( n - 1 \right )! \int_0^1 \text{d} x_1 \text{d} x_2 \ldots \text{d} x_n \delta \left ( \sum_{i=1}^nx_i - 1 \right )\nonumber\\
&\times \int \frac{\text{d}^{2 + 2 n - 2 \epsilon} l}{( 2 \pi )^{2 + 2 n - 2 \epsilon}} \frac{(u\cdot l)^2+ \left(\sum_{i = 1}^{n - 1} \sum_{j = 1}^i x_i (u\cdot p_j)\right)^2}{ \left [ l^2+\Delta\right ]^n},
\end{align}
where $\Delta= \sum_{i < j}^n x_i x_j \left ( \sum_{k = i + 1}^j p_k \right )^2$ as before and all cross-terms have been dropped since they are odd in $l$. The first term integrates to an expression proportional to $u^2$ which is zero by assumption. The remaining terms have the form of the standard integral (\ref{standard}), so we can give a general expression for (\ref{tensorn}) as a integral over Feynman parameters
\begin{align}
&I_n^{d=4-2\epsilon}\left[\left(\mu^2\right)^{n-1}(u\cdot l)^2,\{p_i\}\right] =\frac{i}{(4\pi)^{2-\epsilon}} \frac{\Gamma ( n -1- \epsilon ) \Gamma ( -1 + \epsilon )}{\Gamma ( - \epsilon )} \nonumber\\
                   &\times\int_0^1 \text{d} x_1 \text{d}x_2\ldots \text{d}x_n \delta \left ( \sum_{i=1}^nx_i - 1 \right ) \left(\sum_{i = 1}^{n - 1} \sum_{j = 1}^i x_i u\cdot p_j\right)^2 \left[\sum_{i < j}^n x_i x_j \left ( \sum_{k = i + 1}^j p_k \right )^2\right]^{1-\epsilon}.
\end{align}
As in the scalar case we can give explicit expressions for all $n$ at $\mathcal{O}\left(\epsilon^0\right)$, using the Feynman-parameter integrals \eqref{Feynman1} - \eqref{Feynman4}. With these, we find that the integrated result takes the form
\begin{multline}
\label{eq:1-loop-n-gon-tensor}
I_n^{d=4-2\epsilon}\left[\left(\mu^2\right)^{n-1}(u\cdot l)^2,\{p_i\}\right] = \frac{i}{16 \pi^2} \frac{1}{(n-1)n ( n + 1 ) ( n + 2 ) ( n + 3 )} \\
\times \sum_{i < j}^n \left(\sum_{m = i + 1}^j p_m \right )^2 \left[ \sum_{k < l}^n 2 a_{i j k l} \left(\sum_{m = 1}^k u\cdot p_m\right)\left(\sum_{m = 1}^l u\cdot p_m\right)+ \sum_{k=1}^n b_{ijk} \left(\sum_{m = 1}^k u\cdot p_m\right)^2 \right]\,,
\end{multline}
where $a_{ijkl}$ is as defined above and 
\begin{equation}
b_{i j k} = \left \{ \begin{array}{ll}
2 & \qquad \text{if $i\ne k$ and $j\ne k$} \\
6 & \qquad \text{if $i=k$ or $j=k$}
\end{array} \right .\,.
\end{equation}

%%%%%%%%%%%%%%%%%%%%%%%%%%%%%%
%%%%%%%%%%%%%%%%%%%%%%%%%%%%%%
\bibliographystyle{JHEP}
\bibliography{Unitarity.bib}

%%%%%%%%%%%%%%%%%%%%%%%%%%%%%%
%%%%%%%%%%%%%%%%%%%%%%%%%%%%%%

\end{document}